\documentclass[12pt]{iopart}
\usepackage{graphicx}
\usepackage[sort]{cite}
\begin{document}
\jl{19}
\title[$\mathcal{O}(N)$ Methods]{$\mathcal{O}(N)$ Methods in electronic structure calculations}
\author{D~R~Bowler$^{1,2,3}$ and T~Miyazaki$^{4}$}
\address{$^{1}$London Centre for Nanotechnology, UCL, 17-19 Gordon St,
  London WC1H 0AH, UK}
\address{$^{2}$Department of Physics \& Astronomy, UCL, Gower St, London WC1E
  6BT, UK}
\address{$^{3}$Thomas Young Centre, UCL, Gower St, London WC1E
  6BT, UK}
\address{$^{4}$National Institute for Materials Science, 1-2-1 Sengen,
  Tsukuba, Ibaraki 305-0047, JAPAN}
\eads{david.bowler@ucl.ac.uk}
\eads{MIYAZAKI.Tsuyoshi@nims.go.jp}
\begin{abstract}
  Linear scaling methods, or $\mathcal{O}(N)$ methods, have
  computational and memory requirements which scale linearly with the
  number of atoms in the system, $N$, in contrast to standard
  approaches which scale with the cube of the number of atoms.  These
  methods, which rely on the short-ranged nature of electronic
  structure, will allow accurate, \emph{ab initio} simulations of
  systems of unprecedented size.  
  The theory behind the locality of electronic structure is described
  and related to physical properties of systems to be modelled, along
  with a survey of recent developments in real-space methods which are
  important for efficient use of high performance computers.  
 The linear scaling methods proposed to date can be
  divided into seven different areas, and the applicability,
  efficiency and advantages of the methods proposed in these areas is
  then discussed.  The applications of linear scaling methods, as well
  as the implementations available as computer programs, are
  considered.  Finally, the prospects for and the challenges facing
  linear scaling methods are discussed.
\end{abstract}
\submitted
\maketitle

\section{Introduction}
\label{sec:introduction}

Electronic structure calculation methods based on the density functional theory (DFT) 
have been playing important roles in condensed matter physics for more than forty years. 
In the early stages, DFT calculations were employed mainly for the study of the 
electronic structure of simple solids, using a few atoms in a unit cell, 
with the use of periodic boundary conditions. 
Since then, there has been a huge effort to improve the accuracy and efficiency of the 
calculation techniques. 
In terms of efficiency, after the pioneering work by Car and Parrinello\cite{Car:1985mz}, the size of 
the target systems has increased dramatically and more and more examples of the DFT studies, 
especially on aperiodic systems like surface structures, have emerged. 
DFT calculations on systems containing hundreds of atoms are currently ubiquitous. 
As the system size for DFT calculations has become larger, the variety of materials and phenomena 
investigated by the method has increased.
The information of the total energy and atomic forces calculated by DFT methods can provide 
reliable data independently from experiments, and the methods are nowadays considered as one of 
the established research tools in many fields, like physics,
chemistry, materials science, and many others.
Recently, there have been DFT studies in the complex fields of nano-structured materials 
and biological systems.
In the study of these classes of materials, we need to treat systems containing at least 
thousands of atoms.
However, as is well known, once the number of atoms $N$ in a system reaches around 
one thousand, the cost of standard DFT calculations increases very rapidly as a cube of $N$. 
To overcome this problem, the methods known as linear-scaling or $\mathcal{O}(N)$ DFT methods have been developed\cite{Goedecker:1999pv}. 
The progress of these methods in the last ten to fifteen years is remarkable and 
the purpose of this review paper is to overview the recent progress of $\mathcal{O}(N)$ DFT methods.

We will start with an overview of the conventional DFT method and its advantages.
In the normal DFT approach, we solve for the Kohn-Sham (KS) orbitals $\Psi_{\nu k}(\mathbf{r})$,
which are the eigenstates of the KS equation\cite{Kohn:1965qf}.
\begin{equation}
\label{eq:80}
 \hat{H}^{\rm KS} \Psi_{\nu \mathbf{k}} (\mathbf{r}) = \left[-\frac{\hbar}{2m} \nabla^2 + V_{\rm ext}(\mathbf{r}) + V_H(\mathbf{r}) + V_{\rm XC}(\mathbf{r})\right] 
                                \Psi_{\nu \mathbf{k}} (\mathbf{r}) 
                             =  \epsilon_{\nu \mathbf{k}} \Psi_{\nu \mathbf{k}} (\mathbf{r}) 
\end{equation}
Here $\hat{H}^{\rm KS}$ is the Kohn-Sham Hamiltonian, and $\nu$ and $\mathbf{k}$ are the band index and $\mathbf{k}$ points 
in the first Brillouin zone, respectively.
Hereafter, we omit $\mathbf{k}$ for clarity because we consider large systems and the number of
$\mathbf{k}$ points is small.
$V_{\rm ext}(\mathbf{r})$ is the potential from nuclei, $V_H(\mathbf{r})$ is the Hartree potential, and $V_{\rm XC}(\mathbf{r})$
is the exchange-correlation potential in the Kohn-Sham formalism.
The most accurate DFT calculations often use a plane-wave basis set to
express the KS orbitals:
\begin{equation}
\label{eq:81}
 \Psi_{\nu} (\mathbf{r}) = \sum_{|\mathbf{G}| < G_{\rm max}} c_{\nu} (\mathbf{G}) \exp(i\mathbf{G}\cdot\mathbf{r})   
\end{equation}
A plane-wave basis set has two main advantages. 
First, the accuracy of the basis set can be systematically improved. 
In Eq.~(\ref{eq:81}), $G_{\rm max}$ is obtained from the cutoff energy $E_{\rm cut}$ as 
$\frac{\hbar^2 G_{\rm max}^2}{2m} = E_{\rm cut}$. 
The number of plane-waves, $N_G$, is controlled only by the number $E_{\rm cut}$.
The accuracy of the basis set can be improved simply by increasing $E_{\rm cut}$, and
a variational principle with respect to $E_{\rm cut}$ is satisfied.
The other advantage is that forces can be calculated easily without the Pulay 
correction term because the basis set is independent on atomic
positions (though such basis-set-dependent corrections become
necessary when changing the unit cell size or shape). 
These two advantages make it possible to calculate both energy and forces accurately
with  plane-wave basis sets. 

In order to realise accurate plane-wave calculations, we need to introduce several
theoretical techniques.
First of all, plane-wave calculations rely on the idea of pseudopotentials\cite{Hamann:1979zr}.
With this method, it is possible to work only with valence electrons
and their pseudo-wavefunctions, which are much smoother than the real
wave functions which oscillate 
strongly in the core region, and to replace the nuclear potential and
the core electrons with a pseudopotential.
There have been several kinds of techniques proposed to make
pseudo-wavefunctions smoother\cite{Bachelet:1982ly,Rappe:1990vn,Troullier:1991kx}. 
Using the method of ultra-soft pseudopotentials\cite{Vanderbilt:1990ys}, even the cutoff energy for 
the localised 3$d$ orbitals of transition metals can be reduced dramatically.
With these improvements in theoretical techniques, the total energy converges quickly with 
respect to the cutoff energy and this is essential to make the accurate DFT calculations feasible.
In addition, the major part of the error in the total energy usually comes from the expression of 
KS orbitals in the core region. 
Hence, the relative energetic stability of two states (e.g. two
different atomic structures) can be reproduced without the absolute convergence 
because most of the errors are cancelled in the energy difference. 
Note that it is also possible to reduce the number of plane-waves by using augmentation 
for the wavefunctions in the core region as in the linearised augmented plane-wave (LAPW)
or the projector augmented wave (PAW) method\cite{BlYochl:1994ve}.

It is essential that we reduce the number of plane-waves by introducing the 
pseudopotential or other similar techniques. 
However, even with very smooth pseudo wavefunctions, $N_G$ is typically one hundred times
larger than the number of electrons. 
When we want to diagonalise $<\mathbf{G}|H^{\rm KS}|\mathbf{G}'>$, the required memory scales as $\mathcal{O}(N_G^2)$ 
and CPU time as $\mathcal{O}(N_G^3)$.
Hence it is impossible to employ direct (exact) diagonalisation except for very small systems.
Instead of using exact diagonalisation, we can obtain the Kohn-Sham orbitals by minimising 
the DFT total energy with respect to the coefficients $\{ c_{\nu}(\mathbf{G}) \}$, 
as shown in the work by Car and Parrinello\cite{Car:1985mz}.  
Since we only need the occupied Kohn-Sham orbitals in such iterative methods,
the memory requirement to store  $\{ c_{\nu}(\mathbf{G}) \}$ is proportional to $N_{B}N_{G}$, 
which is roughly 100 times smaller than $N_{G}^2$.
Then, we update the coefficients $\{ c_{\nu}(\mathbf{G}) \}$ by calculating the gradient of the total 
energy with a constraint to keep the KS orbitals orthogonal to each other. 
This is done by calculating $(H^{\rm KS}-\Lambda_{\nu,\nu^{\prime}})$ or $H^{\rm KS}-\epsilon_{\nu}$ 
with Gram-Schmidt orthogonalisation of $\{ c_{\nu}(\mathbf{G}) \}$. 
In the calculation of the Kohn-Sham Hamiltonian, we need to calculate the density $n(\mathbf{r})$.
For this, we first calculate  $n(\mathbf{r})$ as
\begin{equation}
\label{eq:82}
 n(\mathbf{r})= \sum_{\nu} f_{\nu} \Psi^{\ast}_{\nu} (\mathbf{r}) \Psi_{\nu} (\mathbf{r})
\end{equation}
If we perform this in a straightforward way, we need the operations of $\mathcal{O}(N_G^2)$
for each band $\nu$, and the total number of operations needed for the
transformation from $\{ c_{\nu}(\mathbf{G}) \}$ to $\Psi_{\nu} (\mathbf{r})$ is of the order of $N_B N_G^2$,
which is quite expensive. 
However, we can dramatically reduce the cost of the calculation using the
fast Fourier transform (FFT) method, and the number of operations in Eq.~(\ref{eq:82}) becomes 
$N_B N_G \ln (N_G)$. 
Although Eq. (\ref{eq:82}) is still the most expensive part in the calculations of small systems for
many plane-wave DFT codes, the reduction of the computational cost by the FFT method is essential 
for the success of plane-wave DFT calculations. 

The orthogonalisation of the KS orbitals is also an expensive operation,
which includes the calculations $\int \mathrm{d}\mathbf{r} \Psi_{\nu}(\mathbf{r}) \Psi_{\nu' }(\mathbf{r}) $ 
for all pairs of band indices $ \{ \nu, \nu' \}$. 
The total cost of the operations is $\mathcal{O}(N_B^2 N_G)$, but we can see that 
it is only proportional to $N_G$.
As we have seen, in the iterative method with the plane-wave basis set,
there are no operations where the cost increases as fast as $N_{G}^2$. 
This is the reason why we can do efficient calculations even with large $N_G$.
The iterative diagonalisation technique, FFTs, and \emph{ab initio} pseudopotentials used in the plane-wave
calculations are the key factors which make it possible to employ accurate but efficient DFT calculations. 
Using these techniques, with the increase of the computer power, the time for solving KS equations 
has become smaller and smaller, and the system size for the target of DFT studies has become larger
and larger. 
There was a report already in 2002 of DFT calculations on a DNA system including
hydrating water molecules, which consisted of 1194 atoms, including 138 water molecules\cite{Gervasio:2002fk}.

However, this situation changed about 5--10 years ago.
Recently, the growth of computer power mainly comes from the increase of the number 
of processors or cores, while the speed of each core or processor remains unchanged. 
The number of cores of the biggest supercomputers is currently reaching sub-millions. 
The Jaguar machine at Oak Ridge in the US has 224,162 cores, and the
new Japanese supercomputer `K' already has
more than 700,000 cores. To utilize such computing power, 
it is essential to determine whether or not a technique or approach
has good \emph{parallel} efficiency.
In this respect, the FFT method has a serious drawback. 
As is well known, the FFT needs all-to-all communication (i.e. each
core communicating with all other cores) and the time required for 
communications  will grow rapidly with the increase of cores or processors. 
As explained above, we cannot perform efficient plane-wave DFT calculations 
without the FFT technique. Thus, we need to introduce a different type of 
basis set which will be more suitable for parallel calculations. 

In addition, there is another serious obstacle to increasing the system size in DFT calculations. 
When the number of atoms exceeds a few hundred, the orthogonalisation
of the Kohn-Sham eigenstates becomes the most
expensive operation instead of the FFT.
The CPU time for the FFT part is proportional to $N_B N_G \log(N_G)$ and 
it increases as $\mathcal{O}(N^2)$, since both $N_B$ and $N_G$ are proportional to 
the number of atoms $N$.
On the other hand, the CPU time for orthogonalisation increases as $N_B^2 N_G$,
which is $\mathcal{O}(N^3)$. 
Once this part becomes the most expensive part, 
it is very difficult to make the system size larger.

From our brief survey of the field, we can see two key points which must be
overcome to realize efficient DFT calculations on extremely large
systems and massively parallel computers.
\begin{itemize}
\item Develop a method to calculate  electronic structure which is
  suitable for massively parallel
  calculations\cite{Fattebert:2004xi,Iwata:2010yq} 
\item Solve the electronic structure with better scaling than $\mathcal{O}(N^3)$, ideally 
with linear-scaling.
\end{itemize}

We note that there have been considerable efforts in this area within
standard DFT codes, often
focussing on molecular dynamics\cite{Hutter:2005rw,Gygi:2006ys}.
However, there are limits to these efforts, and so we consider alternatives.
For the first point, real-space methods are considered to have an
advantage. There has been a concerted effort to develop practical  
real-space methods and great progress has been achieved in the last decade. 
Several DFT codes using this technique are now available to researchers. 
Here, it is essential to understand whether these new methods keep the advantage 
of plane-wave methods or not, such as high accuracy, ease of calculating 
atomic forces, and systematic convergence.
Regarding the second point, there were already several proposals for $\mathcal{O}(N)$ methods
more than ten years ago\footnote{The recursion method, described in
  Sec.~\ref{sec:recursion}, dates back to the 1970s, while the first
  linear scaling approaches were proposed in the early 1990s, for
  instance divide-and-conquer (Sec.~\ref{sec:divide-conquer}) and DMM
  and OMM methods (Sec.~\ref{sec:direct-iter-appr})}, where the cost of the calculation is only proportional to $N$.
Within empirical tight-binding (TB) methods, there have been a
significant number of applications 
using such linear-scaling techniques\cite{Goringe:1997uq, Galli:1994fk}.
However, there were almost no examples, until very recently, where linear-scaling DFT 
methods were used for the purpose of actual scientific research.
To replace conventional DFT methods with a new method, it is necessary for
the method to have the same accuracy and stability as standard methods, and reasonable efficiency.
Compared with empirical TB methods, DFT calculations are more complex 
and have many potential sources of instability, especially 
in large-scale calculations. In this respect, the success of the first point is 
important also for the second point.  Plane-wave DFT methods have been
under intense development for over twenty five years and are widely
used; competing for efficiency is therefore difficult for linear
scaling methods, except for very large (thousands of atoms) systems.
Identifying problems which require systems of this size can be a
challenge, particularly to researchers used to the constraints of
cubic scaling codes. 

The main purpose of this paper is to review the recent progress of 
the $\mathcal{O}(N)$ methods. However, following our discussion above, 
we first survey recent progress in real-space methods. 
We then turn to the localisation of electronic structure, first for
Wannier functions and then the density matrix.  In the major part of
the review, we survey linear
scaling methods and related developments in seven different areas and
consider extensions to standard DFT.
Technical details (including non-orthogonality, electron number, 
parallelisation and sparse matrices) are dealt with in a separate
section which is mainly intended for practitioners in the field; however, it
is important to note that high parallel efficiency is a key criterion
for a successful linear scaling code.  Finally, we
describe various implementations of the methods, as well as
applications of linear scaling DFT before concluding with a survey of
the challenges facing the field.

\section{Real-space methods}
\label{sec:real-space-methods}

As touched on above, the use of real-space methods both for efficient parallelisation and
for modelling larger systems is well established, and has been
reviewed elsewhere in
detail\cite{Beck:2000qy,Torsti:2006fr,Saad:2010lq}.  It is also used
extensively for combined quantum mechanical/molecular mechanical
(QM/MM) simulations, which is used both in solid state
systems\cite{Bernstein:2009os} and more commonly in biological
systems\cite{Senn:2009lr}. There are many approaches to a real-space
implementation of density functional theory, which can, like
Gaul, be
roughly divided into three parts\footnote{``All Gaul is divided into
  three parts'', Julius Caesar, \emph{De Bello Gallico}, Chapter 1}: finite-difference methods;  finite
element methods; and the use of local basis functions.  In these
methods, the kinetic, pseudopotential and exchange-correlation
energies are found in real space exclusively.  The solution of the
Laplace-Poisson equation for the electrostatic potential, however,
sometimes retains the use of reciprocal space. 

In all cases, the advantage of real-space methods stems from
\emph{spatial locality}, which in turn leads to sparsity of the
Hamiltonian; these ideas will re-appear throughout
Sec.~\ref{sec:line-scal-meth}.  Finite difference approaches represent
the electronic states directly on a fixed grid in real space, with a
finite difference operator for the kinetic energy.  Finite element
(FE) methods\cite{Pask2005a} use the piecewise continuous local basis
of FE analysis.  Local basis functions represent the wavefunctions in
terms of local orbitals, often centred on the atoms (e.g. Gaussians).
This spatial locality can lead to linear scaling Hamiltonian building,
and is also at the heart of the linear scaling electronic solvers
described in Sec.~\ref{sec:line-scal-meth}.  We describe these
approaches in outline below, but without the intention of giving a
detailed review of the methods; other reviews are cited for the
interested reader. 

\subsection{Finite Differences}
\label{sec:finite-differences}

Finite difference methods do away with a basis set entirely, and represent the wavefunctions directly by their numerical values on a grid; the grid spacing is one parameter by which the convergence can be judged.  This approach requires an approximation for the differential operator used in calculating the kinetic energy and in solving the Poisson equation; the simplest approximation is found by expanding a function in positive and negative directions:
\begin{eqnarray}
  \label{eq:1}
  \psi(x_{n+1}) &=& \psi(x_n) + \psi^{\prime}(x_n)h + \frac{1}{2}\psi^{\prime\prime}(x_n)h^2+ \frac{1}{6}\psi^{\prime\prime\prime}(x_n)h^3\ldots\\
  \psi(x_{n-1}) &=& \psi(x_n) - \psi^{\prime}(x_n)h + \frac{1}{2}\psi^{\prime\prime}(x_n)h^2-\frac{1}{6}\psi^{\prime\prime\prime}(x_n)h^3+\ldots
\end{eqnarray}
where $h$ is the grid spacing and $\psi(x_n)$ is the value of the function $\psi(x)$ at a grid point $x_n$.  By adding these two equations, an approximation for $\partial^2\psi(x_n)/dx^2$ can be derived which is accurate to second order in $h$:
\begin{equation}
  \label{eq:2}
  \frac{\partial^2 \psi(x_n)}{\partial x^2} \simeq \frac{1}{h^2}\left[\psi(x_{n+1}) + \psi(x_{n-1}) - 2\psi(x_n)\right] - \frac{1}{12}\frac{\partial^4}{\partial x^4}\psi(x_n) h^2 + \mathcal{O}(h^4)
\end{equation}
The errors can be of either sign, depending on the derivatives and
value of $\psi$, which has the consequence that the FD method is not
variational; the loss of variational nature can make it harder to
converge the parameters used in computational methods. Naturally,
there are higher order approximations for the Laplacian which can be
generated (see, for example, the algorithm in Appendix A of
Ref.\cite{Beck:2000qy}); the order of the expansion is the other
parameter which defines the convergence of these methods.  An
alternative discretisation, the Mehrstellen
discretisation\cite{Briggs:1996fr,Fattebert:2000yr} has also been used
and introduces a non-Hermitian Hamiltonian in exchange for greater
accuracy of representation for a given order.  In all cases, a higher
order discretisation leads to a larger range for the operator, which
impacts on efficiency.  There have been proposals for variational
representations of the kinetic energy operator in
real-space\cite{Maragakis:2001uq,Skylaris:2002kx} which have been
compared in detail\cite{Skylaris:2002kx}; these representations would
alleviate the variational problem in finite difference methods. 

Once the Schr\"odinger equation has been discretised on the grid, it can be written in matrix form, with a size proportional to the number of grid points.  However, the resulting matrices are sparse owing to the locality of the real-space representation.  Indeed, the main source of spread in the matrices is the kinetic energy term (where the order of the approximation chosen will directly affect the locality).  The sparsity of the matrices makes the method ideal for massive parallelisation\cite{Beck:2000qy} and efficient solvers.  Solution of the Poisson equation is often accomplished directly on the grid via multigrid techniques\cite{Bernholc:2008ly}, without recourse to FFTs which can cause problems to parallel scaling of plane-wave codes.

The FD technique has been reviewed extensively\cite{Beck:2000qy,Torsti:2006fr,Saad:2010lq}.  It has been applied by a number of groups\cite{Chelikowsky:1994vn,Zhou:2006eu,Alemany:2007km,Fattebert:2000yr,Iwata:2006kx,Iwata:2010yq,Fujimoto:2010qe,Heiskanen:2001mz} with extension to PAWs\cite{Mortensen:2005rr}.  The solution of the equations can be accelerated by the multigrid method\cite{Press:1992dp,Beck:2000qy}, which has been extensively applied in at least one real-space DFT code\cite{Bernholc:2008ly}.  FD methods can also be combined with localised orbitals, either fixed to the atoms\cite{Hoshi:1997jy,Bernholc:2008ly} or with adaptive localisation regions\cite{Fattebert:2008th}; these localised orbitals are an integral part of linear scaling approaches.

\subsection{Finite Elements And Local Real-Space Bases}
\label{sec:finite-elements}

Finite element (FE) methods are well-known from the engineering field,
and their application to electronic structure methods has a long
history\cite{White:1989fy,Tsuchida:1995wd,pask1999a,Pask2005a}.  The
method uses basis functions which are chosen to be piecewise
polynomials, local in real-space.  The simplest possible FE basis
consists of linear functions which are one on the defining grid point
and zero beyond its nearest neighbours; cubic functions are more
common\cite{White:1989fy,pask1999a} (and indeed the blip functions
mentioned below as a basis for the \textsc{Conquest} $\mathcal{O}(N)$
code\cite{Hernandez:1997ay} are functionally equivalent to finite
element basis functions).  

In the FE method, the unit cell is divided into elements (the simplest of which are cubes, but the shape is in principle arbitrary so long as the simulation cell is filled).  Once the elements and basis functions are chosen, the Schr\"odinger equation can be written as a matrix equation, as was the case for the FD method (and the similarities and differences between the two techniques are elegantly described by Beck\cite{Beck:2000qy}).  In particular, the FE method introduces non-orthogonal basis functions, which leads to a generalised eigenvalue equation\cite{Tsuchida:1995wd} (though this is a familiar problem in electronic structure techniques seen also for ultra-soft pseudopotentials).  The mesh fineness can be different in different areas of the simulation cell, leading to efficiencies when large areas of vacuum are considered, or allowing all-electron calculations to be performed with appropriate resolutions in different areas of the cell.  The method shares the advantages of locality of the FD method while also being variational, and has been applied in various areas\cite{Pask2005a,Tsuchida:1996sp,Tsuchida:1998dp,Tsuchida:2008ai,Wijesekera:2007qc,Feng:2006cq}.

Similar to finite elements are another class of real-space basis functions which are local in real space, with many of the well-defined properties which make plane-waves valuable (e.g. orthonormality and systematic convergence).  Daubechies wavelets\cite{Genovese:2008ya} are a specific class of wavelets with attractive properties for electronic structure (particularly massively parallel or linear scaling implementations): they are orthogonal and local in real and Fourier space; the use of wavelets as a general approach to electronic structure has been discussed in detail elsewhere\cite{Arias:1999lr}.  It is also possible to make a multiresolution implementation\cite{Niklasson:2002dw}.  Wavelets have been implemented as a basis in one major \emph{ab initio} code (\textsc{Abinit}).  

Discrete variable representations\cite{Liu:2003ul} are another of this class of basis, and have been successfully applied to \emph{ab initio} molecular dynamics calculations\cite{Lee:2006qf}.  It is intriguing to note the close relationship between these basis functions and the psinc functions\cite{Mostofi:2002wb} described below and used for a linear scaling code (and earlier used to calculate kinetic energy in a real-space DFT code\cite{Hoshi:1995mw}.  Both of these basis sets have the property that they are non-zero only on one grid point, and zero on all others (known as \emph{cardinality}; wavelets are known as semi-cardinal as they are cardinal only at one resolution).  However, despite their attractive properties, these basis sets are not yet in widespread use.  As computational resources shift towards multi-core processors it may be that their properties make them more attractive than plane-waves.  

Lagrange functions form another cardinal basis set which have been
proposed for electronic structure
calculations\cite{Varga:2004xz,Varga:2006pt}.  A local, grid-based
non-orthogonal basis also used in linear scaling methods are blip
functions (or b-splines)\cite{Hernandez:1997ay}, which can be shown to
be a form of finite element.  They have also recently been used for
linear scaling Quantum Monte Carlo
calculations\cite{Alfe:2004ee}.  The psinc functions mentioned above,
which are periodic bandwidth-limited delta
functions, are another local basis
set\cite{Mostofi:2002wb}; interestingly, almost the same functions
were derived in the context of optimal local basis
sets\cite{E:2010fk}.  One of the first \emph{ab initio}
$\mathcal{O}(N)$ methods proposed\cite{Galli:1992yu} used a plane-wave
basis to represent localised orbitals.

\subsection{Atomic-like Orbitals}
\label{sec:real-space-local}

Functions which mimic the atomic wavefunctions near the ionic core are
a popular choice of basis function, which make sound computational
sense: they provide an excellent solution for much of real space and
are spatially local.  However, to model correctly the changes which
occur to electronic structure on formation of bonds, variational
freedom is required, including both radial freedom for the valence
electrons and angular freedom (often solved by adding orbitals with
higher angular momentum than the valence states).  Numerical atomic
orbitals (NAOs---used for all-electron calculations) or pseudo-atomic
orbitals (PAOs---used with pseudopotentials for convenience) are in
wide use in both standard and linear scaling codes\cite{Soler:2002kn,Blum:2009kw,VandeVondele:2005lp,Ozaki:2003xq,Ozaki:2004qq,Ozaki:2004pt,Kenny:2000hq,Torralba:2008wm,Otsuka:2008ri}, though this is by no means an exhaustive list (other bases include Gaussian-based orbitals\cite{Briddon:2000gb,VandeVondele:2007sp}, muffin-tin orbitals, and augmented plane-waves).  Numerical atomic orbitals have been recently reviewed\cite{Shang:2010rr}.

These basis functions are written as a radial function multiplied by a spherical harmonic (normally the real spherical harmonics): 
\begin{equation}
  \label{eq:3}
  \chi_{nlm}(\mathbf{r}) = R_{nl}(r)Y^l_m(\hat{\mathbf{r}})
\end{equation}
The formalism allows for efficient evaluation of matrix elements.  Analytic operations are possible for the angular terms, while the radial terms are performed in reciprocal space with a very fine mesh\cite{Sankey:1989gf,Soler:2002kn}, or analytically, for Gaussians.  Typically, functions are confined within a sphere which removes extended tails and results in sparse matrices.  Even for conventional codes, this can render the building of Hamiltonian matrices linear scaling (discussed further below, in Sec.~\ref{sec:hamiltonian-building}), reducing a significant cost.

\begin{figure}[h]
  \centering
  \includegraphics[width=0.419\textwidth]{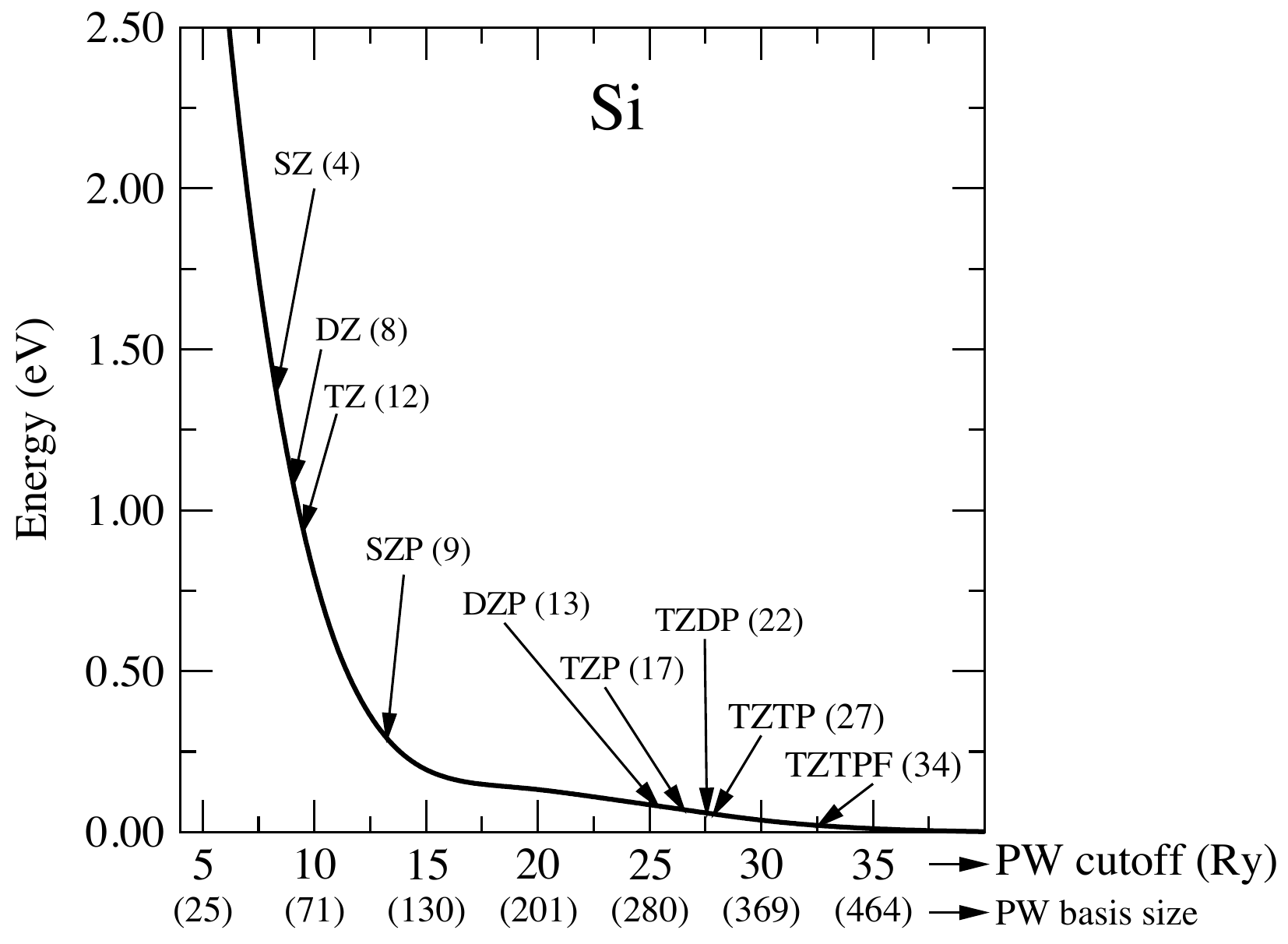}
  \includegraphics[width=0.3\textwidth]{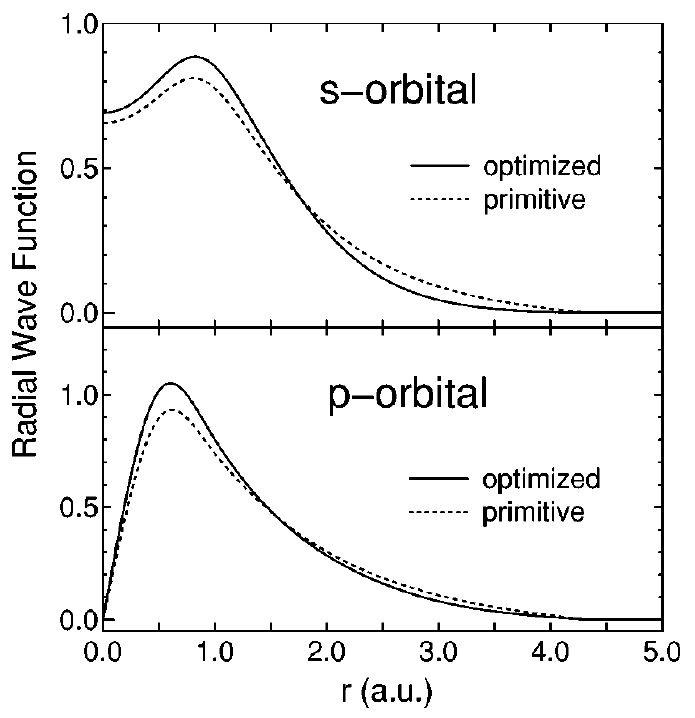}
  \caption{(a) Convergence of energy with PAO basis size for bulk
    silicon in \textsc{siesta}\cite{Junquera:2001fk}; (b) Change in
    shape for atomic orbital following optimisation within
    \textsc{OpenMX}\cite{Ozaki:2003xq}.  Reprinted figures with
    permission from J. Junquera \emph{et al.}, Phys. Rev. B
    \textbf{64}, 235111 (2001) and T. Ozaki, Phys. Rev. B \textbf{67},
    155108 (2003).  Copyright (2001) and (2003) by the American Physical
    Society.}
  \label{fig:PAOs}
\end{figure}

The major drawback with these basis sets is the lack of systematic
convergence: the number of radial functions in a given angular
momentum channel can be increased (often known as multiple zeta or
multiple valence, so that two radial channels are notated DZ or DV)
and extra angular momentum channels can be included but there is no
clear rule as to how functions should be added to systematically
improve the energy.  There have been studies which show that
convergence can be achieved, and which suggest routes to creation of
convergent basis sets\cite{Soler:2002kn,Ozaki:2003xq,Blum:2009kw} but
these schemes lack the simplicity of basis sets with a single
parameter (e.g. the kinetic energy cutoff for plane-waves, or grid
spacings for analytic real-space methods); an example of the
convergence with respect to basis set size is shown in Fig.~\ref{fig:PAOs}(a).

The problem of confinement has generated a number of different
solutions.  The simplest approach is effectively to impose an infinite
potential well of some radius on the atom\cite{Sankey:1989gf}, though
this has the side effect that the derivative of the orbital is discontinuous at the boundary, which can cause problems with the calculation of forces and stresses.  The confinement excites the atom slightly and mimics the effect of condensation into a molecule or condensed matter environment.  However, it is not clear how to confine the atom, particularly as different orbitals will have different ranges.

To avoid the discontinuity produced by an infinite potential, a number of suggestions have been made for an alternative potential (surveyed, along with the methodology known as \emph{ab initio} tight binding, in a review\cite{Horsfield:2000nx}). Confining potentials suggested include: simple polynomials: ($r^2$\cite{Porezag:1995eu}, $r^6$\cite{Horsfield:1997cr}); smoothing the free atomic wavefunctions with an exponential using a cutoff and width over which the smoothing is applied\cite{Kenny:2000hq}; an exponential potential applied between two points\cite{Junquera:2001fk}; a cubic truncation between two points (applied to the bare atom potential)\cite{Ozaki:2003xq}; and a product of an exponential and $1/r^2$ on the free atom (for all-electron calculations)\cite{Blum:2009kw}.  All these different schemes produce a smooth transition to zero in the tails of the orbitals.

As well as methods to confine the orbitals, there are many different
methods to generate the basis sets themselves.  These can be split
into two approaches: first, how to generate a set of either pseudo
atomic orbitals or numerical atomic orbitals (depending on whether or
not a pseudopotential is used); and second, how to use these as basis
functions (either as they are or combined into other functions).  We
will consider these two problems in turn.   

As there is considerable flexibility in deciding, for instance, at
what radius to cut off a function, or how many radial functions to use
for each angular momentum channel, much effort has gone into deciding
how to generate accurate basis sets with minimal computational effort
and human intervention.  An early approach to the cutoff
problem\cite{Artacho:1999gf} was to use the energy change found on
confining individual orbitals (typically in the range 50-300\,meV) as
a single, balanced criterion.  The advantage is that one parameter can
be used for orbitals of different inherent sizes, and this is used
extensively in the \textsc{siesta} code (see
Sec.~\ref{sec:implementations} for more details). Another approach
within \textsc{siesta} is to optimise the orbital shape relative to a
highly converged plane-wave calculation by varying the confinement
potential, but applying a fictitious pressure-like
quantity\cite{Anglada:2002uq} to stop expansion of the orbitals beyond
a reasonable size.  This idea of optimising the confinement has also
been applied to a damping function multiplying the
orbitals\cite{Esfarjani:1999hc}.

Spillage\cite{Sanchez-Portal:1995qe,Sanchez-Portal:1996ai} is an idea
which can be used when optimising atomic orbitals.  In a periodic
system, it is defined as: 
\begin{equation}
  \label{eq:78}
  S = \frac{1}{N_k}\frac{1}{N_B}\sum_{\mathbf{k}}\sum_{i=0}^{N_B}\langle\psi_i(\mathbf{k})|\left[1-P(\mathbf{k})\right]|\psi_i(\mathbf{k})\rangle
\end{equation}
where $N_k$ is the number of k points, $N_B$ is the number of bands,
$|\psi_i(\mathbf{k})\rangle$ is a Bloch state found with a plane-wave
code and the projection operator $P(\mathbf{k})$: 
\begin{equation}
  \label{eq:79}
  P(\mathbf{k})=\sum_{\mu\nu}|\phi_\mu(\mathbf{k})\rangle S^{-1}_{\mu\nu}\langle\phi_\nu(\mathbf{k})|
\end{equation}
with $S_{\mu\nu}$ the overlap between atomic orbitals (see
Sec.~\ref{sec:non-orthogonal-basis} for more details on
non-orthogonality).  The spillage measures how well atomic-like
orbitals can reproduce the wavefunctions from another calculation
(e.g. highly converged plane-wave basis calculations).  The spillage
is used to guide optimisation of atomic-like orbitals and to
investigate the generation of transferrable basis sets.
For instance, basis sets using orbitals from both neutral and
positively charged ions, and optimising cutoff radius to minimise
spillage\cite{Kenny:2000hq} were found to be transferrable and accurate.
Spillage has also been used to optimise pseudo-atomic orbitals with
radial functions are expanded in terms of Bessel
functions\cite{Chen:2010kq}; the functions are fit to converged
plane-wave calculations of dimers. 

Generation of basis sets consisting of atomic-like orbitals can also
be performed in a semi-systematic way.  For instance, generation of
successive solutions of a confined atom with increasing numbers of
nodes for a given angular momentum channel\cite{Ozaki:2003xq}, or
building a large set of functions including different cutoffs, angular
momenta, Rydberg functions and extend basis functions, and ordering
the set by searching for the function which lowers energy most on
addition to the existing set\cite{Blum:2009kw}.  A method for the
direct optimisation of the radial function\cite{Talman:2000bs} within
a self-consistent loop has also been given to yield an optimal minimal
basis (this parallels to some extent the representation of support
functions or generalised Wannier functions by b-splines or psincs
mentioned above in Sec.~\ref{sec:finite-elements}; it has also been
suggested that localised functions can be expanded in terms of
spherical waves\cite{Haynes:1997qe} for free electrons, and recent
analytic developments\cite{Monserrat:2010bh} have simplified and
improved the scaling of this method).  Methods for optimising Gaussian
basis sets are also available\cite{Porezag:1999th,VandeVondele:2007sp}

While a large set of atomic-like orbitals can be used directly to
represent the wavefunctions, it can be more efficient to combine them
into a smaller set of functions.  Polarised atomic
orbitals\cite{Lee:1997tg} are one way to do this: a minimal set of
polarised atomic orbitals is defined in terms of a large basis set of
standard quantum chemistry orbitals.  As is to be expected, the
contraction results in a small increase in total energy, but the
convergence is good, and the error is linear in system size
(indicating size extensive behaviour, and local errors); moreover, the
structural relaxation is reliable.  The idea has been refined to
extract polarised AOs from molecular orbitals\cite{Lee:2000tg} which is closely
related to the extraction of Wannier functions described in
Sec.~\ref{sec:dens-matr-prop}.  It has also been
extended\cite{Berghold:2002fr} so that minimisation of polarised AOs
and the density matrix is separated; this was implemented within an
$\mathcal{O}(N)$ method and shown to be effective.  A related
method\cite{Aquilante:2006bv} uses Cholesky decomposition to extract
localised molecular orbitals from the density matrix.

The \textsc{OpenMX} code mixes large numbers of PAOs (typically six
per angular momentum channel) into a set of orbitals equivalent to a
DZP basis\cite{Ozaki:2003xq} with the PAOs simply generated as
orthonormal functions (by increasing the number of nodes) for a confined
atom; the change in radial function following optimisation is illustrated in Fig.~\ref{fig:PAOs}(b).  An optimised set of PAOs has been generated for calculations of
biological molecules (covering H, C, N, O, P and
S)\cite{Ozaki:2004qq}, and it was suggested that optimisation at each
step of the minimisation is not necessary.
A similar approach is used for the representation of support functions
with pseudo-atomic orbitals in
\textsc{Conquest}\cite{Torralba:2008wm}.  It is important to note,
though, that there are certain symmetry-imposed restrictions on the
number of support functions that can be used: for instance,
trying to represent four support functions only with $l=0$, $l=1$ and
$l=2$ will give incorrect answers in bulk crystals (breaking symmetry).
A brief overview of approaches to PAO mixing, as well as a scheme for
mixing two PAOs (neutral and $2^+$ atom) has been
given\cite{Basanta:2007mz}.  A comparison of the method used in
\textsc{OpenMX} and using spillage\cite{Gusso:2008pl} shows that
comparison to plane-wave results with chemical accuracy can be
achieved with localised orbitals,  though two functions per angular
momentum channel (DZDP) are generally needed (and up to $l=2$).
A method combining gaussians from multiple sites into a minimal basis on each
atom\cite{Rayson:2009vo,Rayson:2010fk} has been proposed.  It uses a filtration algorithm ($cf$
the FOE method below in Sec.~\ref{sec:recursion}) which removes
unwanted high energy components to optimise the orbitals, and may
allow large systems to be calculated on modest resources.

Overall, it should be clear that there is considerable effort being
made to understand and optimise atomic-like orbital approaches.
Given the history of the method, it is perhaps a little surprising
that there is still so much work to do, but that simply reflects the
impossibility of finding a perfect basis set for the diverse
environments modelled by density functional theory.

\subsection{Representing Localised Orbitals}
\label{sec:repr-local-orbit}

When performing $\mathcal{O}(N)$ calculations, many codes represent
the density matrix (described below in Sec.~\ref{sec:solving-dm}) in
terms of localised orbitals, $\phi_{i\alpha}(\mathbf{r})$; for
instance, \textsc{onetep} calls these non-orthogonal generalised
Wannier functions and \textsc{Conquest} calls them support functions.
These functions can be simply individual PAOs, or more generally
represented in terms of a basis set, either the atomic-like orbitals
of Sec.~\ref{sec:real-space-local}, or the local real-space basis sets
described in Sec.~\ref{sec:finite-elements}.

Using atomic-like orbitals is convenient and gives a relatively small
basis size.  In the limit of a single PAO per localised orbital, of
course, there is no optimisation required, which removes a level of
complexity from the calculation; however, it is important to note that
there can be problems for inverting the overlap matrix in the case of
double zeta basis sets (discussed further in
Sec.~\ref{sec:non-orthogonal-basis}).  Atomic-like orbitals also
suffer from basis-set superposition error\cite{Boys:1970qe}; while the
magnitude of the error reduces with basis set size, it is still
significant\cite{Artacho:2003ai}, and significantly worse for more
contracted basis sets\cite{Otsuka:2008ri}, though can be corrected
very successfully.  These basis sets are widely used (e.g. in
\textsc{OpenMX}, \textsc{siesta} and \textsc{Conquest}).

The local, real-space basis sets, such as the psincs used in
\textsc{onetep} and the b-splines used in \textsc{Conquest}, can be
converged systematically to the plane-wave
limit\cite{Hernandez:1997ay,Skylaris:2007lp} and are free from
basis-set superposition error\cite{Haynes:2006qe}.  The resulting
orbitals will be more transferable as they are optimised \emph{in
  situ}, and possibly more accurate.  However, these
basis sets require more computational effort to converge than PAOs,
and calculating small energy differences between structures will
require tight tolerances on the minimisations.

The real question when deciding on a basis set is that of accuracy
versus computational cost.  The minimisation is variational, which
means that less effort will be required once the initial functions
have been converged, and also that library functions could be
calculated and read in as a starting point.  In terms of matrix
multiplications, which lie at the heart of linear scaling codes,
real-space basis sets have a significant advantage: as a minimal
number of orbitals can be used, the multiplies are significantly
faster.  For instance, for Group IV elements such as C and Si, the
computational cost to go from four orbitals (minimal) to nine orbitals
(the smallest number possible when using polarisation functions in a
PAO basis) is a factor of 11, and going to thirteen orbitals (a double
zeta plus polarisation) is a factor of 34.  These factors can offset
the extra time required but there is no single correct answer.

\subsection{Hamiltonian Building}
\label{sec:hamiltonian-building}

Many methods which are not linear scaling in the search for the ground
state nevertheless use localised real-space basis sets, and rely on
this locality to build the Hamiltonian in a linear scaling manner\cite{Challacombe:1997qy,Watson:2004qo,VandeVondele:2005lp,Shao:2006xq}.
The computational effort required for Hamiltonian building is
significant, and for a few hundred atoms with localised orbitals the
resulting matrix can be exactly diagonalised efficiently, with most
effort being spent in the Hamiltonian build.
The Hamiltonian typically is made up of different terms, which can
require different treatments: kinetic energy; electron-ion interaction
(either via bare Coulomb term or pseudopotentials); Hartree energy;
and exchange and correlation energies.  The kinetic energy is
inherently local (and only shows spread for high order finite
difference methods) and will not be considered further.

As individual basis functions are local in space, the integrals
required to form the Hamiltonian can be reduced from
$\mathcal{O}(N^3)$ scaling (the integrals between all pairs of basis
function ($N^2$) must be evaluated over the whole system ($N$)) to
$\mathcal{O}(N)$ scaling.  For non-local pseudopotentials, the
electron-ion interaction can be evaluated using a separable form to
give integrals between localised orbitals and non-local projector
functions, followed by a matrix multiplication.  The local part is
more complex: it is formally written as $H^{\mathrm{local}}_{ij}=\sum_k
\langle \phi_i | V_k | \phi_j \rangle$ for the matrix element between
atoms $i$ and $j$, which involves three-centre integrals and hence
poor efficiency.  The standard solution involves integration on a grid
(making the potential by summing over atoms $k$ \emph{before} the integration is performed),
often using the charge and potential from the neutral (sometimes free)
atom\cite{Sankey:1989gf,Soler:2002kn} to form a smoothly varying
function (which is relatively insensitive to grid spacing).  A method
to make the neutral-atom potential separable has been
proposed\cite{Ozaki:2005dp} which involves expanding the potential in
terms of local functions; this allows the potential to be evaluated in
the same way as the non-local potential.  As the potential is
spherically symmetric, it can be expanded with radial functions and
spherical harmonics (up to $l=6$ was found to be sufficient for convergence\cite{Ozaki:2005dp}).  This
procedure removes any grid dependence apart from a charge density
difference $\delta n(\mathbf{r}) = n(\mathbf{r}) - \sum_i
n_i(\mathbf{r})$ with $n_i(\mathbf{r})$ an atomic density.

The Hartree potential (found as the solution of the Poisson equation)
is often solved using fast Fourier transforms, which are strictly
$\mathcal{O}(N \log N)$; the standard approach to the electrostatic
solution of the ionic problem, the Ewald sum, scales as
$\mathcal{O}(N^{3/2})$, though the use of a neutral-atom potential\cite{Sankey:1989gf}
removes the need for this step.  The other method in common use for
electrostatic problems is the fast multipole method, which may scale
as $\mathcal{O}(N)$ asymptotically (see, for instance,
discussions\cite{Challacombe:1996fj,Challacombe:1997qy,Watson:2004qo,Gan:2004ss,Rudberg:2010tg}).
Other methods used include density fitting\cite{Sodt:2006fk}, FFTs
combined with a wavelet solution for surface
problems\cite{Genovese:2007rc} and combining finite elements and
Gaussians for the direct solution of the Poisson equation
with the fast multipole method for calculation of the boundary
conditions\cite{Watson:2008ff}. 

The calculation of the exchange-correlation matrix is generally
straight-forward, but a number of different approaches have been
given, both exact and
approximate\cite{Sankey:1989gf,Horsfield:1997cr,Challacombe:2000lh,Balbas:2001sp,Torralba:2009hc,Rudberg:2010tg}.  The question of calculating exact exchange within DFT (or
Hartree-Fock) has been studied extensively, and a number of approaches
which scale linearly with system size have been
derived\cite{Schwegler:1996yq,Burant:1996gf,Schwegler:1997mz,Ochsenfeld:1998rt,Schwegler:1999fr,Lambrecht:2005ul,Weber:2006ng,Rudberg:2008fk,Wu:2009kx}.
This ensures accuracy and efficiency are possible within local
orbital, real-space codes.  The route to linear-scaling construction
of the Hamiltonian building is clear.  Now we turn to consider the
solution for the ground state of the system.

\section{Linear Scaling Methods}
\label{sec:line-scal-meth}

As we have noted above, significant savings can be made even for conventional eigenvalue solvers if a basis set which is local in real space is used to represent the wavefunctions.  The Hamiltonian building process becomes linear scaling, leaving the solution for the eigenstates as the most expensive part, and the part with the worst scaling.  It is natural to consider whether this can be improved as well; herein lies the heart of the development of linear scaling codes.

A natural first point to consider is that the Kohn-Sham formulation of
DFT introduced the wavefunctions as an aid to solution, not as an
integral part of the formalism.  Indeed the Hohenberg-Kohn theorems
rely only on the charge \emph{density} of the system; we might ask
whether a search over charge densities might not be a route to finding
the electronic ground state.  This leads to the approach known as
orbital-free DFT (OFDFT), discussed in detail in
Sec.~\ref{sec:orbital-free-dft}; in brief, it requires an
approximation for the kinetic energy which can reduce accuracy, but is
used as an efficient method for calculations on large metallic
systems.

Instead of the charge density, it is more helpful to work in terms of the \emph{density matrix}, which is defined formally in terms of the eigenstates of the system as:

\begin{equation}
\label{eq:4}
\rho(\mathbf{r},\mathbf{r}^\prime) = \sum_n f_n \psi_n(\mathbf{r}) \psi^\star_n(\mathbf{r}^\prime)
\end{equation}
where $n$ indexes the eigenstate and $f_n$ gives the occupancy of the
state.  As we are still within the Kohn-Sham approach, this is the
single particle, two-point density matrix (not to be confused with the
many-body density matrix familiar from statistical mechanics and
quantum information).  In operator notation, the finite temperature density matrix can
also be written as the Fermi function of the Hamiltonian\cite{Goedecker:1994li}:
$\hat{\rho} = 1/\left(1+\exp[(\hat{H}-\mu)/k_BT]\right)$.  Many of the
properties of the density matrix were investigated and summarised by
McWeeny\cite{McWeeny:1960zp}.  A consequence of quantum interference
effects on the density matrix is that it is ranged:

\begin{equation}
\label{eq:11}
\rho(\mathbf{r},\mathbf{r}^\prime) \rightarrow 0, |\mathbf{r} - \mathbf{r}^\prime| \rightarrow \infty
\end{equation}
However, the functional details of how the decay proceeds is rather complex, and can be related to the localisation of Wannier functions.  We consider these ideas in the next section.

\subsection{Density Matrix Properties and Wannier Functions}
\label{sec:dens-matr-prop}

The Bloch states found when solving the Schr\"odinger equation for a
periodic system (as is done in most electronic structure codes) are
delocalised, and spread throughout the unit cell, and hence the entire
system.  There are many advantages to using localised functions, and
the arbitrary phase associated with the Bloch states (which can be
scaled by $e^{-i\phi}$ with no change to the properties of the system)
gives the freedom to do this.  The most important route to
understanding localisation is via Wannier functions.  Wannier
functions have been used extensively in electronic structure
theory\cite{Wannier:1937kk,Kohn:1959tp,Kohn:1973rt,Blount:1962mb}.
They are formally defined for a periodic potential, with Bloch
wavefunctions $|\psi_{n\mathbf{k}}\rangle$ with $n$ labelling a
band.  Then, for the unit cell at $\mathbf{R}$, we can define the
Wannier function as: 

\begin{equation}
  \label{eq:5}
  | w_{\mathbf{R}n}\rangle = \frac{V}{2\pi}\int d\mathbf{k} e^{i\mathbf{k}\cdot \mathbf{R}}|\psi_{n\mathbf{k}}\rangle
\end{equation}

with 
\begin{equation}
  \label{eq:7}
  \psi_{n\mathbf{k}}(\mathbf{r}) = e^{i\mathbf{k}\cdot\mathbf{r}} u_{n\mathbf{k}}(\mathbf{r})
\end{equation}
and $u_{n\mathbf{k}}(\mathbf{r})$ the periodic part of the Bloch function.  The inverse relationship allows us to write the wavefunction in terms of the Wannier functions:
\begin{equation}
  \label{eq:6}
  | \psi_{n\mathbf{k}}\rangle =\sum_\mathbf{R} e^{i\mathbf{k}\cdot\mathbf{R}} | w_{\mathbf{R}n}\rangle
\end{equation}
It is important to note the considerable freedom in the choice of
Wannier functions that the arbitrary phase of the Bloch functions gives.  The localisation of Wannier functions is closely related to the range of the density matrix (considered below)\cite{Kohn:1995mq}; it is also important in considering insulating behaviour against metallic behaviour in condensed matter systems\cite{Kohn:1964vn}.  The equivalent theory for non-periodic systems was developed for localised molecular orbitals\cite{Adams:1971sp} which also can be linked to pseudopotential theory\cite{Weeks:1973th}.

The study of localisation of Wannier functions is far from trivial.  The earliest results\cite{Kohn:1959tp} showed that the decay was exponential with distance for a one-dimensional centrosymmetric crystal, using complex wave vectors.  This was extended to three dimensions\cite{Blount:1962mb} for periodic solids with no degeneracy (with the decay rate related to the distance of the branch surface from the real axis), as well as general one dimensional\cite{Cloizeaux:1964jy} and  three-dimensional\cite{Cloizeaux:1964vo,Nenciu:1983xd} solids (though only in the tight binding limit for the 3D case).  Non-periodic systems in one dimension have been shown to exhibit exponential localisation\cite{Kohn:1973fk}, and in the case of a localised perturbation, such as a defect, the Wannier functions converge to the periodic functions as the distance from the perturbation increases; this result has also been extended to three dimensions\cite{Nenciu:1993uq}.  The general relation between eigenfunctions of a Hamiltonian and localised orbitals (leading to generalised Wannier functions) was investigated thoroughly\cite{Cloizeaux:1963km}.
The exact form of the decay of Wannier functions in 1D has been investigated in detail\cite{He:2001cq}, and it was found that it can be written as a power law multiplying an exponential.  If the rate at which the functions decay is given by $x^{-\alpha}e^{-hx}$ for Wannier functions in 1D,  a value of $\alpha=0.75$ describes orthonormal Wannier functions, while non-orthonormal Wannier functions result in $\alpha=0.5$ or, with careful construction, $\alpha=1.5$.  
A more general, and formal study, of localisation has shown that
Wannier functions demonstrate exponential localisation in insulators
with a Chern number\footnote{The Chern number is related to the Berry
  connection; the Berry phase is an important part of the modern
  theory of polarisation and is discussed extensively elsewhere\cite{Marzari:1997eu,Resta:2002mw}.} of zero (i.e. which are time reversal symmetric) in 2D and 3D\cite{Brouder:2007ul} . 
A recent study of many-body (Quantum Monte Carlo) Wannier functions
and localisation\cite{Hine:2007qq} has shown that localisation (except
for strongly correlated systems) is similar for one-electron and
many-electron systems, with the difference related to the correlation
hole.  This builds on earlier work on natural Wannier functions in
correlated systems (defined in terms of the natural orbitals) where
similar localisation properties were found\cite{Koch:2001kx}.  
An important study\cite{Annett:1995rt} of the number of iterations
required to reach convergence in a given system, and how this number
of iterations scales with system size found that Wannier-representable
insulators can be considered truly $\mathcal{O}(N)$, with the time to the ground
state not dependent on system size (though some of the results of
this paper have been shown to be pessimistic\cite{Kresse:1996fk}).

It is not the intention of this review to cover all aspects of Wannier
function theory and their use, though we summarise results relevant to
linear scaling methods below.  There are excellent reviews on this
subject, particularly as it relates to
polarisation\cite{Resta:1994wd,Resta:2002mw}.  In quantum chemistry,
the equivalent localisation procedure (though without Bloch states) is
known as Boys-Foster localisation \cite{Foster:1960zr}.
The modern approach to polarisation relies on a definition in terms of the expectation values of the position operator in terms of Wannier functions of occupied bands\cite{Vanderbilt:1993lq}.
Much of the modern theory of polarisation\cite{Resta:1994wd}, particularly relating to the Berry phase, is concerned with and overlaps with definitions of localisation, and the difference between insulators and metals; full details can be found elsewhere\cite{Stephan:2000ys,Resta:1999zr,Souza:2000ly,Kramer:1993ve}.  Recent important developments in the field also relate to topological insulators, and the wealth of physics contained therein.

\begin{figure}[h]
  \centering
  \includegraphics[width=0.4\textwidth]{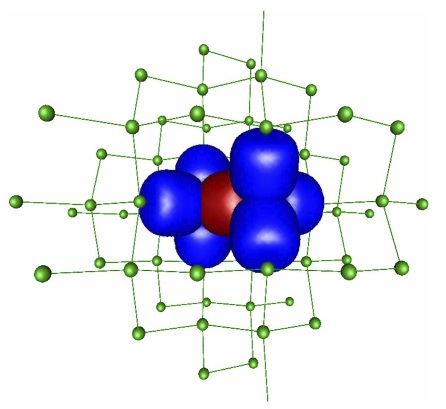}
  \includegraphics[width=0.4\textwidth]{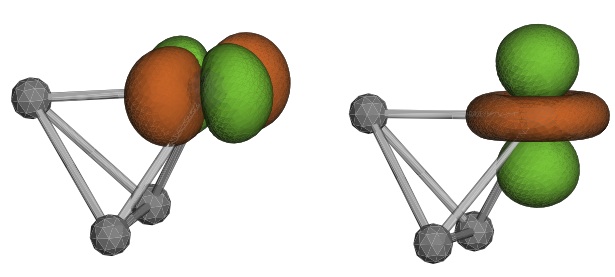}
  \caption{(a) Maximally localized Wannier functions in silicon used in calculating exact
    exchange with linear scaling effort (from \cite{Wu:2009kx}); (b)
    Maximally localized Wannier functions in Cu (from \cite{Souza:2001qe}).  Reprinted
    figures with permission from X. Wu \emph{et al.}, Phys. Rev. B
    \textbf{79}, 085102 (2009) and I. Souza \emph{et al.},
    Phys. Rev. B \textbf{65}, 035109 (2001).  Copyright (2009) and
    (2001) by the American Physical Society.}
  \label{fig:WFillustration}
\end{figure}

Creating Wannier functions is a difficult problem, due to the many possible different definitions of functions themselves, and issues with composite bands.  An early proposal showed that removing the orthogonality constraint created ultra-localised functions \cite{Anderson:1968pd}.  Kohn presented a variational method \cite{Kohn:1973rt} and an efficient iterative method \cite{Kohn:1993xd}, both to find generalised Wannier functions.
A key development in the use of Wannier functions was a reliable
method to produce maximally localised Wannier functions
(MLWF)\cite{Marzari:1997eu}.  The freedom in phase of the Bloch states
results in the Wannier transformation being underdetermined.  The lack
of determination allows extra restrictions to be placed on the Wannier
functions, without loss of generality. The criterion used was to find
the Wannier functions which minimised the functional $\Omega$, defined
as: 
\begin{equation}
  \label{eq:8}
  \Omega = \sum_n \left[\langle r^2 \rangle_n - \langle \mathbf{r}_n\rangle^2\right],
\end{equation}
for a sum over the bands $n$ in the simulation, with
$\langle\mathbf{r}_n\rangle=\langle w_{0n}| \mathbf{r}|
w_{0n}\rangle$ and $\langle r^2 \rangle_n = \langle w_{0n}| r^2|
w_{0n}\rangle$; $w_{0n}$ is the $n^{th}$ Wannier function in the cell
at the origin (cf Eq.~(\ref{eq:5})).  Despite the real-space definitions, the
transformations can be written in terms of the Bloch wavefunctions in
reciprocal space.  This technique has found widespread application
throughout the first principles community, and has been shown to be
effective for disordered systems\cite{Silvestrelli:1998sh} and
entangled bands\cite{Souza:2001qe,Birkenheuer:2005gf}; an efficient,
iterative approach to forming MLWFs has been
given\cite{Berghold:2000kx}.   Examples of MLWFs in
silicon (used for linear scaling evaluation of the exchange potential
and energy\cite{Wu:2009kx}) are shown in
Fig.~\ref{fig:WFillustration}(a), with clear $sp^3$ symmetry.  The MLWFs from
the entangled, narrow d bands in Cu, which overlap with a wide s band, are shown in Fig.~\ref{fig:WFillustration}(b)
following disentanglement; the $d$-symmetry of these functions is
clearly seen.  

There have been many further developments in finding and using Wannier
functions.  Allowing some unoccupied bands to mix with the occupied
bands\cite{Thygesen:2005mq} allows better localisation, and in some
cases a more intuitive picture of bonding (it is important to note
that including unoccupied bands has been proposed
before\cite{Souza:2001qe} for entangled bands; entangled bands overlap
with other bands across the Brillouin zone, as opposed to, for
instance, the isolated valence bands of an insulator).  A linear
scaling technique has been demonstrated for the creation of Wannier
functions\cite{Stephan:1998mj}, which uses projection as others have
before\cite{Cloizeaux:1963km,Cloizeaux:1964vo,Cloizeaux:1964jy,Kohn:1973fk},
while another technique derived separate eigenvalue equations for each
Wannier state\cite{Hoshi:2000uf}; an alternative approach builds
Wannier functions using perturbation theory to correct a simple
initial approximation\cite{Geshi:2003kl}.  Another linear-scaling
approach\cite{Xiang:2006hc} starts from the ground state density
matrix and derives the MLWFs from there.  A dynamic approach allowing
on-the-fly localisation during molecular dynamics shows improvements
in speed of simulation compared to normal band
techniques\cite{Iftimie:2004ys}. Several of the linear scaling methods
for finding the ground state described in
Sect.~\ref{sec:direct-iter-appr} find the Wannier functions for the
system (specifically those methods starting with the work of Mauri,
Galli \& Car\cite{Mauri:1993lf,Kim:1995yi} and Ordej\'on \emph{et
  al.}\cite{Ordejon:1993zk,Ordejon:1995yj}), and this approach has
been used for the calculation of polarisation\cite{Nunes:1994rr},
while non-orthogonal Wannier functions are used for other problems,
for instance for a self-consistent implementation of the DFT+U
method\cite{ORegan:2010fu}.  The problem of finding non-orthogonal
localised molecular orbitals\cite{Adams:1971sp,Liu:2000vn} can be
simplified \cite{Feng:2004lq} by defining centroids based on single,
double or triple bonds\cite{Cui:2010bf}, though this requires some
input and chemical intuition.  A more recent study has investigated
the general localisation properties of bases for eigenvector
problems\cite{E:2010fk}.  An approach similar to Wannier functions to
generate a minimal basis of quasi-atomic orbitals
(QUAMBO)\cite{Lu:2004hs} for post-processing uses occupied and
unoccupied states, and forces the orbitals to be as close as possible
to free-atom orbitals; the method is applicable to metallic as well as
insulating systems.

We also mention that localisation is used in quantum
chemistry (see Sec.~\ref{sec:quantum-chemistry}) to improve the
scaling of perturbative and more accurate methods; for instance, a
recipe to create localised orthonormal orbitals for fast MP2
calculations has been developed\cite{Subotnik:2005ai}.
Wannier functions (and other localised orbitals) are becoming
extremely powerful tools in extending the accuracy of DFT and
Hartree-Fock methods, and we expect to see their use becoming
widespread over the next few years.

The localisation of the Wannier functions for a system is intimately related to the localisation of its density matrix.  This is easily seen as the density matrix can be written in terms of the bands as in Eq.~(\ref{eq:4}), and is unaffected by a unitary transformation of the bands; hence, the Wannier transformation allows the density matrix to be written in terms of the Wannier functions, and the localisation properties follow.  We can write\cite{Ismail-Beigi:1999zk}:

\begin{equation}
  \label{eq:10}
  \rho(\mathbf{r},\mathbf{r}^\prime) = \sum_n \sum_{\mathbf{R}, \mathbf{R}^\prime} W_n(\mathbf{r},\mathbf{R})F_n(\mathbf{R} - \mathbf{R}^\prime)W^\star_n(\mathbf{r}^\prime,\mathbf{R}^\prime)
\end{equation}
where the occupation matrix is defined by the Wannier transform of the occupancy $F_n(\mathbf{R}) = \frac{V}{2\pi}\int d\mathbf{k} e^{i\mathbf{k}\cdot\mathbf{R}}f_{n\mathbf{k}}$.

The general principle that the electronic structure of a system is localised is summarised by Kohn's principle of near-sightedness\cite{Kohn:1996lp}.  The principle is defined in terms of a typical de Broglie wavelength found in the wave function of the ground state, which in turn defines a local volume; the thermal de Broglie wavelength $\lambda = \sqrt{h^2/3m_ek_BT}$ might be reasonable definition to start from\cite{Baer:1997vz}.  Changes to distant parts of the system (such that the distant part is far from all points in the volume) have a negligible effect on the electronic structure in the local volume.
The nearsightedness of electronic structure for systems with a non-zero gap can be expressed as:
\begin{equation}
  \label{eq:9}
  \rho(\mathbf{r}-\mathbf{r}^\prime) \sim e^{-\gamma | \mathbf{r} - \mathbf{r}^\prime|}
\end{equation}
with $\gamma > 0$, though, as noted above, an algebraic prefactor can be defined.  Much work in recent years has sought to relate $\gamma$ to the properties of the system. 

As with the Wannier functions, there has been considerable effort
devoted to quantifying the localisation (or, equivalently, the range)
of the density matrix; an excellent overview is given by
Goedecker\cite{Goedecker:1999pv}.  
The most elegant and appealing, as well as intuitive, results suggests
that the decay rate should depend on the gap.  Kohn\cite{Kohn:1995mq}
showed that the decay rate for Wannier functions is proportional to $\sqrt{m^\star\Delta}$ for
a gap $\Delta$ (see also \cite{Kohn:1959tp}); as $m^\star$ can
be shown to depend on $\Delta$\cite{Ismail-Beigi:1999zk}, this gives
an overall 
dependence on $\Delta$. It was also shown that projection
operators for specific bands localise exponentially at large distances
in the one-electron approximation\cite{Cloizeaux:1964jy} and that in
periodic solids with no degeneracies there is an exponential decay
related to the distance of the branch surface from the real
axis\cite{Blount:1962mb} (extending the earlier work on Wannier function
ranges\cite{Kohn:1959tp}); remembering that the density matrix is a
projection operator for the whole system we see the importance of
these results. 

There is still no complete analytical understanding of the range of
the density matrix.  One study used Chebyshev polynomials\cite{Baer:1997vz} to
explore the properties of the density matrix and the complexity of
different linear scaling methods.  This suggested that the range was
related to the gap (not exponentially necessarily, but $\propto
1/\sqrt{\Delta}$) for insulators; in metals, a finite electronic temperature,
$T$, is required to localise the density matrix, and the range was
found to be $\propto \sqrt{T}$.  However, these results were
subsequently shown to be incomplete.  Using Fourier analysis, it was
shown\cite{Goedecker:1998qw} that metals (again at finite electronic
temperature $T$) show exponential localisation proportional to $k_BT/k_F$ both in real space and also in Fourier space.  
These properties were studied further\cite{Goedecker:1999tv} using
wavelets (see Sect.~\ref{sec:real-space-local}), confirming the Fourier space nearsightedness.
A careful analytic study of simple systems also found that decay is
$\propto \Delta$ for semiconductors and $\propto T$ in metals at low
$T$\cite{Ismail-Beigi:1999zk} (certainly in the weak binding limit);
the study found that some materials show decay $\propto \sqrt{\Delta}$
in the tight binding limit, but the behaviour is complex (depending in
detail on atomic potentials).  At low temperatures, the decay is
$\propto k_BT/k_F$ (in agreement with previous
work\cite{Goedecker:1998qw}) with $\sqrt{T}$ behaviour at high T.  It
was also shown\cite{He:2001cq} for one dimensional systems that the
density matrix decays exponentially (in the same way as Wannier
functions) but with a prefactor of $\alpha=0.5$.   Overall, there is
good evidence to support the exponential localisation of the density
matrix in insulators, and in metals at finite electronic temperature,
though detailed behaviour depends strongly on the system.

\begin{figure}[h]
  \centering
 \includegraphics[width=0.3\textwidth]{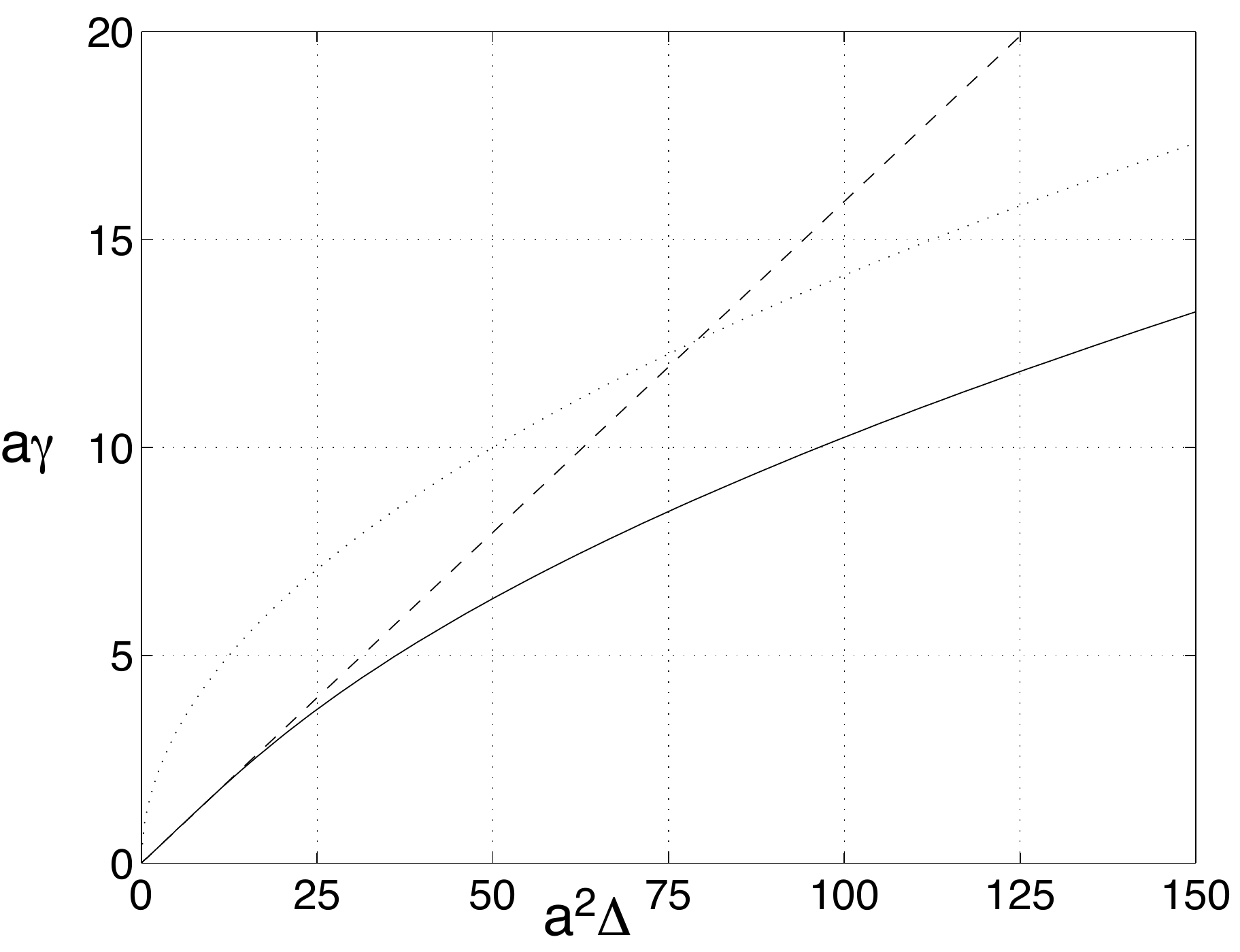}
  \includegraphics[width=0.29\textwidth]{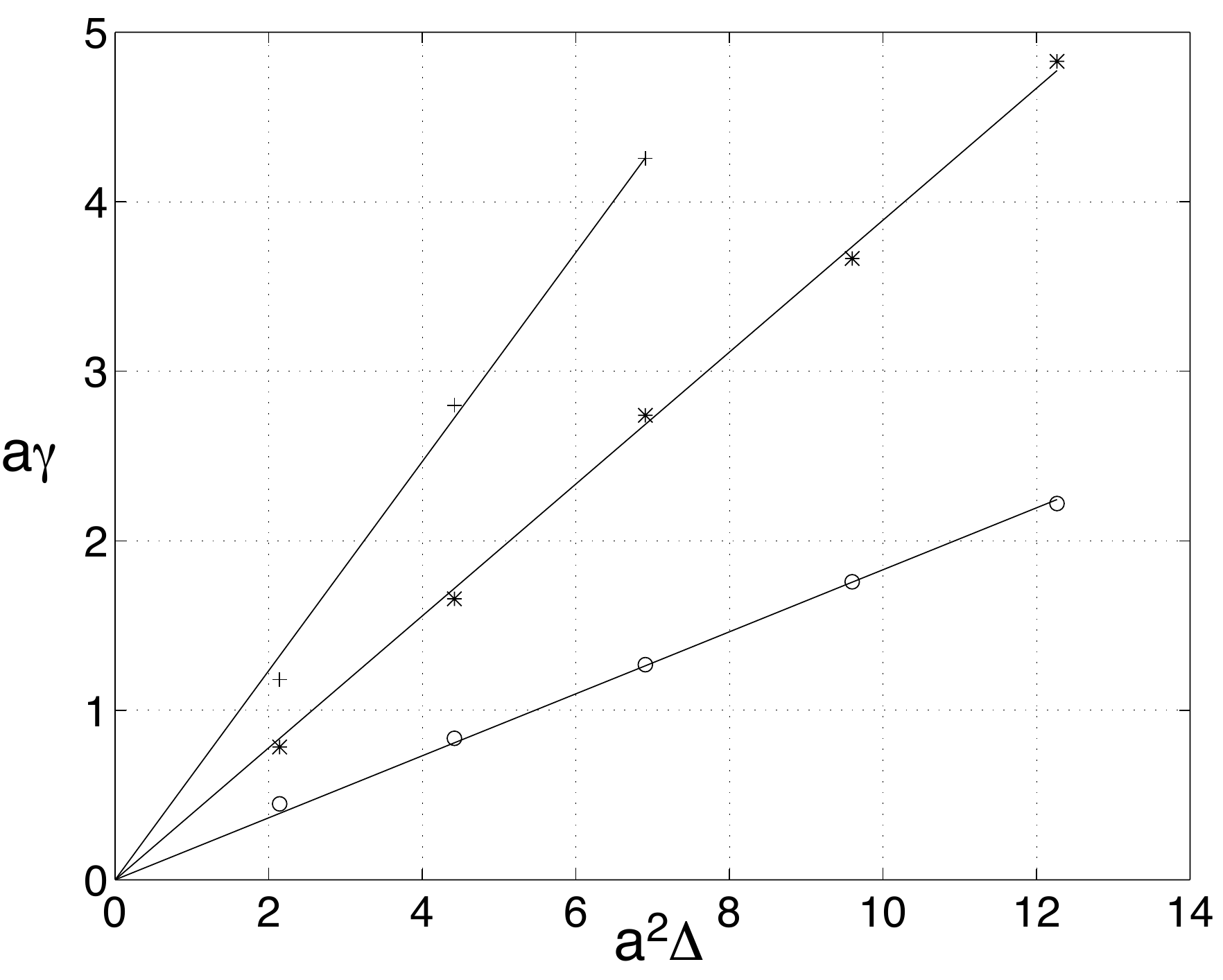}
  \includegraphics[width=0.325\textwidth]{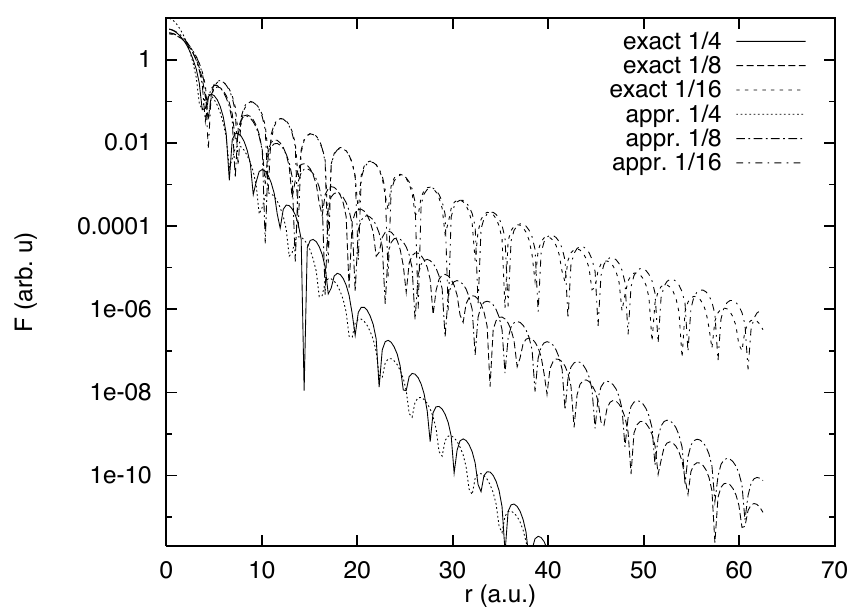}
  \caption{(a) and (b): Density matrix range ($\gamma$) as a function of gap
    ($\Delta$) for (a) a one dimensional insulating system and (b) a
    three dimensional insulating system (from
    \cite{Ismail-Beigi:1999zk}). (c) Density matrix range in a metal
    at a range of temperatures, given as a fraction of $k_F$ (from
    \cite{Goedecker:1998qw}).  Reprinted figures with permission from S. Ismail-Beigi
    \emph{et al.}, Phys. Rev. Lett. \textbf{82}, 2127 (1999) and
    S. Goedecker, Phys. Rev. B \textbf{58}, 3501 (1998).  Copyright
    (1999) and (1998) by the American Physical Society.}
  \label{fig:DMillustration}
  
\end{figure}

Further work on the principle of nearsightedness\cite{Prodan:2005cv}
underlined the earlier results that the decay  is proportional to
$\Delta$ for 1D to 3D systems, and some interacting systems.  There is
a well-motivated suggestion that disorder increases the density matrix
range for insulators and decreases it for metals.  Numerical studies
of the range of the density matrix for the Anderson model with varying
levels of disorder\cite{Sacksteder:2005xw} have shown that coherence
is strongly affected by disorder, with the exponential localisation
depending inversely on disorder.  Density matrix decay for both
Hartree-Fock and DFT has been plotted for different
systems\cite{Rubensson:2011mi}. 
It has also been shown that the correlation between fermionic
operators is exponentially localised at non-zero
temperatures\cite{Hastings:2004kb}.  Figs.~\ref{fig:DMillustration}a
and b
show the behaviour of $\gamma$ as a function of gap ($\Delta$) for 
model insulating systems.  Fig.~\ref{fig:DMillustration}a plots the
behaviour for a periodic one-dimensional potential, period $a$; this is clearly linear 
for a weak potential (small gap), while stronger potentials
are more complex. Fig.~\ref{fig:DMillustration}b shows the behaviour
for a simple cubic array of Gaussian
potentials along different directions in the crystal; the
linear dependence of density matrix decay on gap is clear in all directions.  By
contrast, Fig.~\ref{fig:DMillustration}c shows the spatial decay of
the density matrix in a metal, comparing exact results with a simple
model which simplifies to give decay proportional to $k_BT/k_F$ for
different values of $k_BT$ as a fraction of $k_F$.

Further analytic studies have been performed on specific systems.  A study of a simple cubic TB model\cite{Taraskin:2002bd} allowed the derivation of analytic results, showing both the exponential decay and a power-law prefactor for the density matrix.  The decay length found depended on both the gap and the hopping integrals (specifically, the decay length $\lambda \propto \Delta/t$ for hopping integral $t$); this work was extended to tight binding models of metals\cite{Taraskin:2002sw}, where a power law decay was found. Related earlier work using a simplified approach to density functional theory (a Sankey-Niklewski approach) for Si, C and Al\cite{Zhang:2001fh} found algebraic decay for Al which was very close to the nearly-free electron model, with anisotropic exponential decay in Si and C (with the directions along bonds showing the slowest decay).
A one-body density matrix with TB model for insulators and decay found 1D and 2D analytic results with a universal power law times exponential but second energy scale emerges when hopping modulates, so that the decay is not entirely gap dependent\cite{Jedrzejewski:2004lo,Jedrzejewski:2005if}.   
An extension of this work to 2D anisotropic
hopping\cite{Jedrzejewski:2007ec} showed exponential localisation, but
not with the same form as isotropic hopping; in particular, the
correlation length does not vanish as the gap closes. 

While much progress has been made in extending the analytic results
for density matrix range, particularly based on Wannier functions, it
is clear that there is still work to be done.  The relationship of
localisation to the important solid state problems of disorder and
polarisation (particularly as they extend into the new research areas
of topological insulators) is fascinating.  Furthermore, this area
is relevant to problems in graphene, where defects can have an
extremely strong effect on the electronic structure.  Overall, the
increasing complexity of a system changes the decay properties, and
the whole area is far from simple.

\subsection{Solving for the density matrix}
\label{sec:solving-dm}

Having established that the density matrix, and indeed the electronic structure, of matter is nearsighted, we can now turn to the matter of exploiting this nearsightedness in the search for the ground state.  There have been developments in this field for over thirty years, and as might be expected papers comparing the different methods\cite{Bowler:1997qw,Lippert:2000sj,Jordan:2005bs,Rudberg:2011ul} and general reviews of the subject have already been written\cite{Ordejon:1998ly,Goedecker:1999pv,Galli:1996cr,Galli:2000lq,Stephan:2000qm,Wu:2002ve,Shang:2010rr,Scuseria:1999kb}.  This review will, naturally, build on these excellent surveys.

The fundamental quantity in linear scaling techniques is the density
matrix, and the fundamental property of the density matrix is its
sparsity.  While there are methods which operate using, for instance,
Wannier functions, they still rely on the density matrix as the
fundamental quantity, and the short range of the functions to achieve
linear scaling.  To obtain a linear scaling method, we must impose a
range on the density matrix, which is a controllable approximation.
An appropriate localised basis must be used, which will make matrices
sparse; however, they must be also be stored and operated on as sparse
matrices, which requires significant extra effort.  Once these
preparations have been made, the computational effort required to
reach the ground state should scale linearly with system size.  It is
important to note that the choice of truncation is imposing an
additional constraint on the system, and that there will therefore be
an extra error (or energy difference) compared to an unconstrained
problem. 

The search for the ground state in terms of the density matrix cannot
be made in terms of the original, six dimensional object
$\rho(\mathbf{r},\mathbf{r}^\prime)$ (which is defined simply in terms
of the bands of a system in Eq.~(\ref{eq:4}) above).  The most common
approach is to work in terms of localised orbitals (also called
\emph{support functions}), and to assume that the density matrix is
separable\cite{Hierse:1994ta,Hernandez:1995hc}: 

\begin{equation}
  \label{eq:12}
  \rho(\mathbf{r},\mathbf{r}^\prime) = \sum_{i\alpha, j\beta} \phi_{i\alpha}(\mathbf{r}) K_{i\alpha j\beta} \phi_{j\beta}(\mathbf{r}^\prime),
\end{equation}
where atoms are labelled with roman letters $i$ and $j$ and orbitals
on atoms with greek letters $\alpha$ and $\beta$ (we note that the
density matrix is often notated only in terms of orbitals, as $K_{ij}$).
The only approximation made by this assumption is that the original density matrix had a finite number of non-zero eigenvalues (which is at most the number of local orbitals used).  However, this is not restrictive in the context of electronic structure calculations.  Much of the rest of this section is devoted to discussing different methods for finding $\left\{\phi_{i\alpha}(\mathbf{r})\right\}$ and $K_{i\alpha j\beta}$.  However, there are two further conditions which must be considered:

\begin{enumerate}
\item Correct electron number.  The electron number is given by:
  \begin{equation}
    \label{eq:13}
    N_e = 2\mathrm{Tr}\left[KS\right]
\end{equation}
where $S_{i\alpha j\beta} = \langle
\phi_{i\alpha}|\phi_{j\beta}\rangle$ is the overlap matrix between the
localised orbitals and we assume spin degeneracy, giving the factor of
two.  We will consider methods for imposing the correct electron
number below in Sec.~\ref{sec:pres-electr-numb}. 
\item Idempotency.   The density matrix is a projector onto the
  occupied subspace, and must have eigenvalues of either zero or one.
  Another way of writing this is, in both operator and real-space
  notation: 
  \begin{eqnarray}
    \label{eq:14}
    \hat{\rho}^2 &=& \hat{\rho}\\
    \rho(\mathbf{r},\mathbf{r}^\prime) &=& \int \mathrm{d}\mathbf{r}^{\prime\prime} \rho(\mathbf{r},\mathbf{r}^{\prime\prime}) \rho(\mathbf{r}^{\prime\prime},\mathbf{r}^\prime)
  \end{eqnarray}
This requirement is rather hard to impose exactly, and many approaches adopt a weaker restriction (often known as weak idempotency)\cite{Li:1993lg,Galli:1996cr}, where:
\begin{equation}
  \label{eq:15}
  0\le \lambda_\rho \le 1,
\end{equation}
for the eigenvalue of the density matrix $\lambda_\rho$.  
McWeeny\cite{McWeeny:1960zp} showed how an iterative scheme could be used to force an approximately idempotent matrix to exact idempotency; this will be discussed in Sec.~\ref{sec:direct-iter-appr}.
\end{enumerate}

As alluded to above, there are often situations where the localised
orbitals used to form the density matrix are not orthogonal.  This
leads to some complications in the formalism, which are described
in Sec.~\ref{sec:non-orthogonal-basis}.  Briefly, either the
inverse of the overlap is required (which can be difficult to find
for sparse matrices) or some form of orthogonalisation must be
applied.  If the density matrix for only the \emph{occupied} subspace
is required, then it is equal to the inverse overlap matrix of the
local orbitals (see, for instance,
\cite{Galli:1992yu,Ordejon:1993zk,Ordejon:1995yj,Mauri:1993lf}). 

\subsubsection{Direct and Iterative Approaches}
\label{sec:direct-iter-appr}

In this section, we consider two major approaches to linear scaling
density matrix search, which turn out to share considerable
theoretical background.  We choose to group these methods to emphasise
this shared background and stimulate further work on the development
of effective techniques.

Galli and Parrinello\cite{Galli:1992yu} noted that, instead of writing the density matrix in terms of the occupied eigenfunctions (as seen, for example, in Eq.~(\ref{eq:4}) above), it can be equally well be written in terms of the same number of non-orthogonal orbitals $\phi_i$:
\begin{equation}
  \label{eq:51}
  \rho(\mathbf{r},\mathbf{r}^\prime) = \sum_{ij} \phi_i^\star(\mathbf{r}^\prime) S^{-1}_{ij} \phi_j(\mathbf{r})
\end{equation}
(Note that the matrix $S^{-1}_{ij}$ is the inverse of the overlap matrix for
the orbitals, $S_{ij} = \langle \phi_i | \phi_j \rangle$, which
automatically makes the matrix $\rho(\mathbf{r},\mathbf{r}^\prime)$
idempotent.)
They then impose localisation constraints (in this case, using
bucket-like potentials).  The formulation allows the removal of any
explicit orthogonalisation between eigenfunctions (which leads to the
asymptotic cubic scaling behaviour seen in normal methods).  By taking
advantage of the sparsity of the overlap and Hamiltonian matrices,
linear scaling can also be achieved; the proposed methods for
exploiting locality (involving local volumes where most operations are
performed) have points of contact with both the divide-and-conquer
method (Sec.~\ref{sec:divide-conquer}) and the FFT box method (see
below under the \textsc{onetep} method in Sec.~\ref{sec:implementations}).  This method has been
developed further to include unoccupied states, with a real-space grid
as the basis (using finite differences, as described in
Sec.~\ref{sec:finite-differences})\cite{Fattebert:2000yr}, with an
extension to allow the centres of the localisation regions to adapt
and move during molecular
dynamics\cite{Fattebert:2004xi,Fattebert:2006hf,Fattebert:2008th}.
These implementations are often not truly linear scaling, as the
inverse overlap matrix is calculated exactly, though there are many
methods to remove this final barrier; Galli and Parrinello proposed
solving for the dual basis functions by an iterative application of
$(I-S)$ (described in more detail in
Sec.~\ref{sec:non-orthogonal-basis}). 

\begin{figure}[h]
  \centering
  \includegraphics[width=0.7\textwidth]{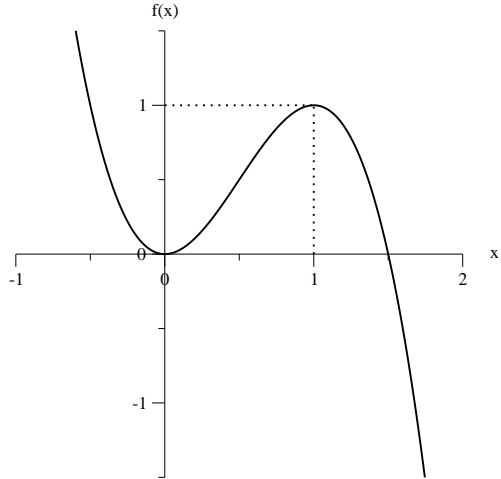}
  \caption{The McWeeny purification function, $f(x) = 3x^2 - 2x^3$}
  \label{fig:McW}
\end{figure}

A variational  approach using the McWeeny purification
transformation\cite{McWeeny:1960zp} has been used in a wide variety of
approaches and
methods\cite{Li:1993lg,Daw:1993so,Nunes:1994pi,Hernandez:1995hc,Hernandez:1996bf,Daniels:1997dw,Millam:1997bu,Bowler:1999if,Challacombe:1999mb,Head-Gordon:2003gy,Li:2003gb
  ,Haynes:2008zr}; we will refer to it as the Density Matrix
Minimisation (DMM) method.  The McWeeny transformation uses the function:
\begin{equation}
  \label{eq:52}
  f(x) = 3x^2 - 2x^3
\end{equation}
which is plotted in Fig.~\ref{fig:McW}.  It has the property that for $-\frac{1}{2} \le x \le \frac{3}{2}, 0\le f(x) \le 1$.  This property is used to impose idempotency during a variational search: if an input matrix has eigenvalues bounded by $-\frac{1}{2}$ and $\frac{3}{2}$, then the output will be bounded by $0$ and $1$.  This is known as \emph{weak} idempotency.  If an auxiliary density matrix is taken as $\sigma(\mathbf{r},\mathbf{r}^\prime)$, then the true density matrix $\rho(\mathbf{r},\mathbf{r}^\prime)$ is defined as:
\begin{equation}
  \label{eq:53}
  \rho = 3\sigma * \sigma - 2 \sigma * \sigma * \sigma
\end{equation}
where $\sigma * \sigma(\mathbf{r},\mathbf{r}^\prime) = \int \mathrm{d}\mathbf{r}^{\prime\prime} \sigma(\mathbf{r},\mathbf{r}^{\prime\prime})\sigma(\mathbf{r}^{\prime\prime},\mathbf{r}^\prime)$.  If the band energy (written as $2\mathrm{Tr}[\rho H]$) is varied with respect to the elements of the \emph{auxiliary} density matrix, then the resulting minimisation is variational and converges to idempotency at the ground state.  It can easily be shown that:
\begin{equation}
  \label{eq:54}
  \frac{\delta E}{\delta \sigma} = 3(\sigma H + H \sigma) - 2(\sigma^2H + \sigma H \sigma + H \sigma^2)
\end{equation}
Provided some method is chosen to account for the chemical potential of the electrons (either adjusting it to keep the number of electrons fixed, or fixing the potential, as described in Sec.~\ref{sec:pres-electr-numb}), minimising the total energy with respect to the auxiliary density matrix elements leads to the ground state in a variational manner.  Imposing sparsity on the density matrix (using  methods described in Sec.~\ref{sec:parallelisation}) and the Hamiltonian results in linear scaling.  

This approach was first proposed for orthogonal tight
binding\cite{Li:1993lg,Daw:1993so} and subsequently extended to
non-orthogonal bases\cite{Nunes:1994pi} (discussed further in
Sec.~\ref{sec:non-orthogonal-basis}) and density functional
theory\cite{Hernandez:1995hc,Hernandez:1996bf} as well as finite
electronic temperatures\cite{Corkill:1996fe}; it is often referred to
as the LNV method (after Li, Nunes and Vanderbilt).  The utility of mixing
an iterative McWeeny process to restore idempotency with the
minimisation has also been
explored\cite{Palser:1998fd,Bowler:1999if,Challacombe:1999mb}.
Implementations are numerous\cite{Hernandez:1995hc,Carlsson:1995bs,
  Hernandez:1996bf,Goringe:1997cy,Daniels:1997dw,Millam:1997bu,Bowler:2002pt,Li:2003gb
  ,Haynes:2008zr} and cover quantum chemistry approaches as well as
density functional theory,  and use minimisation techniques such as conjugate gradients and direct inversion in the iterative subspace.
Implementations of Car-Parrinello molecular dynamics have also been described\cite{Schlegel:2001ez,Li:2009mf}, both for a simple fictitious electronic mass\cite{Schlegel:2001ez} and a variable fictitious mass to optimise convergence \cite{Li:2009mf}.
A closely related method to the DMM uses the sign matrix
\cite{Beylkin:1999fx,Nemeth:2000cy}, which is equivalent to expanding
the Fermi matrix (as in the Fermi Operator Expansion described in Sec.~\ref{sec:recursion}); another approach including \emph{implicit}
purification for finite temperatures has been
derived\cite{Niklasson:2003gd}.

An alternative approach using similar ideas for purification starts by noting that the density matrix and the Hamiltonian should commute (they share eigenvectors, as the density matrix projects onto the occupied subspace).  By starting with an initial density matrix which commutes with the Hamiltonian, and using iterative purification methods, it can be proved that the ground state density matrix is the result\cite{Palser:1998fd}.  This should not be surprising, as the density matrix can be written as a function of the Hamiltonian:
\begin{equation}
  \label{eq:55}
  \mathbf{\rho} = \theta(\mu \mathbf{I} - \mathbf{H}),
\end{equation}
where $\mu$ is the Fermi level and $\theta(x)$ is the Heaviside step function (with $\theta(x) = 1,
x>0$ and $\theta(x) = 0, x<1$).  Approaches to expanding out the step function (or the Fermi function as it becomes at finite temperature) using polynomials and recursion are discussed in Sec.~\ref{sec:recursion}.

The method first proposed\cite{Palser:1998fd} used the original McWeeny transform for grand canonical (i.e. fixed Fermi level) minimisation, with the initial density matrix given by:
\begin{eqnarray}
  \label{eq:56}
  \mathbf{\rho}_0 &=& \frac{\lambda}{2}(\mu \mathbf{I} - \mathbf{H}) + \frac{1}{2}\mathbf{I}\\
  \mathbf{\rho}_{n+1} &=& 3\mathbf{\rho}_n^2 - 2\mathbf{\rho}_n^3
\end{eqnarray}
with $\lambda$ chosen to take the smaller value of
$1/(H_\mathrm{max}-\mu)$ and $1/(\mu-H_\mathrm{min})$, so that the
eigenvalues of $\mathbf{\rho}$ lie between 0 and 1.  The same work also proposed a canonical method with an adapted version of McWeeny's purification function, 
\begin{eqnarray}
  \label{eq:57}
  \mathbf{\rho}_{n+1} &=& \cases{ \left[(1-2c_n)\mathbf{\rho}_n + (1+c_n)\mathbf{\rho}_n^2 - \mathbf{\rho}_n^3\right]/(1-c_n) \mathrm{\ if\ } c_n\le\frac{1}{2}\cr
    \left[(1+c_n)\mathbf{\rho}_n^2 - \mathbf{\rho}_n^3\right]/c_n \mathrm{\ if\ } c_n\ge\frac{1}{2}  }\\
  c_n &=& \frac{\mathrm{Tr}[\mathbf{\rho}_n^2 - \mathbf{\rho}_n^3]}{\mathrm{Tr}[\mathbf{\rho}_n - \mathbf{\rho}_n^2]}
\end{eqnarray}
This function allows the unstable fixed point of the McWeeny function
at $x=c$ (where $f(x) = x$ and $f^\prime(x) \ge 1$) to move away from
$c=\frac{1}{2}$ to lie between $c=0$ and $c=1$.  As a result, the
electron number is conserved throughout the iteration.  It is
important to note\cite{McWeeny:1960zp,Palser:1998fd} that the grand canonical
iteration is equivalent to a steepest-descent minimisation of the
function $f(\rho) = tr[\rho^2(I-\rho)^2]$, which links to the orbital
minimisation method described below.

This basic idea has been further extended and elaborated in numerous ways.  As noted above, this approach has been combined with variational minimisation, both as an initialisation for the density matrix\cite{Bowler:1999if} and to restore idempotency\cite{Challacombe:1999mb}. An iterative purification was introduced as a way of correcting density matrices following Car-Parrinello steps (rather than imposing orthogonality)\cite{Schlegel:2001ez}. A larger set of generic purifications was proposed\cite{Kryachko:2000dz}, based on the equation $T_n(P) = I - (I-P)^n, n \ge 2$, and later extended to systems with unoccupied states\cite{Holas:2001fu} (though it has been suggested that this extension is no more efficient than the original McWeeny transform\cite{Habershon:2002ss}).  Similar high-order polynomials have been derived elsewhere\cite{Niklasson:2002dw}; these methods can be formulated in terms of only a few matrix multiplications (e.g. four multiplies for a ninth order polynomial), but are not  in general use.  The difference between methods which have the same number of filled orbitals as eigenstates, and therefore density matrix eigenvalues of 1 only, and those with more orbitals and hence eigenvalues of 0 and 1 is key in these expansions: an expansion which only has to consider filled states generally requires fewer matrix multiplications (and the form $2P-P^2$ proposed by Stechel\cite{Stechel:1994mk} discussed below is a key example).  

Building on this idea, Niklasson\cite{Niklasson:2002ef} suggested a
trace-correcting approach, using different polynomial expansions for
purification depending on whether the trace of the density matrix is
below or above the correct number of electrons, $N$:
\begin{equation}
  \label{eq:58}
  T_m(x) = \cases{
    1-(1-x)^m[1+mx],\ N_e \ge N\cr
    x^m[1+m(1-x)],\ N_e \le N}
\end{equation}
If $m=2$, then the original McWeeny expansion is recovered.  When $m=1$ the two polynomials are $x^2$ and $2x-x^2$; this has become known as the TC2 (trace-correcting second order) method.  It has been extended to spin-unrestricted methods\cite{Xiang:2005tg} and non-orthogonal bases\cite{Niklasson:2005bq} (discussed more fully in Sec.~\ref{sec:non-orthogonal-basis}; the main change to the algorithm is to require the overlap matrix in density matrix products, so $P_\perp^2$ becomes $PSP$ where $P_\perp$ is the density matrix in an orthogonal basis, and to require the inverse of the overlap for initialisation), and compared to LNV\cite{Jordan:2005bs}, with some advantage found especially for high and low filling (though it is important to note that these iterative methods are not variational, and so forces cannot be calculated using the Hellmann-Feynman theorem).

Two closely-related approaches use both the particle density matrix
and the hole density matrix (defined as $Q = I-P$).
Mazziotti\cite{Mazziotti:2001qc,Mazziotti:2003ec} recasts and combines
the formulae of Ref.~\cite{Holas:2001fu} in terms of the particle and hole matrices (with
the method depending on the order -- hole or particle first), and
shows significant computational speed up compared to the McWeeny
approach.  A similar technique\cite{Kohalmi:2005mq,Szakacs:2008bv}
uses hole and particle density matrices, but includes an
empirically-adjusted parameter to optimise convergence.  

Niklasson\cite{Niklasson:2003mw} also introduced a series of trace
\emph{resetting} algorithms.  This method combines a purification
polynomial, $F(x)$, and a reoccupation polynomial, $G(x)$ which
between them purify the density matrix and keep its trace correct
within a certain domain of applicability.  If the trace falls outside
the domain, then it is reset by application of the TC2 method.  The
following quartic polynomials have been empirically found to be
effective: 
\begin{eqnarray}
  \label{eq:59}
  F(x) &=& x^2(3x-3x^2)\\
  G(x) &=& x^2(1-x)^2,
\end{eqnarray}
leading to the TRS4 algorithm.  Tests on both high- and low-filling
problems (C$_{60}$, a zeolite, chlorophyll  and water clusters) show
that the approach is more efficient and more accurate than the original
Palser-Manolopoulos method; at mid-filling the methods are of similar
efficacy. 

A comprehensive comparison of LNV-based minimisation and iterative
methods\cite{Rudberg:2011ul} finds considerable efficiency gains using
purification, though it should be emphasised that these methods are
not variational, and hence force calculations will be complicated.
Studies of error propagation for the trace resetting method
(TRS4)\cite{Niklasson:2003mw} with magnitude-based truncation (see
Sec.~\ref{sec:parallelisation} for discussion of different methods  of
enforcing sparsity on matrices) as well as the TC2
method\cite{Rubensson:2005vs} show that, applying truncation at
different stages, errors can be rigorously controlled.  A method for
controlling errors within the TC2 method has been proposed and
demonstrated on water clusters\cite{Rubensson:2008jl}.  These
approaches show that it is possible to pursue high accuracy within linear
scaling methods.

Despite this plethora of methods, some fairly simple conclusions can
be drawn.  First, that McWeeny's original proposal has been remarkably
robust, and has led to significant important physics.  Second, that
for variational methods the formulation of McWeeny's purification
algorithm in terms of an auxiliary density
matrix\cite{Li:1993lg,Nunes:1994pi} is the method of choice.  It is
probably the most commonly used method, and is ideal for
distance-based truncation schemes.  Thirdly, for iterative approaches
improvements over the original McWeeny formula can be made (e.g. the
TC2 method) and error control can be introduced.  The iterative
methods are not variational, but are simple and are commonly used in
conjunction with tolerance-based truncation.

Another class of methods, which is closely related to the minimisation
and iteration techniques just described, builds the Wannier functions
of a system by direct minimisation \emph{without constraint}.  The ideas were
proposed independently by Mauri, Galli and Car\cite{Mauri:1993lf} and
Ordej\'on \emph{et al.}\cite{Ordejon:1993zk}, though the method is
generally referred to as MGC; given that the heart of the method is
orbital minimisation, we suggest that the method be known as the
Orbital Minimisation Method (OMM),
following~Ref.~\cite{Tsuchida:2007kq}.  In both cases, the density
matrix is defined in terms of $N/2$ localised Wannier functions which
tend to orthogonality as the minimisation progresses.  By introducing
an approximation to the inverse overlap (which coincides with the
density matrix if only occupied states are considered), we can write\cite{Mauri:1993lf}: 
\begin{eqnarray}
  \label{eq:60}
  \rho(\mathbf{r}) &=& \sum_{ij} \phi_i(\mathbf{r}) Q_{ij} \phi_j(\mathbf{r})\\
  \label{eq:62}
  Q &=& \sum^K_0 (I-S)^K
\end{eqnarray}
where $K$ is an odd integer.  When $K=1$, then the approximation is $Q = 2I -S$.  This same formula was derived by Ordej\'on \emph{et al.}\cite{Ordejon:1993zk} starting from a Lagrange multiplier approach to enforce orthogonality (adding a term $\sum_{ij} \Lambda_{ij}(S_{ij} - \delta_{ij})$ to the band energy) and substituting the value of the multipliers \emph{at the minimum} ($\Lambda_{ij} = H_{ij}$) for all values of the orbitals.  After a little re-arrangement, this yields the same functional; the family of polynomials can also be derived by rearranging the original equation to $\Lambda = H + (I-S)\Lambda$ and treating it as a recurrence relation\cite{Ordejon:1995yj}. (The Ordej\'on \emph{et al.} approach has some commonality with a method due to  Wang and Teter described in Sec.~\ref{sec:penalty-functionals}, though is more flexible.) It can be proved\cite{Mauri:1993lf,Mauri:1994kc,Ordejon:1993zk,Ordejon:1995yj} that the energy is a minimum at orthogonality provided that the Hamiltonian is negative definite (which can be enforced by applying a rigid shift to the Hamiltonian of an amount $\eta$).  The functional often quoted for these methods is then:
\begin{equation}
  \label{eq:61}
  E\left[Q,\left\{\phi\right\}\right] = 2\left( \sum_{ij}^{N/2} Q_{ij}T_{ij} + F[\rho]\right) + \eta\left(N - \int d\mathbf{r} \rho(\mathbf{r})\right),
\end{equation}
where $\rho$ is defined in Eq.~(\ref{eq:60}) and $Q$ is defined in Eq.~(\ref{eq:62}).  Given a suitable basis set (e.g. localised atomic orbitals, as described in Sec.~\ref{sec:real-space-local}), the minimisation expresses the orbitals $\left\{\phi_i\right\}$ in terms of this basis, and seeks the minimum energy in terms of the expansion coefficients, while also applying a localisation criterion to the orbitals.  At the minimum, the resulting orbitals will be orthogonal, by construction.

However, as described, the method has a serious problem: there are
large numbers of local minima, leading to severe convergence
problems\cite{Mauri:1994kc,Ordejon:1995yj,Kim:1995yi}.  While the
method can be used, and has been implemented in this form, practical
solutions require some input guess for the orbitals based on chemical
intuition; a more general solution to this problem is required.  Kim
\emph{et al.}\cite{Kim:1995yi} showed that generalising the original
formulation so that more orbitals than just those spanning the
occupied subspace could be used.  This means that the orbitals are not
orthogonal at the energy minimum, but has the advantage that local
minima are avoided.  Hierse and Stechel\cite{Hierse:1994ta} also
generalised the original OMM approach,  using a different approximation for the
trial density matrix, using Eq.~(\ref{eq:12}).  Yang\cite{Yang:1997nx}
used a variational approach to derive a general functional of this class which
can use unoccupied orbitals and which only requires a Hermitian matrix
(as opposed to a positive definite matrix required in previous work\cite{Hierse:1994ta}).
It is interesting to note that Kohn suggested a method for building
orthogonal Wannier functions\cite{Kohn:1973rt} which may be seen as a
precursor to these methods; he noted that, unless the starting
functions were reasonably close to the final functions, there were
multiple minima.

There are numerous implementations of these methods: the original
papers\cite{Mauri:1993lf,Mauri:1994kc,Ordejon:1993zk,Ordejon:1995yj};
the generalised
versions\cite{Kim:1995yi,Hierse:1994ta,Sternberg:1999bs,Sasaki:2006pd};
in parallel\cite{Itoh:1995yq,Shellman:2003rt}; with ultra-soft
pseudopotentials\cite{Hoshi:1997jy}.  A real-space implementation of
similar ideas\cite{Fattebert:2006hf} uses exact inversion of the
overlap, thus not forming a strict linear scaling method (though the
cubic scaling part will have a small prefactor).  A method for projecting localised functions onto the occupied subspace\cite{Stephan:1998lt} using the Fermi Operator Expansion described in Sec.~\ref{sec:recursion} showed improvements in convergence, but does not solve the problem of initial functions.  

More recently, Tsuchida\cite{Tsuchida:2007kq,Tsuchida:2008ai} has
proposed an \emph{augmented} orbital minimisation method to overcome
the convergence problems.  The essence of the method is to define
highly localised kernel functions (which contain the centre of a
localisation region, and are typically around 1\,$a_0$ in radius); the
localised orbitals are then forced to be orthogonal to the kernel
functions.  This change is sufficient to make the method stable and
rapidly convergent.  Indeed, the applications described in
Sec.~\ref{sec:applications} suggest that this is one of the most
successful linear scaling approaches.

At one level the density matrix-based methods appear quite different
to the orbital minimisation methods: in the first case, the elements
of the density matrix are the variables, while in the second the
orbitals themselves are the variables.  However, we can make a close
connection to the methods described above, by considering alternative
ways of writing the density matrix; these demonstrations have been
developed by a number of
authors\cite{Nunes:1994rr,Hernandez:1996bf,Galli:1996cr,Ordejon:1998ly}.
If the OMM functional $Q$ is written in terms of density matrices
(compare Eq.~(\ref{eq:60})), then we find:
\begin{equation}
  \label{eq:63}
  \tilde{\rho}(\mathbf{r},\mathbf{r}^\prime) = 2\rho (\mathbf{r},\mathbf{r}^\prime) - \rho^2(\mathbf{r},\mathbf{r}^\prime)
\end{equation}
where $\rho (\mathbf{r},\mathbf{r}^\prime) = \sum_i \chi_i(\mathbf{r})\chi_i(\mathbf{r}^\prime)$ is a trial density matrix, and $2\rho - \rho^2$ acts as a purification transformation.  This can be generalised if the localised orbitals $\chi_i(\mathbf{r})$ are expanded in a basis:
  \begin{equation}
    \label{eq:64}
    \chi_i(\mathbf{r}) = \sum_\alpha b_{i\alpha} \phi_\alpha(\mathbf{r})
  \end{equation}
We can then write the trial density matrix in exactly the form of Eq.~(\ref{eq:12}), with $L_{\alpha\beta} = \sum_i b_{i\alpha}b_{i\beta}$.  The purification transformation is then written $2L - LSL$, with $S_{\alpha\beta} = \langle \phi_\alpha|\phi_\beta\rangle$.  Setting $L=I$ gives $S$ as the density matrix, and recovers the original OMM form.

\begin{figure}[h]
  \centering
  \includegraphics[width=0.7\textwidth]{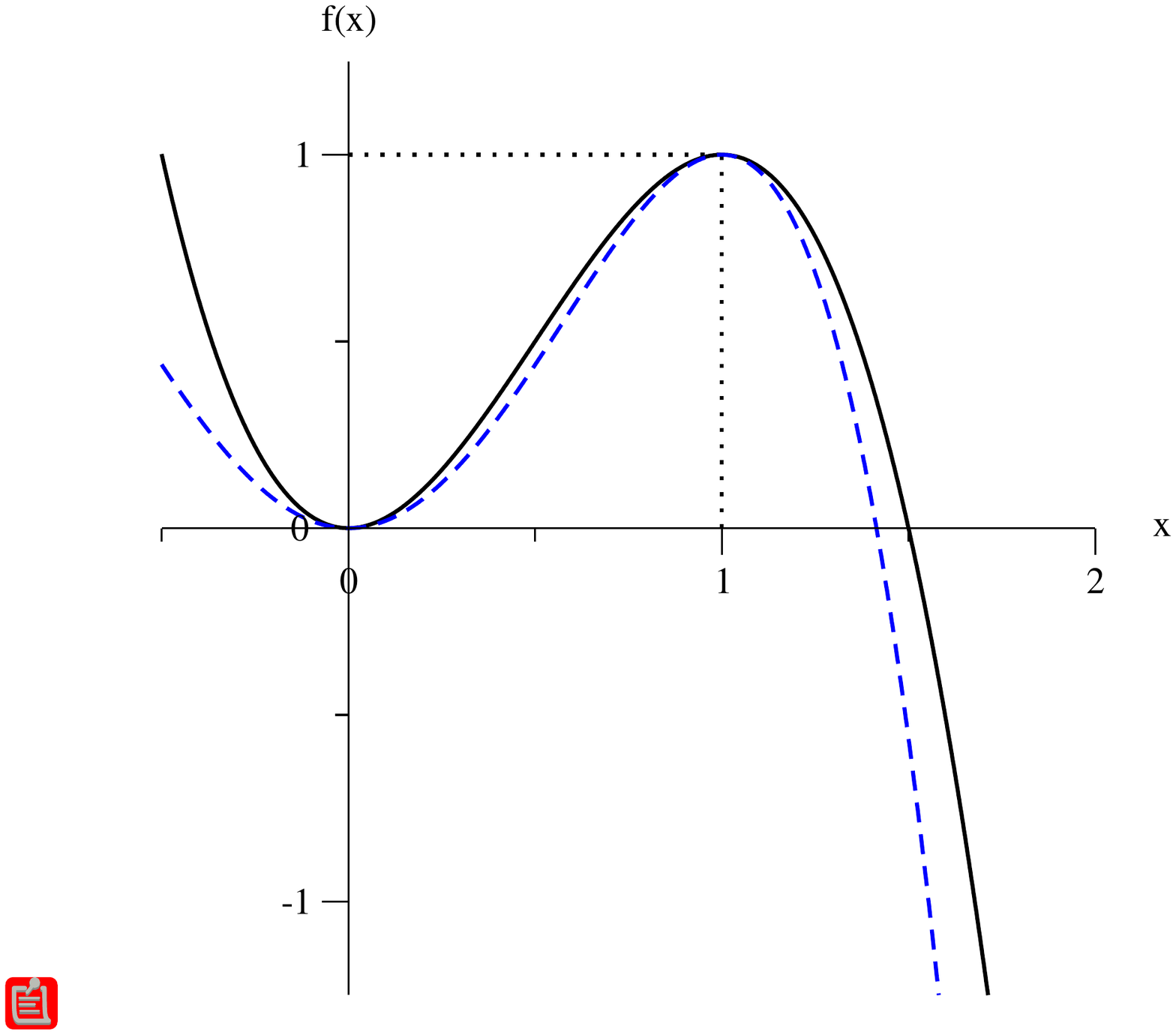}
  \caption{The McWeeny purification function (solid black line) plotted with the implicit purification function of the orbital minimisation methods (dashed blue line).  (Note that $x$ is taken to mean $\kappa_\rho$ in the latter case.)}
  \label{fig:McWMGC}
\end{figure}

Returning to Eq.~(\ref{eq:63}), we note that $\rho$ is required to be positive semidefinite, and that the eigenvalues can thus be represented by\cite{Hernandez:1996bf}:

\begin{eqnarray}
  \label{eq:65}
  \lambda_\rho &=& \kappa^2_\rho\\
  \label{eq:66}
  \lambda_{\tilde{\rho}} &=& \lambda_\rho(2-\lambda_\rho) = \kappa^2_\rho(2-\kappa^2_\rho)
\end{eqnarray}
where $\kappa_\rho$ is real.  The quartic function will lie in the range $[0,1]$ when $-\sqrt{2}\le \kappa_\rho \le \sqrt{2}$, with turning points at $\lambda_{\tilde{\rho}}=0$ and $1$.  This function is plotted along with the original McWeeny function in Fig.~\ref{fig:McWMGC}.  It is clear that the two methods have a very similar form between zero and one, but the quartic potentially will introduce more local minima.  The practical convergence rates for the methods (distinguishing between the MGC and Ordej\'on \emph{et al.} methods) has been examined\cite{Lippert:2000sj}, with the convergence rate in the simple systems (bulk Si and C) found to be dominated by the spectral properties of the Hamiltonian rather than the method.

The DMM and OMM methods share a fundamental connection, but the DMM is
far more commonly implemented.  This is in part due to simplicity and
stability, and also to the single minimum present in the functional.
The augmented OMM promises significant improvements and suggests that
a new resurgence of OMM applications might be seen; either way, these
two methods are the most commonly used and applied linear scaling
methods (see Sec.~\ref{sec:implementations} for a description of
implementations in specific codes).

\subsubsection{Divide and Conquer}
\label{sec:divide-conquer}

The divide and conquer
method\cite{Yang:1991wk,Yang:1991fi,Yang:1992bx} is conceptually the
simplest of the linear scaling approaches.  Taking advantage of the
nearsightedness of electronic structure, the system is divided into
separate, small subsystems whose electronic structure is solved
exactly, for example by diagonalisation.  (A similar method was
proposed almost simultaneously \cite{Cortona:1991ov}, which combines
the idea of dividing up the cell into subsystems with the ideas of
orbital-free DFT (see Sec.~\ref{sec:orbital-free-dft}); however, this
method has not proved significant.) A partition function $p_\alpha$ is
defined for each subsystem, such that $\sum_\alpha
p_\alpha(\mathbf{r}) = 1$ for all points in the system.  If we rewrite
the charge density, then we have: 

\begin{eqnarray}
  \label{eq:16}
  n(\mathbf{r}) &=& 2\langle\mathbf{r}|\theta(\epsilon_F - \hat{H})|\mathbf{r}\rangle\\
  &=& 2\sum_\alpha p_\alpha(\mathbf{r})\langle\mathbf{r}|\theta(\epsilon_F - \hat{H})|\mathbf{r}\rangle = \sum_\alpha n_\alpha(\mathbf{r})\\
  n_\alpha(\mathbf{r}) &=& 2p_\alpha(\mathbf{r})\langle\mathbf{r}|\theta(\epsilon_F - \hat{H})|\mathbf{r}\rangle
\end{eqnarray}
where $\theta(x)$ is the Heaviside step function.  
However, this requires some approximation of the system-wide Kohn-Sham Hamiltonian.  Making a \emph{local} approximation to the Hamiltonian requires some way to determine $\epsilon_F$; Yang suggested writing $\tilde{n}_\alpha(\mathbf{r}) = n_\alpha(\mathbf{r})$, with:
\begin{equation}
  \label{eq:17}
  \tilde{n}_\alpha(\mathbf{r}) = 2p_\alpha(\mathbf{r})\langle\mathbf{r}| f(\epsilon_F - \hat{H}_\alpha)|\mathbf{r}\rangle 
\end{equation}
where $\hat{H}_\alpha$ is the approximation of the Kohn-Sham Hamiltonian in the subsystem.  Yang assumed non-orthogonal basis functions purely localised within the subsystem, $\left\{\phi^\alpha_j(\mathbf{r})\right\}$ and defined:
\begin{equation}
  \label{eq:18}
  \hat{H}_\alpha = \sum_{ij} |\phi^\alpha_i\rangle H_\alpha^{ij} \langle \phi^\alpha_j|
\end{equation}
where $ H_\alpha^{ij} = \sum_{kl}(S^{-1}_\alpha)^{ik} (H_\alpha)_{kl}(S^{-1}_\alpha)^{lj}$, $S^{-1}$ is an inverse matrix for the overlap $(S_\alpha)_{ij} = \langle \phi^\alpha_i|\phi^\alpha_j\rangle$ and $(H_\alpha)_{ij} = \langle \phi^\alpha_i|\hat{H}|\phi^\alpha_j\rangle$.  The only system-wide constraint is the electron number, which gives a value for $\epsilon_F$.  A finite temperature is used to ensure that the electron number is monotonic and continuous as $\epsilon_F$ is varied.  The approximate band energy is given by:
\begin{equation}
  \label{eq:19}
  \tilde{E} = 2\sum_\alpha \sum_i f(\epsilon_F - \epsilon^\alpha_i)\epsilon^\alpha_i \langle \psi^\alpha_i| p^\alpha | \psi^\alpha_i \rangle
\end{equation}
Once a partitioning of the system, and a local basis set for each subsystem, has been chosen, the solution for each is found by direct diagonalisation of $H^{ij}_\alpha$ and $(S_\alpha)_{ij}$.  The system-wide Fermi level is fixed by the requirement on electron number, which is then used to construct a charge density, and fed back into the local Hamiltonians.  A self-consistent solution is thus found for the entire system.

The formalism has been extended to solid state systems\cite{Zhu:1996ct}: here, the idea of buffer atoms is required.  For each subsystem, a buffer zone of a certain width from the main system is retained to remove surface effects.  Comparing the error as a function of the size of the buffer, the authors conclude that cohesive energy is converged to 0.1 eV when there are 40-50 atoms in buffer, but the electronic structure (in particular the density of states, or DOS) is much slower to converge with buffer size.  Forces have been derived for the method\cite{Zhao:1995ya}, though the choice of applying the divide and conquer approach to the total system force (rather than differentiating the divide and conquer energy) may lead to slight discrepancies between the force and energy gradients.  A more recent study of forces\cite{Kobayashi:2011wd} showed that it is possible, though time-consuming, to derive an exact gradient, and proposed another approximate solution.

A further refinement based the partitioning and calculations on the system density matrix  rather than the wavefunctions\cite{Yang:1995fb}: the charge density is built from the density matrix.  A Mulliken-like partitioning is used for assigning the density matrix to subsystems; the partition \emph{matrix} is defined as:
\begin{equation}
  \label{eq:20}
  p^\alpha_{ij} = 
  \cases{
    1 &if $i, j \in \alpha$\\
    1/2 &if $i \mathrm{\ or\ } j \in \alpha$\\
    0 &if $i, j \notin \alpha$
  }
\end{equation}
and then we have:
\begin{eqnarray}
  \label{eq:21}
  K_{ij} &=& \sum_\alpha p^\alpha_{ij} K_{ij} = \sum_\alpha K^\alpha_{ij}\\
  K^\alpha_{ij} &=& 2p^\alpha_{ij}\sum f(\epsilon_F - \epsilon^\alpha_m)C^\alpha_{im}C^\alpha_{jm}
\end{eqnarray}
This method has the advantage that it removes the need to perform integrals between subsystem eigenstates and the projector function. An alternative partitioning\cite{Dixon:1996ta} based on the number of subsystems a density matrix occupies has also been proposed.

The divide and conquer method has been implemented a number of times,
for instance in the \textsc{OpenMX} code\cite{Ozaki:2006uq} (in this
case in combination with a recursion method, as described in
Sec.~\ref{sec:recursion}), in the \textsc{siesta}
code\cite{Cankurtaran:2008uq} and in an extremely large scale
approach\cite{Shimojo:2005wm,Nakano:2007ez,Shimojo:2008gf} based on
hierarchical real-space grids, which scales over multiple HPC centres.
An implementation using the QUAMBO approach to localised orbitals also
exists\cite{Yao:2009lj}.  The divide and conquer approach has also
been applied to quantum chemistry (up to the CCSD
level)\cite{Kobayashi:2008ga,kowski:2010kc}, and implementation of
exact exchange interactions has been described\cite{Akama:2007gd}
(where the accuracy is very dependent on the size of the buffer
region).  An assessment of the computational time required for
different systems\cite{Prodan:2005cv} with divide and conquer has been
made, and relates the time needed to the required accuracy, as well as
physical attributes of the system. 

The main approximation in the divide and conquer method is in
projecting the Kohn-Sham method onto small areas; this means that
there is no variational principle, and the exact result will depend on
the choice of partitioning unless extreme care is taken. Moreover, the
convergence with subsystem size can be very slow, so that good
convergence is hard.  Formally exact partition
theory\cite{Elliott:2009cr,Elliott:2010dq} suggests a route forward in
putting these approaches on a more exact footing. The key idea is to
transform a set of interacting fragments of a molecule or other system
into a set of \emph{non-interacting} fragments in an effective
potential (in exact parallel of Kohn-Sham theory).  However, as with
KS theory,  the approach does not have an analytic functional giving
the effective potential; intriguingly, the obvious local
approximations are strongly related to orbital-free DFT (discussed
below in Sec.~\ref{sec:orbital-free-dft}). 

There are a number of methods related to divide and conquer.  The 3D
fragments\cite{Wang:2008zs,Zhao:2008ve} builds on an earlier charge
patching method\cite{Wang:2002pd,Vukmirovic:2008be}, and gives a good
route to modelling large semiconductor systems, though it requires
passivation of surfaces of the fragments. The key idea is that each
fragment is part of several differently sized fragments.  For
simplicity, consider two dimensions and square fragments: then each
fragment forms one small area (1$\times$1), and is part of four
rectangular areas (2$\times$1 and 1$\times$2) and four large square
areas (all 2$\times$2).  These are combined (with differing signs) to
yield the total energy. The method as formulated is variational,
making the forces easy to calculate.  The Mosaico
method\cite{Seijo:2004wq} combines aspects of divide and conquer and a
Wannier function method (the method seeks local molecular orbitals
within specific areas); it has been implemented and applied within
\textsc{siesta}\cite{Seijo:2007rw} 

The Fragment Molecular Orbital (FMO) method\cite{Kitaura:1999cr} uses
a similar idea to the divide and conquer method (dividing a system up
into small pieces) and applies it to the create an approximate method
for proteins and related biological molecules.  The total energy of
the whole molecule is calculated from the energy of fragments and
pairs of fragments without solving the molecular orbitals of the whole
molecule.  In the calculation of the molecular orbitals of fragments,
they introduce a special technique for treating the bonds at the
boundary of the fragments.  The method has found a wide variety of
applications\cite{Fedorov:2007dq}.  There is a closely related method
(the molecular tailoring approach\cite{Ganesh:2006cw} which introduces
a different method for defining the fragments.  There are also
proposals to use the result of an FMO-like calculation as the input
density for a conventional or linear scaling
method\cite{Szekeres:2006sh}.

Two recently proposed, related methods\cite{Varga:2010uq,Ozaki:2010bh} are related to the main idea of divide and conquer, but can yield exact results (though not in linear scaling time).  Domain decomposition in a quasi-1D system (i.e. one with two dimensions much smaller than the other) shows linear scaling  and good efficiency\cite{Varga:2010uq}, though the generalisation to 3D is not given.  Once the system is subdivided, a block-diagonal Hamiltonian can be built and either the KS eigenstates in each block can be used as a basis for creating a structured H or an $LDL$ factorisation can be used to generate the Green's function.  A similar but more general method\cite{Ozaki:2010bh} uses contour integration of the Green's function to get the density matrix; after re-ordering the Hamiltonian to create a structured matrix, $LDL^T$ factorisation is used recursively to find the Green's function. The method, though not explicitly linear scaling, is efficient: it is $N(\log_2 N)^2$ in 1D, $N^2$ in 2D and $N^{7/3}$ in 3D, and is exact.  

Other related approaches have been proposed for metallic systems.  A
KKR method where the system is divided up into local interaction zones
(LIZ)\cite{Wang:1995uq} then solves for the Green's function within
the zone using multiple scattering theory.  Improvements include
embedding in a local
environment\cite{Abrikosov:1996xq,Abrikosov:1997pt}.  This contrasts
with the screening in k-space found to lead to linear scaling in the
number of layers for multi-layers and surfaces\cite{Zeller:1995vn
  ,Zeller:1997kx}.  Recent work suggests that combining the two
approaches, so that real-space screened KKR is used for large
imaginary energies and k-space screened KKR for small imaginary
energies, reduces scaling to $\mathcal{O}(N^{1+\epsilon})$, where
$\epsilon<0.2$\cite{Smirnov:2001ay,Smirnov:2002aq}.  The method can be
made truly linear scaling by using iterative
minimisation\cite{Zeller:2008bh}; screened full-potential KKR has
also been developed\cite{Ogura:2009rr}.

The simplicity of this type of approach makes it extremely
attractive.  It is easy to implement and has an obvious relation to
the shortsightedness of electronic structure.  However, the rate at
which convergence to the exact result is achieved relative to subsystem size, and the lack of
a variational principle, make these methods unsuitable for
quantitative calculations with full DFT accuracy.

\subsubsection{Recursive and Stochastic Approaches}
\label{sec:recursion}

The electronic structure of a system can be evaluated from the density of states as well as from the eigenvalues; for instance, the band energy can be written:
\begin{equation}
  \label{eq:67}
  E_{\mathrm{band}} = \int^{E_f}_{-\infty} dE n(E) E
\end{equation}
where $n(E)$ is the density of states and $E_f$ is the Fermi level (practically, the lower bound on the integral is normally taken to be the bottom of the band of occupied states).  The potential to use this in linear scaling methods becomes clear when the density of states is written as a sum over \emph{local} densities of states (LDOS).  If a localised basis set such as pseudo-atomic orbitals is used, with $\left\{\phi_{i\alpha}(\mathbf{r})\right\}$, then we can write:
\begin{equation}
  \label{eq:68}
  n(E) = \sum_{i\alpha} n_{i\alpha}(E)
\end{equation}

As with any function, the LDOS can be described by its moments\cite{Heine:1980if} (for instance: its mean, width, skew, etc which correspond to the first, second, third etc moments).  But in a local orbital basis, these moments can be related to powers of the Hamiltonian; the $p$th moment of the LDOS $n_{i\alpha}$, $\mu^{(p)}_{i\alpha}$, is given by:
\begin{eqnarray}
  \label{eq:69}
  \mu^{(p)} &=& \int dE E^p n_{i\alpha}(E) = \langle i\alpha|\hat{H}^p|i\alpha\rangle\\
  &=& \sum_{j_1\beta_1, \ldots, j_{p-1}\beta_{p-1}} H_{i\alpha j_1\beta_1}H_{j_1\beta_1 j_2\beta_2}\ldots H_{j_{p-1}\beta_{p-1} i\alpha}
\end{eqnarray}
where $H_{i\alpha j\beta} = \langle i\alpha|\hat{H}|j\beta\rangle$.  In a simple tight binding picture, this corresponds to hopping around a lattice following closed loops of length $p$.

However, this is, in general, a rather unstable way of building the DOS for a system.  One stable technique is the recursion method\cite{Haydock:1980fx}, which is a Green's function method.  The LDOS is written as:
\begin{equation}
  \label{eq:70}
  n_{i\alpha}(E) = -\frac{1}{\pi} \lim_{\eta \rightarrow 0} \mathrm{Im}\left\{G_{i\alpha,i\alpha} (E+i\eta)\right\}.
\end{equation}
This Green's function can be written in terms of a continued fraction, 
whose components are related to the elements
of the tridiagonalised Hamiltonian of the system\cite{Haydock:1980fx}
The element $G_{00}(Z)$ (where $G_{nm}(Z) = 
\langle U_n| \hat{G}(Z)|  U_m\rangle$) can be found from:
\begin{equation}
G_{00}(Z) = {1\over\displaystyle Z-a_0 - {\strut b_1^2
              \over\displaystyle Z-a_1 - {\strut b_2^2
              \over\displaystyle Z-a_2 - {\strut b_3^2
              \over\displaystyle \ \ \ddots}}}},
\label{eq:G00}
\end{equation}
with $a_n$ the diagonal element and $b_n$ the off-diagonal element.
But most Hamiltonians are not tridiagonal. 

The Lanczos recursion algorithm\cite{Lanczos:1950ud} is an efficient scheme for 
tridiagonalising a matrix. Consider a matrix {\bf H}, which corresponds to the
operator $\hat{H}$. Let the tridiagonal matrix be ${\bf H}^{\prime}$, 
whose diagonal
elements are given by $a_n$ and whose off-diagonal elements are given by $b_n$.
If the states which tridiagonalise the matrix are $| U_n\rangle$, then:
\begin{equation}
H^\prime_{mn}=\langle U_m|\hat H| U_n\rangle =
\cases{a_n ,& if $m=n$;\cr
       b_n ,& if $m=n-1$;\cr
       b_{n+1} ,& if $m=n+1$;\cr
       0 ,& otherwise.\cr}
\label{eq:recrels}
\end{equation}
and the condition that the tridiagonalising states are orthonormal
($\langle U_n| U_m \rangle = \delta_{n,m}$).  The method can be extended to non-orthogonal basis vectors with a bi-orthogonal approach\cite{Ballentine:1986vn}.

The overall procedure in a recursion method is therefore: choose an initial starting state; apply the Hamiltonian repeatedly to this (generating a Krylov subspace, which will be discussed in detail below); from the resulting tridiagonal Hamiltonian, construct the Green's function, density of states and density matrix as required.  The application of recursion to tight binding has been described in detail elsewhere\cite{Goringe:1997cy,Haydock:1980fx}, though it was used as the basis of one of the first DFT-based linear scaling methods\cite{Baroni:1992qc}, with application to silicon using a finite difference approach.

The recursion method as described has a major drawback: the recursion
for diagonal elements of the Green's function, which are required for
the energy, are stable and easy to evaluate, but the
\emph{off-diagonal} elements, which are required for force
calculations, are hard to evaluate with a tendency to numerical
instability. The early techniques used to work around these problems
(using the difference between bonding and anti-bonding orbitals for
inter-site elements, and matrix recursion) did not adequately resolve
convergence problems. The Bond Order Potential (BOP)
method\cite{Aoki:1993rw,Horsfield:1996dz,Horsfield:1996pd} is a
mathematically complex solution to this problem, which involves
redefining the procedure in terms of a new basis, using sum rules for
the Green's function.  A block BOP\cite{Ozaki:1999wl,Ozaki:2000ix}
uses a simpler basis to provide an efficient route to energies and
forces via recursion.  In all these cases, by limiting the range of
the recursion to a cluster of specified size, and truncating the
moments considered, the method becomes linear scaling.  The block BOP
method has been demonstrated with DFT\cite{Ozaki:2001fk}, using a dual
basis approach.  However, BOP methods are more commonly used to bridge
between tight-binding and empirical methods\cite{Pettifor:2002fr}.  
Ozaki\cite{Ozaki:2006uq}] has shown how the divide-and-conquer method
(see Sec.~\ref{sec:divide-conquer}) can be combined with recursion to
create a stable, easily extensible linear scaling method.  This will
be discussed further below. 

A large class of methods use Chebyshev polynomials, which can be defined recursively:
\begin{equation}
  \label{eq:72}
  T_{j+1}(H) = 2HT_j(H) - T_{j-1}(H),
\end{equation}
with $T_0(H) = I$ and $T_1(H) = H$.    The Fermi Operator Expansion (FOE) method\cite{Goedecker:1994li, Goedecker:1995di,Goedecker:1995pb} writes the finite temperature density matrix (or Fermi matrix) as:
\begin{equation}
  \label{eq:71}
  F_{\mu,T} = f\left(\frac{H-\mu I}{k_B T}\right)
\end{equation}
where $f(x)  = 1/(1+\exp(-x/k_B T))$ is the Fermi function, and then
expands the Fermi function as a Chebyshev polynomial.  Each column of
the matrix can be found by a recursive procedure, starting with a
localised orbital, and applying the Hamiltonian repeatedly.  Without
any truncation, this yields an $\mathcal{O}(N^2)$ method (which has been used
for plane-wave DFT\cite{Jay:1999ht}); with truncation of the elements
in a column of the Fermi matrix, the method scales linearly with the
atoms in the system.  It can be shown that the polynomial expansion
should have a degree of the order of $n \simeq
(\epsilon_{\mathrm{max}} - \epsilon_{\mathrm{min}})k_B T$, to represent
the DOS adequately.  The early
applications were to tight binding, but the method can be extended to
DFT\cite{Goedecker:1995pb}.  The forces as originally
calculated\cite{Goedecker:1995di} are not exact derivatives of the
energy; an exact expression for the forces has been
derived\cite{Silver:1996kx,Voter:1996pl} but it involves significant
computational cost.   

There are a number of methods which are closely related to FOE.  
The Kernel Polynomial Method\cite{Silver:1996kx,Voter:1996pl,Roder:1997zv} differs only in the way that the DOS is reconstructed: Gibbs factors are used to weight each term in the polynomial, reducing oscillations which result from the truncation of the polynomial at a finite level. (Note that this is related to the use of a terminator in a continued fraction in the recursion method.)  Liang \emph{et al.}\cite{Liang:2003ud} give various improvements to the original method: they demonstrate fast resumming of polynomials; they explore different approximations to the Fermi function, concluding that the complementary error function is the best to use; and they show how the inverse temperature, $\beta$, can be related to the accuracy required and the system properties (in particular band width and gap).
The use of Chebyshev polynomials to select part of the spectrum has been extended to $\mathcal{O}(N)$ methods\cite{Garcia-Cervera:2009cf}, where the key scaling issue is inversion of the overlap matrix (which is performed iteratively; overlap matrices are discussed in greater detail in Sec.~\ref{sec:non-orthogonal-basis}).  
A method using maximum entropy techniques to extract the moments of a density of states by applying Chebyshev polynomials of the Hamiltonian to random vectors\cite{Drabold:1993vq} uses importance sampling to select vectors which give an idempotent density matrix.  This idea was extended\cite{Sankey:1994wa} to project the random vectors on the occupied subspace as a starting point for the recursion method.  Wang\cite{Wang:1994yb} similarly defines moments of the density of states in terms of Chebyshev polynomials, and then calculates these using random wavefunctions and maximum entropy methods with a plane-wave basis set to generate the density of states and optical absorption spectra.  

Random vectors have also been used in a stochastic approach to invert
the exponential of a Hamiltonian\cite{Krajewski:2005ta}.  Noting that
the correct thermodynamic potential for spinless fermions is given by: 
\begin{equation}
  \label{eq:74}
  \Omega = -\frac{1}{\beta}\log\mathrm{det}\left(I+e^{\beta(\mu-H)}\right)
\end{equation}
the exponential is then rewritten as:
\begin{equation}
  \label{eq:75}
  \Omega = -\frac{1}{\beta}\sum_{l=1}{P} \log \mathrm{det}\left(M(l)\right)
\end{equation}
where $M(l) = 1+\exp(i\pi(2l-1)/P)\exp(\beta/P)(\mu-H)$, with the
exponential written as a Chebyshev polynomial or a Trotter expansion.
It can be shown\cite{Krajewski:2005ta} that expectation values can be
written in terms of a sum over the inverses of $M(l)$, so that the
main computational effort is in inverting these sparse matrices.  The
inverse is found by selecting a random vector $\psi$, solving the
linear equation $M(l)\phi = \psi$ and calculating $M(l)^{-1} =
\langle\phi\psi^\dag\rangle$, with the expectation value taken over
the stochastic process.  The original application was to a very simple
system, but it was developed to tight-binding applications in one
dimensional systems\cite{Krajewski:2006ov,Krajewski:2006tg}, using
Langevin dynamics and noisy forces.  The decomposition of the grand
potential has innate scaling $\mathcal{O}(N)^{3-2/d}$ for a $d$ dimensional
system.  However, by rewriting the partition function in terms of both
electrons and ions, and using a careful sampling of the Boltzmann
distribution an efficient $\mathcal{O}(N)$ method can be developed which is
valid for metals\cite{Krajewski:2007fc}, though only demonstrated so
far on tight binding systems (in this case, metallic carbon
nanotubes).  The method has been further developed with an exploration
of efficient decompositions of the Fermi operator\cite{Ceriotti:2008vn}.

The energy renormalisation group (ERG) method is an approach which takes a different view of the density matrix.  Conventionally, the density matrix is viewed as the zero-temperature limit of the Fermi matrix defined above, and as the temperature is lowered, the correlations become longer ranged (particularly in small gap systems and metals) so that the matrix becomes non-sparse.  The ERG method instead writes the density matrix as a telescopic sum of terms, with the temperature in each term decreasing by a factor $q>1$:
\begin{eqnarray}
  \label{eq:73}
  \hat{\rho} &=& \hat{F}_{\beta_0} + (\hat{F}_{\beta_1} - \hat{F}_{\beta_0}) + (\hat{F}_{\beta_2} - \hat{F}_{\beta_1})  + \ldots\\
  &=& \hat{F}_{\beta_0} + \hat{\Delta}_1 + \hat{\Delta}_2+\dots
\end{eqnarray}
The first term in the series is a high-temperature Fermi matrix which
is short-ranged.  Each successive term gradually corrects for lower
temperatures, so that $\beta_n = q^n\beta_0$.  The Fermi matrix is
written as a Chebyshev polynomial of the Hamiltonian, just as in the
FOE method described above.  The expansion is substituted into
expressions for expectation values (rather than calculating a
long-range density matrix); a recursive approach is given to
coarse-grain the Hamiltonian in each successive space, using the
Chebyshev polynomials.  It is developed
elsewhere\cite{Kenoufi:2004ys}, but has not shown true linear scaling
beyond 1D systems and has not been applied beyond tight binding
implementations.  Recursive bisection of the density
matrix\cite{Rayson:2007if} gives a method which scales linearly for
one-dimensional systems without truncation, but is highly efficient;
given density matrix truncation, it would scale linearly with a small
prefactor.  A recursive bisection of real space\cite{Gygi:2009vk}
allows the truncation of Kohn-Sham wavefunctions to be controlled, and
suggests a possible route to localised orbitals without the need to
specify centres and extents \emph{a priori}.

The essence of a subspace method is the repeated application of the
Hamiltonian operator to a vector to form a new set of vectors, which
span a space; this is often referred to as a Krylov subspace method,
with the Krylov subspace spanned by the set $\left\{|\phi\rangle,
  H|\phi\rangle, H^2|\phi\rangle, H^3|\phi\rangle, \ldots, H^{n-1}|\phi\rangle\right\}$.  For instance, a Lanczos procedure generates basis vectors in a Krylov subspace which are orthogonal.  It has been proposed\cite{Takayama:2004pf} that diagonalising within the subspace (similar, in effect, to the method of Ozaki\cite{Ozaki:2006uq} described above) will give an efficient linear scaling method; they find that around 30 vectors is sufficient.  A more efficient variant of this method\cite{Takayama:2006by} solves linear equations:
\begin{equation}
  \label{eq:76}
  (z-H)|x_j\rangle = |j\rangle
\end{equation}
for given basis vectors $|j\rangle$ to generate vectors $|x_j\rangle$,
from which the Green's function elements $G_{ij} = \langle i |
x_j\rangle$ can be found; the basis vectors are the Krylov subspace
vectors.  The technique used is the conjugate-orthogonal conjugate
gradient method, which generates a set of residual vectors
\emph{independent} of the energy shift $z$.  Once the vectors
$|x_j\rangle$ have been generated for one energy, further energies are
almost trivial, and certainly scale as $\mathcal{O}(1)$.  An overview and
summary of applications using tight binding have been
given\cite{Hoshi:2006yt,Fujiwara:2008qf}, and the method has been
extended to non-orthogonal orbitals\cite{Teng:2011fk}. 

The recursion methods were the first set of linear scaling methods
proposed, and Lanczos approaches are widely used in many areas of
physics and mathematics.  The most successful \emph{ab initio}
recursion methods are based around FOE-like methods, mainly because of
the simplicity of Chebyshev polynomials and the overall approach.  The
convergence and general lack of variational properties have made these
methods less successful than other approaches.
 
\subsubsection{Penalty functionals}
\label{sec:penalty-functionals}

An alternative approach to imposing idempotency was suggested by Kohn\cite{Kohn:1996lp}: add a term to the energy functional which penalises non-idempotent density matrices.   This way, as the ground state is sought, idempotency is automatically included. The original method defines Hermitian trial functions $\tilde{n}(\mathbf{r},\mathbf{r}^\prime)$, in terms of which the density matrix $\bar{n}$ is written as $\bar{n} = \tilde{n}^2$.  The ground state search is written in terms of a functional:
\begin{equation}
  \label{eq:30}
  Q\left[ \tilde{n}(\mathbf{r},\mathbf{r}^\prime)\right] = E\left[\bar{n}\right] - \mu N\left[\bar{n}\right] + \alpha P\left[\tilde{n}\right]
\end{equation}
where $E\left[\bar{n}\right]$ is the usual Kohn-Sham energy functional, $\mu$ is a chemical potential, $N\left[\bar{n}\right]$ is the number of electrons and $P\left[\tilde{n}\right]$ is a penalty functional,
\begin{equation}
  \label{eq:31}
  P\left[\tilde{n}\right] = \left[ \int \mathrm{d}\mathbf{r}\tilde{n}^2\left(1-\tilde{n}\right)^2\right]^{1/2}
\end{equation}

For idempotent $\tilde{n}$, $P\left[\tilde{n}\right] = 0$, so that the penalty functional has no effect.  The number of electrons is set by choosing $\mu$, leaving only $\alpha$ as a parameter.  It can be shown\cite{Kohn:1996lp} that, for a given $\mu$ and $\alpha$ larger than a critical value $\alpha_C$, the correct, idempotent ground state density matrix is found.  However, $\alpha_C$ cannot be predicted exactly; too small a value of $\alpha$ will yield local minima while too large a value will slow convergence.

A lower bound on $\alpha_C$ can be derived\cite{Haynes:1998yk} for a slightly different functional ($Q\left[ \tilde{n}(\mathbf{r},\mathbf{r}^\prime)\right] = E\left[\tilde{n}\right] - \mu N\left[\tilde{n}\right] + \alpha P\left[\tilde{n}\right]$ where the trial function is used as the density matrix):
\begin{equation}
  \label{eq:32}
  \alpha_C > 2 \max_i |\epsilon_i-\mu|
\end{equation}
However, it was also shown\cite{Haynes:1998yk} that, due to the square-root form of the penalty functional, there is a branch point at the minimum which prevents standard minimisation approaches such as conjugate gradients from being effective.  A corrected functional was proposed\cite{Haynes:1999qt} which removes these problems:
\begin{equation}
  \label{eq:33}
  Q\left[ \tilde{n}\right] = E\left[\tilde{n}\right] - \mu N\left[\tilde{n}\right] + \alpha P\left[\tilde{n}\right]^2
\end{equation}
where $\tilde{n}$ is now taken as the density matrix throughout.  This
approach does not impose idempotency exactly (nor weak idempotency),
with occupancy errors which can be written as $\delta f_i =
-(\epsilon_i - \lambda)/\alpha$ where $\epsilon_i$ is a Kohn-Sham
eigenvalue at the minimum; the occupancies are such that occupied
bands have more than one electron and unoccupied bands have negative
occupancies, which also gives an $\alpha$-dependent error in the total
energy.  A correction to the energy can be applied as
$\mathrm{Tr}[\rho(1-\rho)^2(1-2\rho)]$ for occupied bands, which can
be evaluated in $\mathcal{O}(N)$ steps; the unoccupied bands require a
more complex correction.  Following correction, the method gives a
total energy independent of $\alpha$.  An alternative approach uses
this functional as the heart of an iterative method via an augmented
Lagrangian method\cite{Adhikari:2001qf}. 

The penalty functional method is used within the \textsc{onetep}
code\cite{Haynes:2008zr}, as part of a cascading sequence of methods
(from canonical purification, through penalty functional and finally
LNV).  We note that there was an early proposal which used a penalty
functional with linear scaling\cite{Wang:1992mg}; the method sought
highly localised Wannier functions in a basis of atomic tight-binding
orbitals without explicit orthogonalisation, but a penalty applied for
non-orthonomality (using a simple sum over all atoms $i$ and their
neighbours $j$ of the form $\lambda \sum_i \sum_j |\langle \psi_i |
\psi_j\rangle|^2$).   Overall, penalty functional methods have not
been widely taken up.

\subsubsection{Orbital-free DFT}
\label{sec:orbital-free-dft}

Orbital-free DFT\cite{Wang:1992qe ,Pearson:1993eu} returns to the original spirit of density functional theory, and seeks only the ground state charge density and not the associated orbitals, giving a huge advantage in terms of simplicity and speed.  However, to find the energy and hence the ground-state density, a functional for the energy in terms of the charge density is required; this follows standard DFT methods, except for the kinetic energy, and much of the difficulty in orbital-free DFT lies in finding a kinetic energy functional for the charge density.  An overview of different approaches can be found in a recent review article\cite{Garcia-Aldea:2007ij}.

For a uniform electron gas, the Thomas-Fermi functional gives:
\begin{equation}
  \label{eq:24}
  T_{TF} = c_k \int \mathrm{d}\mathbf{r}n(\mathbf{r})^{5/3}
\end{equation}
where $c_k = \frac{3}{10}(3\pi^2)^{2/3}$.  However, this is not sufficient for systems where the charge may vary.  Rapidly varying perturbations can be represented by the von Weizs\"acker  functional:
\begin{equation}
  \label{eq:25}
  T_{vW} = \frac{1}{8}\int \mathrm{d}\mathbf{r}|\nabla n|^2/n
\end{equation}
These two functionals represent limits of behaviour, and early work on
orbital-free DFT combined the two to form a single functional.  While
the functionals form the main topic of research, there is a further
problem regarding pseudopotentials, which are in common use throughout
physics.  As the charge density is a local quantity (a function only
of one position, $\mathbf{r}$) the pseudopotentials used must be local
only. This introduces a significant restriction, as much of the
transferability and accuracy of modern pseudopotentials comes from the
angular freedom given by non-local potentials (where dependence on two
positions $\mathbf{r}$ and $\mathbf{r}^\prime$ allows different
potentials to be used for different angular momentum).  Progress has
been made in creating local pseudopotentials suitable for bulk
use\cite{Watson:1998wd}, though this still causes problems, and
restricts the transferability of the method.  The most successful
potentials are for metallic systems, and in particular those closest
to the nearly-free electron model. 

Most of the work beyond this conceptual starting point is in creating new functionals\cite{Smargiassi:1994rq,Foley:1996kh,Wang:1998ss}, with extra freedom found by including density dependence in the kernel\cite{Wang:1999pr} and non-local functionals\cite{Chai:2007mw,Ho:2008ma}.  An approach to allow the kernel to be calculated in non-periodic systems\cite{Choly:2002ul} has enabled development of methods which combine finite-element modelling and coarse graining (with the properties of each element based on a single representative atom modelled with OFDFT)\cite{Gavini:2007pd}.
Extension to covalent systems and semiconductors have been
introduced\cite{Zhou:2005fh ,Huang:2010ai}; the key step forward was
the use of a non-local kinetic energy (KE) functional, involving two parameters which
are transferrable within environments with similar coordination
numbers.  There is a freely-available implementation of orbital-free
DFT, which has been demonstrated a calculation on one million atoms of
Al\cite{Ho:2008fs,Hung:2009yl}.  Very recently, a new theoretical
approach has been suggested\cite{Cangi:2011qy} which writes the
density as a functional of the potential (reversing the standard
approach); this appears to give a well-defined route to kinetic energy
functionals, though its impact on the field is yet to be seen.

While this family of methods gives a good way to model large metallic
systems, there is still the concern that the kinetic energy functional
is not exact, and the pseudopotentials are limited, despite
significant effort.  The recursion methods (Sec.~\ref{sec:recursion})
are a viable alternative, though not widely used, and a detailed
investigation of the relative accuracies of the different approaches,
and their convergence, would be a valuable contribution.

\subsubsection{Expansion of the density matrix and tensorial approaches}
\label{sec:expans-dm-tens}

A recent class of approaches has emerged which use a change of variables (generally an exponential parameterisation) to impose the idempotency of the density matrix, or the orthogonalisation of the Kohn-Sham orbitals implicitly.  These methods offer efficient routes to optimisation of conventional methods, as well as potential for linear scaling.

The first proposal within linear scaling was an exponential parameterisation of the density matrix\cite{Helgaker:2000ck}, $\mathbf{D}$.  We write:
\begin{equation}
  \label{eq:26}
  \mathbf{D}(\mathbf{X}) = \exp(-\mathbf{X}\mathbf{S})\mathbf{D}\exp(\mathbf{S}\mathbf{X})
\end{equation}
where $\mathbf{X}$ is an anti-symmetric matrix, and $\mathbf{S}$ is
the overlap matrix between basis functions, as usual.  This
transformation preserves both the idempotency and trace of the density
matrix, and gives a set of variables which allow a search for the
ground state; note that the starting density matrix is key.  The
exponential can also be written in terms of a Baker-Campbell-Hausdorff
(BCH) expansion: 
\begin{eqnarray}
  \label{eq:27}
  D(X) &=& D + [D,X]_S + \frac{1}{2}[[D,X]_S,X]_S+\ldots, \\
  \label{eq:28}
  [A,X]_S &=& ASX - XSA,
\end{eqnarray}
where the commutator in a non-orthogonal basis is notated $[,]_S$.  From this formalism, both Hartree-Fock and DFT can be made linear scaling, and an implementation has been given, with details on preconditioning of the minimisation\cite{Larsen:2001kl}.  A similar idea has been independently derived from geometric considerations of the minimisation\cite{Van-Voorhis:2002zr}, though without linear-scaling application. A related, orthogonal basis method using curvy steps has demonstrated linear scaling\cite{Shao:2003il}, and its generalisation to non-orthogonal bases is discussed in detail below.  The method has also been shown to be effective as an extended Lagrangian approach to Car-Parrinello molecular dynamics\cite{Herbert:2004kl}. This type of approach, using an exponential transformation as a unitary transform,  has been used a number of times in the past in conventional methods (e.g. for Car-Parrinello Molecular Dynamics\cite{Hutter:1994rp}).

Another example of the unitary transformation leads to an approach to
optimisation of the density matrix with curvy
steps\cite{Head-Gordon:2003gy}.  Here, the unitary transformation is
used to preserve idempotency and electron number, with the unitary
operator written as an exponential, via a BCH expansion. Much care is
devoted to tensorial correctness (as discussed below in
Sec.~\ref{sec:non-orthogonal-basis}).  The transformed density matrix
is written as a power series following the exponential expansion,
which is again written in terms of an anti-Hermitian operator,
$\hat{\Delta}$, with $\hat{U} = e^{\hat{\Delta}}$.  Using tensor
notation (explained below in Sec.~\ref{sec:non-orthogonal-basis}), they write:
\begin{eqnarray}
  \label{eq:22}
  \tilde{P}^\mu_{\ \nu} &=& e^{-\Delta^\mu_{\ \lambda}}P^\lambda_{\ \sigma} e^{-\Delta^\sigma_{\ \nu}}\\
  &=& \sum_{j=0}\frac{1}{j!}\left(P^{[j]}\right)^\mu_{\ \nu}\\
  \left(P^{[j+1]}\right)^\mu_{\ \nu} &=& \left[P^{[j]},\Delta\right]^\mu_{\ \nu} = \left(P^{[j]}\right)^\mu_{\ \alpha} \Delta^\alpha_{\ \nu} - \Delta^\mu_{\ \alpha} \left(P^{[j]}\right)^\alpha_{\ \nu}
\end{eqnarray}
While the matrices in these equations are mixed, fully contravariant matrices, such as the density matrix, are simpler to work with, giving symmetric commutators.  The gradient can be found as $S^{-1}HK-KHS^{-1}$ in the contravariant notation; taking a step in this direction and expanding out the energy in terms of the density matrix gives a polynomial search to some order, rather than the usual linear assumption.  If the expansion for the expansion is truncated at linear order, then the gradients for the LNV method are recovered.  As the expansion is truncated, Idempotency is not exactly conserved, and must be re-applied using McWeeny iterations.  The main advantage of the method is that it allows longer steps to be taken than would be possible with a linear method.

A method for the direct minimisation of the wavefunction coefficients
while remaining on the Grassmann manifold\cite{Raczkowski:2001fv} has
also been developed (the Grassmann manifold arises when the energy of
the system depends only on the space spanned by the orbitals, and not
on the orbitals themselves: the transformation from eigenstates to
Wannier functions remains on the Grassmann manifold, for
instance). The basis set coefficients are written as a matrix $\Phi$,
which is defined as a transformation of the bands $\Psi$.  Then the
overlap of the basis functions is incorporated into a covariant matrix
$\bar\Phi^\dag = (\Phi^\dag \mathbf{S}\Phi)^{-1}\Phi^\dag$, with the
application of $(\Phi^\dag S\Phi)^{-1}$ as a metric.  Writing the
energy as $E = \mathrm{Tr}\left[\bar{\Phi}^\dag \mathbf{H}
  \Phi\right]$ and minimising with respect to $\Phi$ gives a method which
automatically includes the constraint of idempotency, and stays on the
Grassmann manifold to first order.  
The inverse overlap matrix, $S^{-1}$, must be applied for tensorial correctness; the authors suggest finding this via inversion of a submatrix of S made only from the orbitals within the localisation region\cite{Stechel:1994mk}.  The resulting method is closely related to Wannier function methods described above in Sec.~\ref{sec:direct-iter-appr}, particularly Ref.\cite{Stechel:1994mk}.

Another application of the parameterised unitary transform is an approach to sparsity\cite{Weber:2008gg}. The $\ell_1$-norm (defined for a vector $\mathbf{x}$ as $\ell_1 = \sum_j |x_j|$, compared to the more usual $\ell_2 = \sqrt{\sum_j x^2_j}$) is used as a sparsity measure for the wave-function coefficients.  The key idea of the method is to perform unitary transformations on the orbital coefficients so as to maximise the sparsity of individual columns, using the $\ell_1$-norm.  The resulting method has a slight restriction, in that only gradients for steepest descent have been defined, but it shows promise, and is intended for linear scaling.

Unitary transformations and sparsity have also been used within the  CP2K code\cite{Vandevondele:2003gd,Weber:2008lt}.  The normal constrained problem (with the constraint being orthogonality of eigenstates) is transformed to a locally unconstrained one; it is suggested that linear scaling behaviour should result for sparse problems\cite{Weber:2008lt}.  The method is based on an  orbital transformation\cite{Vandevondele:2003gd}: the wavefunction coefficients, $C$, are transformed to new variables $x$:
\begin{eqnarray}
  \label{eq:29}
  \mathbf{c}(\mathbf{x}) &=& \mathbf{c}_0 \cos(\mathbf{U}) + \mathbf{x}\mathbf{U}^{-1}\sin(\mathbf{U})\\
  \mathbf{U} &=& (\mathbf{x}^T\mathbf{S}\mathbf{x})^{1/2}\\
  \mathbf{c}_0^T\mathbf{S}\mathbf{c}_0 &=& \mathbf{I}\\
  \mathbf{c}(\mathbf{x})^T\mathbf{S}\mathbf{c}(\mathbf{x}) &=& \mathbf{I}\ \forall\ \mathbf{x}
\end{eqnarray}
Then $\mathbf{x}$ can be used in a minimisation, and remain linear (compared to previous ideas).  The constraint applied is that $\mathbf{x}^T\mathbf{S}\mathbf{c}_0 = 0$.  In the first-proposed form, a  diagonalisation of $\mathbf{x}^T\mathbf{S}\mathbf{x}$ is required to get eigenvectors.  The linear scaling method\cite{Weber:2008lt} uses iterative refinement for the orbital transformation (OT/IR) --- a function is defined such that $f_n(Z) \sim f(Z)$ such that $Z^TSZ=1$.  The authors propose to use the method by Niklasson\cite{Niklasson:2004cv}, with fourth order found to be particularly efficient:
\begin{equation}
  \label{eq:23}
  f_4(Z) = \frac{1}{128}Z(315 - 420Y+378Y^2-180Y^3+35Y^4)
\end{equation}
where $Y = Z^TSZ$.  The iterative refinement uses $f_n(\ldots
f_n(Z)\ldots)$.  To achieve linear scaling, a Taylor expansion is made
for matrix functions rather than diagonalising; after a conjugate
step, iterative refinement is used to reimpose the constraint.  The
main drawback with the method is that the  preconditioners required
are based on dense algebra, giving $\mathcal{O}(N^3)$ scaling, but this may be
lifted.

These methods are not yet widely used, but sit at an interesting
junction between standard approaches and linear scaling ones.  The
mathematical identities underlying the methods may well find wider use
in linear scaling applications, if they can be translated efficiently.

\subsubsection{Quantum Chemistry}
\label{sec:quantum-chemistry}

The general area of wavefunction-based methods (starting, for
instance, with Hartree-Fock wavefunctions and adding correlations via
perturbative methods such as M\o ller-Plesset (MP) methods or
coupled-cluster approaches) tends to be known as quantum chemistry.
Many of the approaches in this area scale prohibitively with system
size: the simplest, MP2, scales asymptotically with $N^5$;
coupled-cluster single double (CCSD) methods scale as $N^6$ and CCSD
with perturbative triples, CCSD(T), scales with $N^7$.  Local
correlation methods\cite{Sherrill:2010fj} can reduce scaling, and
various other methods have been proposed which build on the ideas of
locality for correlation.  This is an area of sufficient complexity
which has been reviewed
elsewhere\cite{Ochsenfeld:2007ee,Sherrill:2010fj}; below, we will
briefly summarise some key ideas.

The MP2 approach in its canonical form involves pairs of integrals
between occupied and virtual molecular orbitals.  Local approaches to
correlation reduce the scaling, but it is possible to produce linear
scaling methods.  The key step in one approach was the development of
Laplace MP2, where the exact energy is written in terms of
non-canonical orbitals, via a Laplace transform of the original energy
denominator.  This can be extended to use atomic orbitals, which gives
asymptotic $N^2$ scaling; by defining spherical interaction domains
centred on atoms, efficient linear scaling has been
demonstrated\cite{Ayala:1999tw}. Another implementation of the same
approach uses multipole-based integral estimates for
screening\cite{Doser:2009yg}. An alternative approach builds on the
local correlation methods\cite{Schutz:1999bv}, and uses density
fitting and explicitly correlated wave functions (which depend on
inter-electron distance, and improve the convergence of the method)
\cite{Werner:2006fj}.  It is also possible\cite{Lee:2000qo} to work in
terms of atomic orbitals, and truncate the excitations based on the
number of atoms involved in excitations (yielding a hierarchy of
methods which can lead to linear scaling).  This has been extended
with a method for forming localised orbitals\cite{Subotnik:2005ai},
with related work leading to a coupled cluster algorithm which scales
nearly linearly\cite{Subotnik:2006ku}.  The implementation of some of
these methods in a standard code has been
discussed\cite{Shao:2006xq}. 

These ideas have also been extended beyond MP2: linear scaling CCSD
has been demonstrated using non-orthogonal localised molecular
orbitals confined to fragments\cite{Flocke:2004xw} and by expanding
the coupled cluster wavefunction in a local basis formed from a
divide-and-conquer approach (see Sect.~\ref{sec:divide-conquer}); the
localisation is adapted dynamically to ensure error
control\cite{kowski:2010kc}. MRSD-CI has also demonstrated linear
scaling by using local correlation and integral
screening\cite{Chwee:2008lm}.

The divide and conquer method itself (Sect.~\ref{sec:divide-conquer})
has also been extended to MP2\cite{Kobayashi:2007le} and
CCSD\cite{Kobayashi:2008ga}.  In both cases, the \emph{full} HF
orbitals from the subsystem are used (as opposed to other quantum
chemistry approaches which typically localise the molecular orbitals
or use atomic orbitals).  Quantum chemistry is showing great promise
in the area of increasing system size; at present, calculations on
many tens of atoms are possible, and this trend should continue even
for the most expensive methods. 

\subsubsection{Extensions}
\label{sec:extensions}
While much of the work on linear scaling methods has been devoted to finding the ground state efficiently, there is also effort being put into going beyond the ground state to model the response of large systems to perturbations, in particular excitations.  In this section, we briefly survey this work, though it is not a comprehensive list; other reviews cover parts of the ground in more detail\cite{Ochsenfeld:2007ee,Shang:2010rr}.  

One obvious route for extending DFT is to perform real-time propagation of the density matrix (instead of the wavefunctions) within the framework of time-dependent DFT (TDDFT).  It can be easily shown that the time variation is given by:
\begin{equation}
  \label{eq:77}
  i\hbar\frac{\partial \rho(t)}{\partial t} = \left[H(t),\rho(t)\right]
\end{equation}
though there has yet to be any comprehensive investigation of the effect of truncation on the accuracy of propagation.  This has been implemented\cite{Yam:2003qe} following earlier work on time-dependent Hartree-Fock\cite{Yokojima:1998eu,Yokojima:1999fq,Yokojima:1999jf}.  It has been extended to density functional tight binding\cite{Wang:2007kx} and applied to calculating spectral properties\cite{Wang:2007vn} using Chebyshev polynomials to expand the exponential of the Hamiltonian.  However, the authors note that the amount of effort required is still high (and propagation for around 35fs is needed for 0.1eV energy resolution in medium to large systems, though the size is not quantified).  Another approach to real-time propagation TDDFT\cite{Iitaka:1997yq,Iitaka:2000et} uses random vectors and a projection method (as described in Sec.~\ref{sec:recursion}) to calculate response functions.  Using an empirical pseudopotential, it has been applied to calculation of optical properties of silicon nanostructures\cite{Nomura:1999mz}.2

Standard implementations of TDDFT normally avoid real-time propagation, and instead search for solutions at the linear response level.  The time propagation, Eq.~(\ref{eq:77}), can be recast in terms of a Liouvillean superoperator, $\mathcal{L}$, whose eigenvalues represent vertical excitation energies.  If the full Liouvillean is used, this is the random phase approximation (RPA), while if certain off-diagonal terms are neglected, the Tamm-Dancoff approximation (TDA) results.  By using projectors onto the occupied subspace (the density matrix, $\mathbf{P}$) and its complement ($\mathbf{Q} = \mathbf{I} - \mathbf{P}$), the eigenvalues of the Liouvillean can be found\cite{Tretiak:2009ys}; while Krylov subspace approaches (closely related to recursion methods described in Sec.~\ref{sec:recursion}) are efficient, a direct variational method based on Rayleigh quotients\cite{Lucero:2008zp} is also tested, and has been extended\cite{Challacombe:2010uq} and applied to carbon nanotubes and polymers.  A different approach to speeding up TDDFT also uses recursion\cite{Baroni:2010zr}.  Again, the method avoids explicit representation of the virtual orbitals (a common theme in methods to improve speed and convergence of TDDFT and many-body perturbation theory), though it is not linear scaling in its present form.  Another efficient, though not yet linear scaling, TDDFT method uses Lagrange functions (see Sec.~\ref{sec:real-space-local}) and domain decomposition (see Sec.~\ref{sec:divide-conquer})\cite{Goncharov:2011qa}.

Density matrix perturbation theory\cite{Niklasson:2004nh} (which has
been extended to non-orthogonal basis
functions\cite{Niklasson:2005bq}) uses the trace-correcting TC2 method
to generate a sequence of density matrices ($X^{(0)}_n$ for the unperturbed Hamiltonian).  The expansion $X_n = X^{(0)}_n+\Delta_n$ allows a recursive expression for $\Delta_{n+1}$ to be derived in terms of $\Delta_n$ and $X^{(0)}_n$.  The method is easily extended to different levels of perturbation theory.
The first application\cite{Weber:2004bh} was to calculations of polarizability of water clusters, going to 150 water molecules with a 6-31G$^{**}$ basis set.
Other groups have built on this method for calculating polarisation.  One approach\cite{Xiang:2006fd} uses only the non-diagonal part of perturbation to find polarizability and the Born effective charge in solids.  A different approach to the coupled-perturbed equations\cite{Ochsenfeld:1997wn} allowed linear scaling calculation of the derivative of the density matrix with respect to a parameter (e.g. atomic position or electric field) using the McWeeny expansion; this method has been made numerically more stable and applied, among other things, to calculation of NMR\cite{Kussmann:2007km}.  It has recently been reformulated in terms of a Laplace transform\cite{Beer:2008pw} (in the same way that MP2 calculations, described in Sec.~\ref{sec:quantum-chemistry} were reformulated) for greater efficiency.  A less complex version of this approach has also been given, based on the projection properties of the density matrix; it has been shown to be competitive with other linear scaling methods\cite{Liang:2005dq}.
A post-hoc method for calculating polarizability with linear scaling\cite{Izmaylov:2006hm} uses another variant of perturbation theory.  A further approach to molecular response\cite{Coriani:2007ai} uses an exponential parameterisation (as described in Sec.~\ref{sec:expans-dm-tens}) to find excitation energies and polarizabilities.

Finally, it is clear that band edges can be found from linear scaling
methods (for instance, see the discussion in the work by
Stechel\cite{Stechel:1994mk}). Recent developments\cite{Xiang:2007kv}
have made the search more efficient.  The maximum eigenvalue of $\rho
H$ is sought \emph{after} finding the ground state, using the density
matrix as a projection operator; equivalently, the minimum eigenvector
of $(I-\rho)H$ is sought for the LUMO.  The simplest solution uses the
Lanzcos algorithm for an extreme eigenval.  The method has been
applied to the case of a doped semiconductor where there are mid-gap
or band-edge states.  Band-edge states can also be found efficiently
using iterative purification methods\cite{Rubensson:2008qy}.

\section{Technical Details \& Parallelisation}
\label{sec:technical-details-}

\subsection{Non-orthogonal Basis Functions}
\label{sec:non-orthogonal-basis}

In general, the localised orbitals used as basis functions in most
linear scaling methods are non-orthogonal; it has been shown that
these functions are more contracted\cite{Liu:2000vn} and give computational advantages
in various systems (e.g. silicon\cite{Mortensen:2001kx} and organic
molecules\cite{Feng:2004lq}).  This introduces complications in
maintaining the correctness of tensors, and in defining different
types of operator representation.  The clearest notation uses
covariant and contravariant  notation (lower and upper indices), which
was introduced for recursion methods with an excellent
overview\cite{Ballentine:1986vn}.  There is also a general explanation
of the notation, and an application to second
quantisation\cite{Artacho:1991px}.  The key implication for linear
scaling methods is the need for a good approximation to the inverse of
the overlap matrix, or its decomposition, which can be found in linear
scaling time.  We summarise the basics below, and urge the interested
reader to find more in the references given. 

If a non-orthogonal basis is defined, $\left\{|e_\alpha\rangle \right\}$, then the overlap matrix has elements defined by:
\begin{equation}
  \label{eq:34}
  S_{\alpha\beta} = \langle e_\alpha|e_\beta\rangle = S^\star_{\beta\alpha},
\end{equation}
where we have assumed that the basis is real.  Any matrix represented in terms of the original basis and notated with two lower indices is called \emph{covariant}; it is actually a tensor.  There also exists a \emph{dual} basis, $\left\{|e^\alpha\rangle\right\}$, which satisfies the relation:
\begin{equation}
  \label{eq:35}
  \langle e^\alpha|e_\beta\rangle = \delta^\alpha_{\ \beta} = \langle
  e_\alpha|e^\beta\rangle = \delta_\alpha^{\ \beta}
\end{equation}
A matrix represented in terms of the dual basis and notated with two upper indices is called \emph{contravariant}; it is also a tensor.  It is easy to show that, for a complete basis, we can write:
\begin{equation}
  \label{eq:36}
  \sum_\alpha |e_\alpha\rangle\langle e^\alpha| = \sum_\alpha |e^\alpha\rangle\langle e_\alpha| = \mathcal{I}
\end{equation}
where $\mathcal{I}$ is the identity.  The overlap between elements of the dual basis also forms the \emph{inverse} of the overlap for the overlap matrix in the original basis; confusingly, different authors notate this in different ways.  If we write $T^{\alpha\beta} = \langle e^\alpha|e^\beta\rangle$, then 
\begin{eqnarray}
  \label{eq:37}
 T^{\alpha\gamma}S_{\gamma\beta} &=& \delta^\alpha_{\ \beta}\\
  \label{eq:38}
 T^{\alpha\beta}|e_\beta\rangle &=& |e^\alpha\rangle\\
 \label{eq:39}
 S_{\alpha\beta}|e^\beta\rangle &=& |e_\alpha \rangle\\
 \label{eq:40}
 \sum_{\alpha\beta}|e_\alpha\rangle T^{\alpha\beta}\langle e_\beta| &=& \sum_{\alpha\beta}|e^\alpha\rangle S_{\alpha\beta}\langle e^\beta| = \mathcal{I}
\end{eqnarray}
where we have used the Einstein summation convention in Eqs.~(\ref{eq:37})-(\ref{eq:39}).  Some authors write $T^{\alpha\beta} = \left(S^{-1}\right)^{\alpha\beta}$, while others write $T^{\alpha\beta} = S^{\alpha\beta}$.  Some care has to be taken when considering different representations: as well as fully covariant and contravariant forms, there are mixed forms, such as $\langle e^\alpha|\hat{A}|e_\beta\rangle = A^\alpha_{\ \beta}$.  A Hermitian operator is represented by a Hermitian matrix in the co- or contravariant forms, but not in the mixed form:
\begin{eqnarray}
  \label{eq:43}
  H_{\alpha\beta} &=& \left(H_{\beta\alpha}\right)^\star\\
  H^{\alpha\beta} &=& \left(H^{\beta\alpha}\right)^\star\\
  H_{\alpha}^{\ \beta} &=& \left(H^{\beta}_{\ \alpha}\right)^\star\\
\end{eqnarray}
This necessitates careful notation, with the position of the indices (i.e. whether the upper or lower index comes first) being significant.

It is when considering differentials and the difference between
classes of tensor that the notation and choice of metric becomes
important.  If the Hamiltonian is represented in terms of the original
basis then it forms a covariant tensor, and the density matrix is a
contravariant tensor.  But the differential of the energy with respect
to the density matrix (as used in methods such as the LNV technique
described in Section~\ref{sec:direct-iter-appr}) is covariant, and
should be scaled by an appropriate metric before it can be added to
the density matrix\cite{White:1997hc}; this metric is
$T^{\alpha\beta}$ (we note that it is possible to proceed with a
different metric, which is equivalent to the original
formulation\cite{Nunes:1994pi}).  If we write the auxiliary matrix
$\sigma(\mathbf{r},\mathbf{r}^\prime) =
\sum_{\alpha\beta}\phi_\alpha(\mathbf{r})L^{\alpha\beta}\phi_\beta(\mathbf{r}^\prime)$
and search for a minimum energy
with respect to $L$ matrix elements, for instance, then one search
step in the minimisation might be written: 
\begin{eqnarray}
  \label{eq:41}
  \frac{\partial E}{\partial L^{\alpha\beta}} &=& \sigma_{\alpha\beta}\\
  \sigma^{\alpha\beta} &=& T^{\alpha\gamma}\sigma_{\gamma\delta}T^{\delta\beta}\\
  L^{\alpha\beta} &=& L^{\alpha\beta}_0 + \lambda\sigma^{\alpha\beta}
\end{eqnarray}
where $\lambda$ is a varying step length.  Of course, the problem can be formulated in terms of a mixed representation, as suggested by Stephan\cite{Stephan:2000qm}, where all matrices are mixed.  This requires that the Hamiltonian be premultiplied on the left by $T^{\alpha\beta}$, which connects closely with inverse preconditioning approaches to minimisation, such as the AINV method described below; we also note that this goes back to observations on non-Hermiticity for localised molecular orbitals\cite{Weeks:1973th}.  The search step is now written:
\begin{eqnarray}
  \label{eq:42}
  \frac{\partial E}{\partial L^{\alpha}_{\ \beta}} &=& \sigma^\alpha_{\ \beta}\\
  L^\alpha_{\ \beta} &=& \left(L^\alpha_{\ \beta}\right)_0 + \lambda\sigma^\alpha_{\ \beta}
\end{eqnarray}
An alternative approach is to effectively orthogonalise the basis, which can be achieved by transforming the matrices to an orthogonal representation.  Some decomposition of the overlap matrix is required for this; either a symmetric one (using $S^{-1/2}$, the L\"owdin transformation) or a non-symmetric one (e.g. a Cholesky decomposition\cite{Millam:1997bu}, where $S = UU^\dag$ and transforming with $U^{-1}$).  Either way, we require either the inverse overlap matrix or the inverse of a decomposed matrix; naturally, these must be found using a linear scaling method\footnote{We note that various authors use cubic scaling methods to find the inverse overlap, on the grounds that the prefactor for this operation is rather small; while a pragmatic approach, it is not a scalable one.}.  Cholesky decomposition can be made to scale linearly with system size for sparse matrices\cite{Millam:1997bu ,Schweizer:2007kb}, though these techniques are extremely hard to parallelise efficiently \cite{Grote:1997fj}.

There is therefore a need to find the inverse or decomposed inverse of
the overlap, for sparse matrices with linear scaling time (and ideally
good parallelisation).  An excellent overview of approaches from a
computational science stance\cite{Benzi:2002jl} makes the important
point that the  sparsity pattern of the inverse of a matrix may not be
same as that of original matrix.  This raises the problem of how
sparsity is imposed, which is discussed fully below in
Sec.~\ref{sec:parallelisation}.  We note that some of the methods in
Sec.~\ref{sec:direct-iter-appr} effectively converge on the inverse
overlap matrix.  The range and sparsity pattern of $S^{-1}$ are
extremely important, as is the condition number of $S$.  It can be
shown that, for a localised $S$, $S^{-1}$ is exponentially
localised\cite{Nunes:1994pi}, though the range will depend on the
spread of eigenvalues of $S$.  The condition number of the overlap
matrix (the ratio of the largest to the smallest eigenvalue) will
determine how easily an inverse can be found; the condition number
will depend on the basis and the number of support functions/localised
orbitals\cite{Maslen:1998dt}. 

An iterative approach, known as Hotelling's method or Schultz's method\cite{Press:1992dp}, is extremely effective:
\begin{equation}
  \label{eq:44}
  X_{n+1} = 2X_n - X_nAX_n
\end{equation}
will converge quadratically on the inverse of $A$, so that $X_\infty =
A^{-1}$.  The iteration must be started  from a suitable initial guess
(which can be shown to be $B_0 = \epsilon A$, with $\epsilon$ a small
number).  This formula appears in a number of places: this iterative
approach to updating an inverse from a previous step (or close to
convergence) was suggested\cite{Stechel:1994mk}.  The close relation
to iterative purification methods (Sec.~\ref{sec:direct-iter-appr})
should not be surprising, as the inverse overlap coincides with the
density matrix when only occupied states are considered.  The main
drawback of this approach is that the iteration stalls when the
truncation error is similar to the change in inverse at a single step.

A divide-and-conquer-like algorithm was first suggested in the context of recursion methods\cite{Gibson:1993zt}: in order to form the matrix elements $M^\alpha_{\ \beta}  = \left(S^{-1}\right)^{\alpha\gamma}H_{\gamma\beta}$ for neighbours $\beta$ of an atom $\alpha$, a series of linear equations in clusters centred on each atom $\alpha$ in the system are solved: $H = MS$.  A similar idea was put forward using conjugate gradients\cite{Stechel:1994mk}: the problem can be written as $\sum_j S_{ij}D_{jk} = \delta_{ik}$, and conjugate gradients is then used to solve a least-squares problem with ($\sum_j\lambda^{(k)}_j\delta_{jk}-[SD]_{jk}^2$ for each localised orbital $k$, with $\lambda^{(k)}_j = S_{jk}^2$, or set to one only if within range).  Enforcing symmetry and idempotency ($D = DSD$) for the diagonal elements was found to improve stability.   An initial guess for the inverse is taken to be $\delta_{jk}$ or $2\delta_{jk} - S_{jk}$. Linear scaling follows from enforcing sparsity on the matrices. 

Defining the generalised inverse of the overlap matrix, $S^-$, by $SS^-S =S$\cite{Yang:1997nx,Burger:2008kx}, a truncated approximation can be found by minimising:
\begin{equation}
  \label{eq:45}
  \mathrm{Tr}[BS^-] = \min_{X} \mathrm{Tr}[B(2X-XSX)]
\end{equation}
where $X$ is constrained to be Hermitian and $B$ is \emph{any}
positive definite matrix.  It was shown that a variational
expression for the energy can be written for any number of localised
orbitals (i.e. including unoccupied states) if the density matrix is
taken to be $K = 2X - XSX$ and the energy is minimised with respect to
$X$. (A similar expression was found before\cite{Stechel:1994mk},
though without the variational derivation.)

The AINV method\cite{Benzi:1996xr} has been used by
Challacombe\cite{Challacombe:1999mb}) as a route to form $S^{-1}H$.
The approximate sparse inverse is constructed from a sparse incomplete
factorisation of the overlap; while this is an effective method, it is
hard to parallelise.  If the overlap is assumed to be decomposed as $S
= LL^T$, traditional methods find $L$ and then create $Z=L^{-1}$ by an
incomplete linear solution.  This can introduce inaccuracies, and the
AINV method avoids these by solving directly for $Z = L^{-1}$.  Once
$Z$ has been found, it is possible to create $Z(Z^TA) \equiv S^{-1}A$
without ever creating the inverse\cite{Challacombe:1999mb}, which may
be dense or have unusual sparsity patterns. 

Ozaki has proposed using the recursion method, normally applied to
finding the density matrix, to solve for the inverse
overlap\cite{Ozaki:2001dd}.  He applies the block BOP method
(described above in Sec.~\ref{sec:recursion}) to the resolvent: 
\begin{equation}
  \label{eq:46}
  R(Z) = (S-ZI)^{-1}
\end{equation}
for $Z$ a complex number.  It is clear that $S^{-1} = \mathrm{Re}[R(0)]$; however, for a method implemented with finite ranges, the matrix must be symmetrised to ensure stability.  The method is also compared to three other approaches described above: the divide-and-conquer-like method\cite{Gibson:1993zt}; Hotelling's method; and a Taylor expansion approach (as suggested in unconstrained minimisation methods\cite{Mauri:1993lf,Ordejon:1995yj} described above in Sec.~\ref{sec:direct-iter-appr}).  All methods are effective for diamond, though the recursion method is slightly faster, while the Hotelling method yields larger errors for fcc Al.  The choice of Taylor expansion (simply using a polynomial) precluded testing on the systems, as it did not converge.  A dual-basis recursion method has also been suggested\cite{Ozaki:2001fk}, with the dual basis calculated simultaneously with the basis (once a starting state has been defined).

A method for \emph{improving} an approximate factorisation of an inverse\cite{Niklasson:2004cv} (or an inverse itself, by decomposing $S^{-1} = IS^{-1}$) iterates $Z_{n+1} = Z_n(\sum_{k=0}^ma_kX^k_n)$ with $X_n = Z^T_nSZ_n$; this method is related to another iterative approach\cite{Baer:1998jw}.  The obvious drawback is the lack of an initial factorisation, but this has been solved\cite{Rubensson:2008tw}  by starting with a recursive decomposition of the $S$ matrix, which allows a convergent calculation of the inverse.  An extremely similar method has been derived for the symmetric square root\cite{Jansik:2007hw}; both discussions also point out that convergence can be improved by scaling the overlap so that the eigenvalues lie within a convergent radius.

In all these approaches, there is the problem of whether the inverse
(or the decomposed inverse) should be sparse at all, or share the same
sparsity patterns as the overlap; in this case, a sparsity algorithm
based on drop tolerances may give some advantages.  VandeVondele
\cite{VandeVondele:2007sp} deliberately optimised basis sets to yield
overlap matrices with small condition numbers, and show revealing data
on the sparsity of inverse overlaps with basis size: the sparsity
decreases as basis increases.  Plots of number of non-zero matrix
elements for different truncation thresholds and different
systems\cite{Rubensson:2011mi} also show that the inverse is less
sparse than the overlap. It is not yet unambiguously clear whether
convergence can be achieved for ill-conditioned $S$ matrices. 

\subsection{Preserving Electron Number}
\label{sec:pres-electr-numb}

When varying the density matrix or localised orbitals to find the ground state, as well as maintaining idempotency (which has effectively occupied most of the methods discussed so far), the correct number of electrons must be maintained, as mentioned above.  It is also possible to work at a fixed Fermi level (chemical potential for electrons, as suggested in one of the earliest methods\cite{Li:1993lg}) though this is often a less physically reasonable approach.  A grand potential is often defined:
\begin{eqnarray}
  \label{eq:47}
  \Omega &=& E_{\mathrm{Tot}} - \mu N_e\\
    &=& \mathrm{Tr}\left[ K\left(H - \mu I\right)\right] \mathrm{\ if\ } \langle \phi_\alpha | \phi_\beta \rangle = \delta_{\alpha\beta}\\
    &=& \mathrm{Tr}\left[ K\left(H - \mu S\right)\right] \mathrm{\ if\ } \langle \phi_\alpha | \phi_\beta \rangle = S_{\alpha\beta}
\end{eqnarray}

An early tight-binding approach\cite{Qiu:1994kr} included the derivative of the electron number in the gradient of the energy with respect to $L$ matrix element, with the chemical potential from the previous step used for $\mu$.  This is similar to the approach used in the \textsc{Conquest} code\cite{Hernandez:1996bf}, though instead of using the previous value of $\mu$, the \textsc{Conquest} code projects out the direction of electron change ($\partial N_e/\partial L_{\alpha \beta}$) from the search direction.  After the line minimisation for energy, a separate search for the correct electron number should be performed\cite{Qiu:1994kr,Hernandez:1996bf}.

An alternative is to treat the auxiliary density matrix $L$ as the real density matrix, and use the McWeeny purification to alter the gradient\cite{Millam:1997bu}.  In this case, the electron number becomes $N_e = \mathrm{Tr}[LS]$ in a non-orthogonal basis, and it can be shown that a traceless gradient can be found by defining a slightly different grand potential, and treating $\mu$ as a parameter:
\begin{eqnarray}
  \label{eq:48}
  \Omega(L) &=& \mathrm{Tr}\left[KH\right] + \mu\left(\mathrm{Tr}\left[LS\right] - N_e \right)\\
  \label{eq:49}
  \mu &=& -\mathrm{Tr}\left[3(HLS+SLH)\right.\\
      &&-\left.2(HLSLS+SLHLS + SLSLH)\right]/N_e\nonumber
\end{eqnarray}
The key advantage of working with this functional is that the search
for the ground state does not perturb the electron number, so that given a
starting point with the correct electron number, only $\mu$ needs to
be updated.  However, the density matrix so defined will be less
idempotent than $K$ will be, imposing an additional approximation. 

The final approach taken to maintain correct electron number is simply
to rescale the density matrix, either after each line
minimisation\cite{Li:2003gb} or continuously\cite{Haynes:2008zr}.  If
the auxiliary density matrix is used as the density matrix, then a
scaling factor $L \rightarrow N_e L /\mathrm{Tr}[LS]$ is applied at
each step.  Alternatively, the purification transformation can be
adapted to form a purification and normalisation
transformation\cite{Haynes:2008zr}: 
\begin{equation}
  \label{eq:50}
  K = N_e \frac{3LSL - 2LSLSL}{\mathrm{Tr}[(3LSL - 2LSLSL)S]},
\end{equation}
where $N_e$ is the number of electrons in the system.  This transformation potentially introduces multiple minima, though is reported\cite{Haynes:2008zr} not to adversely affect convergence.

\subsection{Parallelisation and Sparse Matrices}
\label{sec:parallelisation}

In this section we consider two important technical problems: parallelisation of linear scaling codes, and the implementation of sparse matrix methods.  Sparse matrices are key to linear scaling computational time and storage, while efficient parallelisation is needed for access to systems of more than ten thousand atoms or so.  We will consider sparse matrices first, and then turn to strategies taken for parallelisation.

We must consider both the technology of sparse matrices and how
sparsity is imposed, and what errors different approaches will impose;
the two main approaches to sparsity (or truncation of a matrix) are:
first, to consider a distance-based criterion (appealing to the
results of Sec.~\ref{sec:dens-matr-prop}, so that an element is set to
zero and neglected if the distance between atoms or localisation
centres is greater than some cutoff); and second, to drop an element
if its value falls below some threshold.  The first approach tends to
give a clear, well-defined sparsity pattern while the second requires
application of the tolerance after each operation.  It is, however,
easier to estimate and control an overall error due to sparsity with
the second method.  A recent analysis of matrix sparsity and how the
function of a sparse matrix decays\cite{Benzi:2007fu} provides an
excellent, in-depth understanding of sparsity.  It is highly
recommended for all those developing methods in this area.

An early implementation of a tight-binding orbital minimisation
method\cite{Ordejon:1993zk,Itoh:1995yq} used distance-based
truncation, and stored matrices in a compressed row format, following
a parallelisation described below.  The communication pattern used for
matrix multiplication followed a synchronous, cartesian-based
technique (for a processor at the centre of a cube, communicating with
26 other processors, communicate with neighbouring processors in the
sequence: $(1,0,0); (1,1,0); (0,1,0); (-1,1,0);$ etc).  
Similar approaches occur in other methods based on the orbital minimisation family\cite{Canning:1996ve,Shellman:2003rt}.

The first approaches to drop tolerance used a criterion based on the individual matrix elements\cite{Millam:1997bu,Daniels:1997dw}.  Challacombe\cite{Challacombe:1999mb} introduced a sparse matrix algebra based on atom blocks which are dropped if the Frobenius norm of the \emph{block} (defined as $A_F = \sqrt{A_{IJ}^TA_{IJ}}$ for atoms $I$ and $J$) is smaller than a specified tolerance.  The tolerance is re-applied after each matrix operation (e.g. addition or multiplication).  The reasoning for using blocks rather than elements is that a change of basis is less likely to change sparsity patterns and associated errors. As the number of electrons on an atom can be related to the trace of the on-site block, this seems reasonable.  A sparse-approximate matrix multiply (SPAMM)\cite{Challacombe:1999mb ,Challacombe:2010pd} has also been introduced: the contraction over two atom blocks in a multiply ($C_{IJ} = A_{IK}B_{KJ}$) is neglected if $B_F$ is smaller than the threshold divided by the number of blocks and $A_F$.  This leads to small elements being treated more approximately than large ones, and potential computational savings.  The method has been generalised\cite{Challacombe:2010pd} to a recursive approach on successively smaller sub-matrices; this bears some resemblance to an interesting approach to sparsity, that of hierarchical or $\mathcal{H}$-matrices\cite{Hackbusch:1999kl}, though these have not been used in linear scaling methods to our knowledge.

A sparsity analysis\cite{Maslen:1998dt} of matrices in electronic structure methods (pointed towards quantum chemistry methods) concentrated on linear alkanes, and applied drop tolerances.   For the Hartree-Fock method, provided that a well-conditioned set of localised \emph{occupied} orbitals exist, they prove both that the density matrix is localised, and that the overlap has a localised inverse.  However, it does not follow that a localised density matrix can always be found (nor that localised orbitals can be found); a demonstration of the inverse of S was shown before\cite{Nunes:1994pi}, and its significance is discussed above in Sec.~\ref{sec:non-orthogonal-basis}.

\begin{figure}[h]
  \centering
  \includegraphics[width=0.5\textwidth]{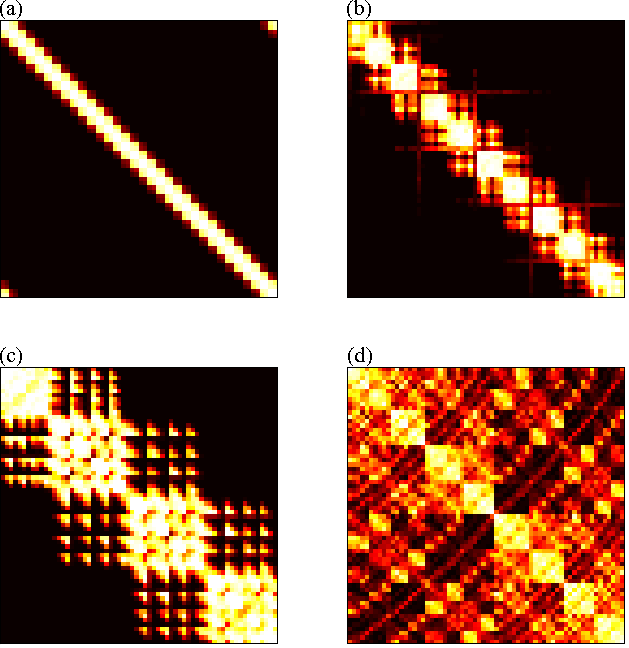}
  \caption{Segment-by-segment filling factors of sparse matrices in
    typical large systems divided over P=64 processes. Matrices of the
    sparsity pattern 􏰔KS (the product of the density kernel and
    overlap matrices) are shown for: (a) a (10,0) zigzag nanotube
    (4000 atoms),􏰔(b) a 64 base-pair sequence of B-DNA (4182 atoms),
    (c) a H-terminated wurtzite-structure GaAs nanorod (4296 atoms),
    and (d) 8 (8$\times$8) supercells of eight-atom cubic unit cells
    of crystalline Si (4096 atoms).  Each pixel represents a segment,
    whose color shows the fraction of matrix elements in that segment
    which are nonzero: black segments contain no nonzero elements,
    through red, then yellow, to white segments containing all nonzero
    elements. The nonzero elements are seen to be clustered near the
    diagonal of the matrix (though less so with increasing periodicity
    and complexity of the structure). The space-filling curve ensures
    that in a given column, there are nonzero overlaps only with rows
    of atoms on nearby processes, so the nonzero elements form a broad
    band centered on the diagonal. This is clearest for the simple
    structure of the nanotube, but even for the crystalline solid,
    there are segments in which there are few or no nonzero
    elements.  Reprinted with permission from N. D. M. Hine \emph{et
      al.}, J. Chem. Phys. \textbf{133}, 114111 (2009)
    \cite{Hine:2010fj}. Copyright 2009, American Institute of
    Physics.}  
  \label{fig:DMStruc}
\end{figure}

\textsc{Conquest}\cite{Goringe:1997cy,Bowler:2001yo} truncates
following distance-based criteria, and uses atom-blocks and compressed
row storage, with each on-site block stored in the place of a matrix
element. An intermediate level of organisation between atoms and the
unit cell (called partitions, and described in detail below) is used
to distribute storage and communications for multiplies.  Special
matrix-multiplication routines\cite{Bowler:2001yo} have been
developed, with a matrix kernel isolated to allow optimisation.  There
are two multiply routines, depending on the radius of final matrix
compared to the initial matrices (\emph{extension} when the final
matrix has a larger radius than the other two, and \emph{reduction}
when the final matrix has a smaller radius).  A similar approach to
division of the unit cell and communications is used for the grid
points (where the charge density is found on a real-space grid).
Matrix multiplication efficiency has been shown for a number of
systems, including large bulk silicon cells of up to 16,384
atoms\cite{Bowler:2001yo} and recently perfect linear scaling and
efficiency has been shown for systems containing millions of
atoms\cite{Bowler:2010uq}. 

\textsc{onetep}\cite{Hine:2009yb} uses atom-blocking, with data for
columns of matrices stored on the atom-responsible processor. Hand-coded
multiplies for block sizes relating to numbers of valence states
(1,4,5,9,10) are used for efficiency.  The developers suggest that
sparsity of  90\% or more is needed to benefit from routines
(especially when going to product matrices without truncation); their
analysis shows that the matrix $KSK$ is 73\% dense even for an 8,000 atom system.
More recent work\cite{Hine:2010fj} allows combined dense/sparse
operations.  Matrix columns are again divided into columns, and then
further into process segments after assigning columns to processors.
Each segment is designated dense (stored in full) or sparse based on
fraction of non-zero elements and a threshold. Plots of density matrix
structure for different systems are shown in Fig.~\ref{fig:DMStruc},
illustrating the different sparsities found in varying simulations.
They show an analysis of performance vs sparsity threshold (which is
typically set around 0.3 fraction), and see up to factor of two
improvement over the original approach.  Calculations have been
demonstrated on systems with up to 36,000 atoms of silicon (as well as
16,000 atoms of DNA and 3,200 atoms of alumina). 

A multi-atom blocked approach to
sparse matrix multiplication\cite{Saravanan:2003ye} has been developed for quantum
chemistry-based approaches.  The method divides the cell along
cartesian dimensions in a binary way if the cell dimension is larger
than some cutoff, $R_c$. This gives the multi-atom blocks, which
necessarily include some zeroes, but allow use of efficient BLAS
calls.  The block size depends on the basis, from about 30 for a
minimal basis to about 70-80 for larger bases; the overall method is
significantly faster than element-by-element sparse multiplies. 

A study of truncation methods\cite{Rubensson:2005vs} suggested that
distance-based truncation does not allow error control, and proposed a drop
tolerances method based on sub-matrix magnitude using a norm, as well
as implementing sparsity control \emph{during} matrix multiplication
for efficiency.  In this original paper, the 1-norm was proposed.  A
hierarchical approach to matrix storage evolved out of this
work\cite{Rubensson:2007qq}, which subdivides a matrix into
sub-matrices, each of which can be further divided into more
sub-matrices.  At the lowest level, a matrix consists of real numbers;
they found that five levels was enough for 36,000$\times$36,000 matrices. The
method permits blocks which are not related to atoms for performance
reasons, and specifies uniform block sizes (32$\times$32 in the
examples given) at the lowest level. The developers use their own
algorithms for symmetric square and inverse Cholesky based on
symmetry, and show good performance and reduced storage relative to
optimised libraries. The sparsity of the density matrix was studied
for different molecules\cite{Rubensson:2011mi}. The effects of
truncation were analysed, and possible ways of truncating matrices to
a given tolerance were examined (so that the error introduced by
truncation is controlled).  A Euclidean norm-based method is accurate
but computationally expensive, while calculation of Frobenius norms
scales poorly with system size; they suggest a mixed norm based on the
Euclidean norm of blocks, where the Frobenius norm of the block is
found.  A related idea\cite{Hoshi:2007uq} uses a tolerance based on
either the number of atoms within a localisation region, or a
dynamically updated number of atoms based on the residual of the
localised functions.  We note, however, that a drop tolerance approach
can be less scalable than a distance based approach, as the sparsity
pattern changes with each iteration. The distance-based truncation is
variational, while the drop-tolerance truncation may not be (and is
often used in non-variational methods such as the purification
methods).

The parallelisation of linear scaling techniques requires considerable
care: the balance between communications and calculation will affect
efficiency, and significant numbers of operations or variables which
require storage or work on all atoms in the system on all processors
will lead to $N^2$ scaling, as well as to memory problems for large
systems.  Nevertheless, owing to the spatial locality inherent in the
methods, they are natural candidates for parallelisation, and
impressive scaling results can be achieved.  We recollect the two
types of scaling in common use: weak scaling, where the load per
processor is kept fixed and the system size and number of processors
are increased simultaneously; and strong scaling, where the system
size is fixed, and the number of processors is increased.  Efficiency
for strong scaling is harder to achieve than for weak scaling, though
weak scaling may well be the mode of operation chosen for linear
scaling codes.

An early TB method\cite{Itoh:1995yq} divided space into a number of
parallelepipeds equal to the number of processors, and assigned all
atoms in a parallelepiped to a processor.  The processor is then
responsible for calculating forces and positions of atoms in the
parallelepiped, and for storing matrix rows corresponding to those
atoms.  As this method also uses localised Wannier functions (LWF)
with well-defined centres not necessarily associated to an atom, the
processor is also responsible for all LWF whose centre lies within its
parallelepiped, and any matrices indexed purely by LWF.  Scaling on up
to 512 processors and 85,000 atoms was demonstrated, and an extensive
analysis of scaling was made, noting that as the volume assigned to
each processor decreases relative to a boundary area (due to
localisation radii) the amount of communication will change from
depending on the number of processors (as N$^{-1/3}_{\mathrm{proc}}$)
to depending on the volume of the boundary.   The same approach has
been used for an implementation of orbital
minimisation\cite{Kim:1995yi} within an \emph{ab initio} tight binding
method\cite{Shellman:2003rt}, though MPI and OpenMP parallelisation
are shared; the resulting code was demonstrated on up 1,024 processors
and 6,000 atoms. 

Another implementation of orbital minimisation\cite{Canning:1996ve}
spent considerable time and effort on sparse matrices and
parallelisation to allow efficient molecular dynamics. Matrices were
stored as orbital blocks and neighbours of atoms (described as four
dimensional storage).  The merits of particle vs spatial distribution
of atoms between processors was discussed; in particular, list
calculation (using the link-cell method) relies on locality for
efficiency.  They chose particle/orbital division for simplicity (and
as the method was communications limited).  The problem of symmetric
matrices and distribution is also considered: if symmetry is used to
reduce storage, even distribution becomes more complex.  Load
balancing is achieved dynamically, by subdividing the system into 3D
blocks.  The blocks are ordered by $x, y, z$, with the atoms in each
block ordered by $z, y, x$. The assignment to processors is balanced
to even work (the time per site is calculated roughly). The resulting
method showed good weak scaling, and reasonable strong scaling (the
normal problem when going to few atoms per processor resulting in
communications dominating). 

The approach in the \textsc{MondoSCF} code (now called \textsc{FreeOn})\cite{Challacombe:2000vh} is to order the atomic coordinates using a space-filling curve, so that atoms which are close in space are close on the curve.  An overlap of communications and computation is achieved by posting a series of non-blocking receives (\texttt{MPI\_Irecv}) and using blocking sends (\texttt{MPI\_Send}).  It is suggested that this arrangement (rather than a series of non-blocking sends followed by blocking receives) is more efficient, and less likely to lead to communication imbalance. Load balancing in the code is achieved by minimising the imbalance of a characteristic matrix (typically the Fock or density matrix) based on the distribution of numbers of non-zero elements between processors.  The resulting tests show reasonable parallelisation up to 16 processors and sustained performance and efficiency increases up to 95 processors.
Significant effort has been put into the linear scaling calculation of
the Fock matrix\cite{Gan:2004ss,Gan:2003bd}. The exchange-correlation
matrix is calculated by hierarchical
cubature\cite{Challacombe:2000lh}: a cartesian grid is divided into
blocks, with load balancing done dynamically to balance times. The key
assumption is that the computational time is proportional to the total
charge in box. Good scaling is found, and the approach is linear
scaling.  A similar approach is taken to the Coulomb problem (as
described in Sec.~\ref{sec:hamiltonian-building}).

\begin{figure}[h]
  \centering
  \includegraphics[width=0.5\textwidth]{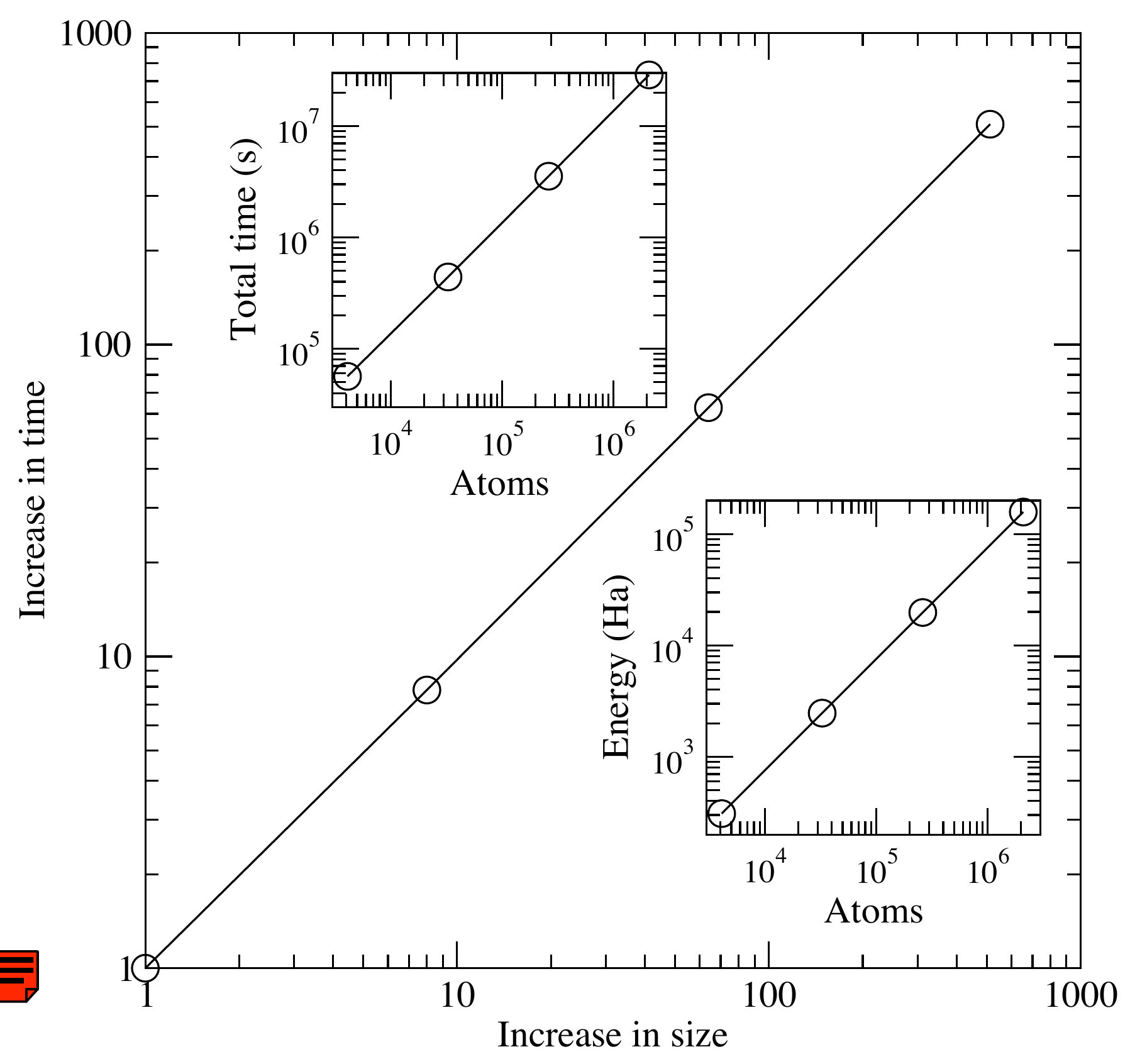}
  \caption{Linear and parallel scaling for bulk silicon on 512--4096
    cores. The insets show total time and total energy (made positive
    to enable log plot) while main graph shows increase in time with
    system size.  From Ref.\cite{Bowler:2010uq}. }
  \label{fig:ConquestScaling}
\end{figure}

\textsc{Conquest}\cite{Bowler:2001yo,Brazdova:2008pz,Goringe:1997cy}
sub-divides the unit cell into small orthorhombic  partitions of atoms
and blocks of grid points (which are not necessarily the same shape
and size), which are then distributed between processors using
either space-filling curves\cite{Brazdova:2008pz} or an optimisation
procedure which considers both communication and computation time.  Communication is
performed by small group (leading to a compromise between local
storage of unneeded elements and communications
efficiency)\cite{Goringe:1997cy}.  The indexing and searching rely on
different sets\cite{Bowler:2001yo}: the primary set (the atoms or blocks for which a
processor is responsible); the halo (all groups within range of a
primary set); the covering set (an orthorhombic set of groups within
range of a primary set for searching over).  Processors are
responsible for group of atoms (a bundle) and grid points (a domain)
which should ideally overlap to reduce communications overhead.
Scaling up to 4,096 processors and 2,000,000 atoms has been
demonstrated\cite{Bowler:2010uq}, as illustrated in
Fig.~\ref{fig:ConquestScaling}.  The code shows perfect weak scaling 
from 8 to 4,096 cores, and reasonable strong scaling (hut clusters of
Ge on Si(001) with either 11,620 or 22,746 atoms showed excellent
speed up from 16-128 cores and still 20 times faster going from 16 to
512 cores)\cite{Bowler:2010uq}.

There have been significant developments recently in the
parallelisation of \textsc{onetep}\cite{Skylaris:2006bu}.  The first scheme
used for parallelisation of matrix multiplication\cite{Hine:2009yb}
started on the diagonal and looped over processors using the modulus of the number of processors.
This approach was much more efficient than the original communications
which used an inefficient \texttt{MPI\_Bcast} call.  The localised
orbitals (non-orthogonal generalised wannier functions or NGWFs) are
distributed by sending lists of parallelepipeds and psinc coefficients
on the grid points in each parallelepiped.  A batch system is used on
columns (to get as many NGWFs as will fit
in memory), with the most intensive operation being the Hamiltonian acting on
the NGWFs. Using this scheme, scaling up to 27,000
atoms and 256 cores was demonstrated, though the speed-up when
increasing number of cores by a factor of four was only 2.5. A new
method\cite{Hine:2010fj} uses segments in each matrix to identify blocks,
and each processor sends only contributing blocks. Rather than a
round-robin approach (where core N sends to
N+1, N+2, etc) the code now has an on-demand communication pattern.
Scaling has been demonstrated again up 256 cores, with
a three times speed up for increasing the core count by four times.
\textsc{onetep} switches between sparse and dense matrix algebra
depending on the filling of the matrices in question.

\textsc{siesta} can be run in parallel, though in its normal implementation only on small numbers of processors (scaling up to 32 or so).  However, recent work\cite{Sanz-Navarro:2011cq} has changed the parallelisation.  The code originally divided up the cell by columns on the integration grid; the new implementation uses recursive bisection and weighting based on the number of neighbours.  It also schedules communications to avoid unnecessary all-to-all communication. The resulting code shows an improvement for scaling 262 water molecules on 8-128 processors from 24\% speed-up (old scheme) to 52\% (new scheme---for an inhomogeneous distribution).  

The divide and conquer method has been parallelised\cite{Vincent:1998ri} following the obvious route: the individual subsystems are assigned to processors, with the limiting parallelisation given by subsystem size.  In the implementation described, the system data was copied to all processors, with the intended aim being small numbers of processors.
The method has also been included in a massively multi-scale MD approach with divide and conquer as the embedded quantum method\cite{Shimojo:2005wm,Shimojo:2008gf}.  This approach has scaled to systems with 1.2 million quantum atoms and billions of classical atoms\cite{Nakano:2007ez}.
 
Orbital-free DFT has also been parallelised\cite{Watson:2000lo}, by dividing up the real-space grid evenly between processors.  Since the real-space data is purely real, only the positive half-sphere of reciprocal space is needed; these points are also divided evenly between processors, but the sparse grid involves a map. Load balancing requires consideration of a compromise: if it is performed by points, this gives ideal balance, but a communications penalty; by line gives better communications, but worse balance; by plane gives good communications by potentially poor load balance.

\subsection{Structure Relaxation}
\label{sec:structure-relaxation}

Before considering details of a subset of the applications which use linear scaling methods, we note that the most common task required of a code is to relax the atoms to their lowest energy configuration. A point which is becoming more widely discussed is the scaling of relaxations: just because the electronic ground state can be found in linear scaling time does not mean that a relaxed atomic structure can be found with the same scaling.  Indeed, the difficulty of relaxing a system with low and high frequency phonon modes (or equivalently with very different length scales in the curvature of the energy), will depend on the ratio of the largest and smallest eigenvalues of the Hessian.  It seems likely that, in general, the number of iterations to find a relaxed structure will increase with system size if relaxation is implemented naively; simple arguments indicate that, for conjugate gradient relaxations, the number of iterations with increase with the largest linear dimension in the system\cite{Goedecker:2001gl}.

There have been a number of proposals to alleviate this problem.
Preconditioning or relaxing low frequency relaxations using insights
from elasticity theory has been
suggested\cite{Goedecker:2001gl,Fernandez-Serra:2003fr}.  One
approach\cite{Goedecker:2001gl} transforms the atomic forces onto a
discrete grid (by smearing them) and uses multigrid approaches to
solve for the Hessian without diagonalising.  The resulting method
shows that the number of force evaluations (and hence time per atom required to find relaxed structure) is independent of system size on scaling from 500 to 800,000 atoms of silicon with an interatomic potential.  An alternative approach\cite{Fernandez-Serra:2003fr} uses a model Hessian either to precondition conjugate gradients or as the input Hessian for a quasi-Newton method (in both following exact diagonalisation or inversion, though this could presumably be made linear scaling if needed).  The model Hessian is effective at improving convergence, though is not tested on increasing system size.

The obvious alternative approach, particularly for large molecules, is to use internal coordinates (bond lengths, angles and dihedral angles) instead of cartesian coordinates.  A linear-scaling transformation to and from internal coordinates has been proposed\cite{Nemeth:2000ol}, which uses Cholesky decomposition on a sparse transformation matrix (which is shifted to avoid problematic zeroes) and iterated.  This method has been harnessed to allow a linear scaling relaxation method\cite{Nemeth:2004fj} where the overall problem is decomposed into a set of 3N independent relaxation problems (using a weighted fit).  The method is applied to a protein (with 263 atoms) and is efficient, though scaling is not tested explicitly.

Relaxation can be made more efficient, and less dependent on system size, if each step requires a shorter time to relaxation.  A method to allow extrapolation of the density matrix from a previous step in a relaxation to the current one\cite{Niklasson:2010hq} uses the trace-correcting formalism to extrapolate.  The method can be applied very efficiently to a local perturbation, with only the overlap matrix affected (and the density matrix extrapolated); it also improves convergence for standard geometry optimisation.

The final approach extends the Fragment Molecular Orbital method (Sec.~\ref{sec:divide-conquer}) to geometry optimisation.  Similarly to the perturbation approach just described, the system is divided into frozen, polarisable and active domains: the frozen domain is calculated just once, at the start, and not updated; the electronic structure of the polarisable domain is updated in response to changes in the active domain, but the atomic positions are frozen; while the both the atomic and electronic degrees of freedom for the active domain are updated at each step (note that this is a form of embedding\cite{Bowler:2002kh}).  The method has been applied to prostaglandin synthase in complex with ibuprofen, with over 19,000 atoms.

\section{Implementation and Applications}
\label{sec:appl-impl-perf}

\subsection{Implementations}
\label{sec:implementations}

Implementations of linear scaling methods can be split into two camps: first, those that build on existing codes and methods and simply add a new solver to the self-consistent loop (these often include quantum chemistry-based codes and methods, such as the fragment methods described in Sec.~\ref{sec:divide-conquer}); and second, those that create an entirely new implementation, often involving careful parallelisation (as described in Sec.~\ref{sec:parallelisation}).  At the start of this section, we will briefly survey the second class, though this is not an all-inclusive list, and more details can be found in papers referenced above describing approaches to linear scaling.  

The main codes that we are aware of with linear scaling functionality
are, in alphabetical order: \textsc{Conquest}\cite{Conquest};
\textsc{ErgoSCF}\cite{ErgoSCF}; 
\textsc{femteck}\cite{Tsuchida:1998dp,Tsuchida:2007kq}; \textsc{FreeON}\cite{FreeON}; \textsc{onetep}\cite{onetep}; \textsc{OpenMX}\cite{OpenMX}; \textsc{profess}\cite{Hung:2010ve}; and \textsc{siesta}\cite{Siesta}.  We will discuss the basis sets used to represent the localised orbitals (or support functions or Wannier functions) as well as the linear scaling kernel.

\textsc{siesta}\cite{Ordejon:1998ly,Artacho:1999gf,Ordejon:2000mz,Soler:2002kn} was one of the earliest linear scaling codes made widely available, and is still in widespread use today, though most of the applications use exact diagonalisation to find the ground state.  The code uses pseudo-atomic orbitals  (Sec.~\ref{sec:real-space-local}) as a basis, and the method for calculating forces and stresses is clear\cite{Soler:2002kn}.  The linear scaling kernel used is the Kim, Mauri and Galli functional (Sec.~\ref{sec:direct-iter-appr}) which ameliorates the convergence problems of the OMM functional, and recently\cite{Cankurtaran:2008uq} a divide-and-conquer kernel has been implemented.  Two different linear scaling implementations of the vdW-DF functional for van der Waals corrections\cite{Roman-Perez:2009fk,Gulans:2009vh} have been developed within \textsc{siesta}.  The code is freely available for academic use.

The \textsc{onetep}
code\cite{Skylaris:2005ai,Skylaris:2005wq,Haynes:2006rw,Skylaris:2006bu,Mostofi:2007le,Haynes:2008zr,Skylaris:2008xz,Hine:2009yb}
represents the density matrix in terms of non-orthogonal generalised
Wannier functions (NGWFs)\cite{Skylaris:2002xi}, which are in turn represented by periodic
sinc functions (which are periodic, bandwidth-limited delta functions)
on a fixed, real-space grid.  The kinetic energy (and preconditioning)
are calculated using local Fourier transforms, known as the FFT box
method\cite{Skylaris:2001gs}.  The linear scaling
kernel\cite{Haynes:2008zr} is the LNV
method\cite{Li:1993lg,Nunes:1994pi}, and the method used for
calculating forces has recently been published\cite{Hine:2011gf} (it
is worth noting that the contribution from the local pseudopotential
scales with $N^2$, but with a small enough prefactor that it only
starts to become significant around 15,000 atoms in the published
scaling tests).  Empirical van der Waals interactions have been
implemented\cite{Hill:2009fv} and extensive tests comparing to
plane-waves have been carried out\cite{Skylaris:2007lp}.  The code is 
commercial, though can be obtained for a modest fee by academics.

The \textsc{OpenMX} code\cite{Ozaki:2000ix,Ozaki:2006uq} uses a
carefully constructed set of pseudo-atomic
orbitals\cite{Ozaki:2004pt,Ozaki:2004qq} as a basis for the density
matrix, and, similarly to \textsc{siesta}, concentrates on exact
diagonalisation and other efficient routes to the ground
state\cite{Ozaki:2010bh}, though recursion-based linear scaling
functionality is available\cite{Ozaki:2006uq}.  In particular, the
code uses an ingenious combination of divide and conquer and recursion
methods. An extensive set of extensions has been implemented,
including non-collinear spin, constrained DFT, LDA+U, Wannier function
construction and polarisation calculations.  The code is available freely.

The \textsc{FreeON} (formerly MondoSCF) code\cite{Schwegler:1996yq,Challacombe:1997qy,Schwegler:1997mz,Challacombe:1999mb,Schwegler:1999fr,Challacombe:2000lh,Challacombe:2000vh} comes from the quantum chemistry community, using standard basis sets to represent the density matrix, with linear scaling coming from trace-correcting methods\cite{Niklasson:2003mw}.  Forces are calculated within the standard framework\cite{Weber:2006lq,Weber:2006rr}.  Recent extensions have considered perturbation theory and time-dependent DFT.  The code is freely available.

The \textsc{ErgoSCF}
code\cite{Rudberg:2010tg}
also comes from the quantum chemistry community, using Gaussian basis
sets to represent the density matrix and trace-correcting methods to
find the ground state.  The code is parallelised for shared memory
machines, and is freely available under the GNU Public Licence.

The \textsc{Conquest}
code\cite{Goringe:1997cy,Bowler:2000fx,Bowler:2000xd,Bowler:2001yo,Bowler:2002pt,Bowler:2006wa,Bowler:2010uq,Brazdova:2008pz,Gillan:2007fb,Otsuka:2008ri,Torralba:2008wm}
can use either pseudo-atomic orbitals or systematically improvable
blip (B-spline) functions to represent the density matrix, and the LNV
approach to linear scaling, following initialisation with McWeeny
iteration; exact diagonalisation via \textsc{ScaLAPACK} has also been
implemented.  The forces are calculated as exact derivatives of the
energy with linear scaling time\cite{Miyazaki:2004ee}. The code
implements a number of standard features, including GGA with
non-self-consistent forces\cite{Torralba:2009hc}, and recently spin
and exact exchange.  It also includes constrained DFT, which has been
shown to converge in a linear scaling manner\cite{Sena:2011fc}.  The
code is in late-stage beta release, and will be freely available
during 2012. 

The \textsc{femteck} code\cite{Tsuchida:1998dp,Tsuchida:2007kq} uses
finite elements to represent the Wannier functions, and the augmented
OMM method\cite{Tsuchida:2007kq} to find the ground state in linear
scaling time.  The code has been applied to liquid
ethanol\cite{Tsuchida:2008ai} and a fast ion
conductor\cite{Ikeshoji:2011fk} with stability and efficiency.

The \textsc{profess} code\cite{Ho:2008fs,Wang:1998ss,Watson:2000lo,Hung:2009yl} is an orbital-free DFT method (Sec.~\ref{sec:orbital-free-dft}) which represents the electron density directly on the grid, and has implemented a number of different kinetic energy functionals.  As with all OFDFT codes, only local pseudopotentials can be used, which, along with the limitations on functionals, restricts its use to simple metals and some properties of semiconductors.  Nevertheless, extremely large systems can be addressed: recent developments in ion-ion and electron-ion calculations have allowed the efficient parallelisation of OFDFT, with a benchmark calculation on a million atom sample of perfect bulk Al demonstrated\cite{Hung:2009yl,Hung:2010ve}. The code is freely available.

\subsection{Applications}
\label{sec:applications}

In this section, we survey the applications of the linear-scaling methods explained in previous sections.
There are many research areas where large-scale DFT calculations are
expected to play an important role, 
and recently calculations for actual scientific research have been emerging.
However, the applications of linear-scaling methods are still rather limited.
It is necessary, at this stage in the development of the methods,  to
investigate the availability, accuracy and efficiency of the
techniques employed in each study.
It is also not obvious what kind of quantities can be calculated by such large-scale DFT studies.
So far, different methods have been used depending on the system or
the phenomena of the research area. 
From the examples of the applications introduced here, we would like
to summarise what has been found so far,
such as the size of the systems which can be treated, robustness of the calculations, 
and the accuracy of the calculation method, including the quality of the basis sets
used in the calculations.

\subsubsection{$\mathcal{O}(N)$ calculations on biological systems}
\label{sec:on-calc-biol}
One of the most promising targets for $\mathcal{O}(N)$ DFT study are biological systems. 
In spite of the complex structures of large biomolecular systems, they provide atomically controlled systems 
for realising surprisingly sophisticated functions. 
With the rapid increase of experimental information, the demand for accurate modelling of large biological systems is also growing. 
It is a challenge to understand the mechanism of such phenomena from the atomic scale, especially with quantum mechanics.

So far, most theoretical studies on biological systems from atomic scale have been done using classical force fields. 
Although these methods are powerful tools to investigate various phenomena in biological systems, they have a serious 
problem that the calculated results sometimes depend strongly on the parameters used for interatomic potentials. 
Different force fields or even different version of the same force field can show qualitatively different results. 
In addition, it is quite difficult for the methods to treat the phenomena of bond forming or breaking properly.
Thus, hybrid approaches like ONIOM or QM/MM (quantum mechanics / molecular mechanics) methods are often used for the 
study of chemical reactions, like enzyme reactions in biology.
With these methods, the important region where chemical reactions take place is treated
by a method based on quantum mechanics, and the dynamics or mechanics of the atoms in 
the surrounding region is calculated using a classical force field. 
However, it is sometimes uncertain how to define these two (or more) regions and it is not clear how accurate 
the method is, especially when the QM region is not large enough.  
Obviously DFT calculations on the entire or the sufficiently large region of complex 
biological systems are of great importance and linear scaling DFT methods are expected to answer this request.

\begin{figure}[h]
  \centering
  \includegraphics[width=0.5\textwidth]{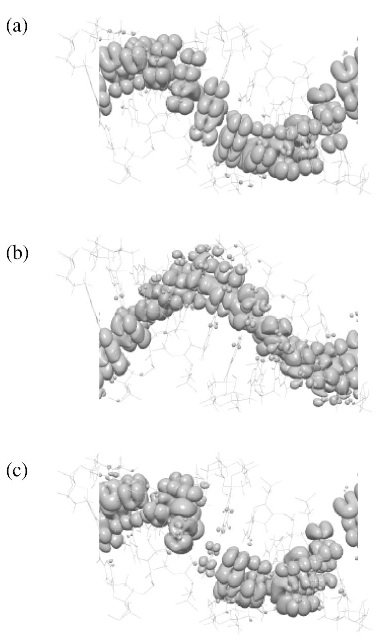}
  \caption{Isosurfaces of the electron density of dry DNA calculated using
    \textsc{siesta}\cite{Pablo:2000ys} for: (a) the eleven
    highest-occupied molecular orbitals; (b) the eleven
    lowest-unoccupied molecular orbitals; and (c) the eleven highest
    occupied orbitals following a single mutation.  Reprinted figure
    with permission from P. J. de Pablo \emph{et al.},
    Phys. Rev. Lett. \textbf{85}, 4992 (2000).  Copyright (2000) by
    the American Physical Society.}
  \label{fig:DNA0}
\end{figure}

With such demands, there was already a report in 2000, showing the
$\mathcal{O}(N)$ DFT calculations on a dry DNA model system 
containing 715 atoms\cite{Pablo:2000ys}.
The system is a periodic double helix DNA consisting of eleven guanine-cytosine (G-C) base pairs in the unit cell.
For this system, $\mathcal{O}(N)$ calculations were performed using the Siesta code to employ the structure relaxation
mainly with a double-$\zeta$ basis sets.
Using the relaxed structure, a conventional diagonalisation technique was also performed to calculate the Kohn-Sham 
energy and orbitals, and to confirm the accuracy of the forces with the $\mathcal{O}(N)$ method.
The results for the simple polyG-polyC system show that the topmost valence bands are made by the eleven 
highest occupied molecular orbitals of Guanine bases, and they are
connected along the direction of the DNA chain; the eleven highest-occupied
molecular orbitals and the eleven lowest unoccupied molecular orbitals are illustrated in Fig.~\ref{fig:DNA0}. 
They further investigated the effects of a defect structure by introducing the mutation of 
one G-C base pair, aiming to mimic the electronic structure of $\lambda$-DNA, which has a random 
sequence of the DNA bases.
Following the same procedure, the orbitals of this system were calculated and they found that
the orbitals showed cleavage at the point where the swap was introduced. This suggests
that the resistivity of $\lambda$ DNA should be very high, consistent with the measurements of the 
electron transport of the system. 
Although some of the results relied on a conventional method, this pioneering work 
clearly shows that $\mathcal{O}(N)$ DFT study will be powerful in the study of biological systems.
Similar hybrid works, combining $\mathcal{O}(N)$ and conventional calculations on DNA 
have also been performed\cite{Bondesson:2007fj,Ozaki:2010bh}

\begin{figure}[h]
  \centering
  \includegraphics[width=1.0\textwidth]{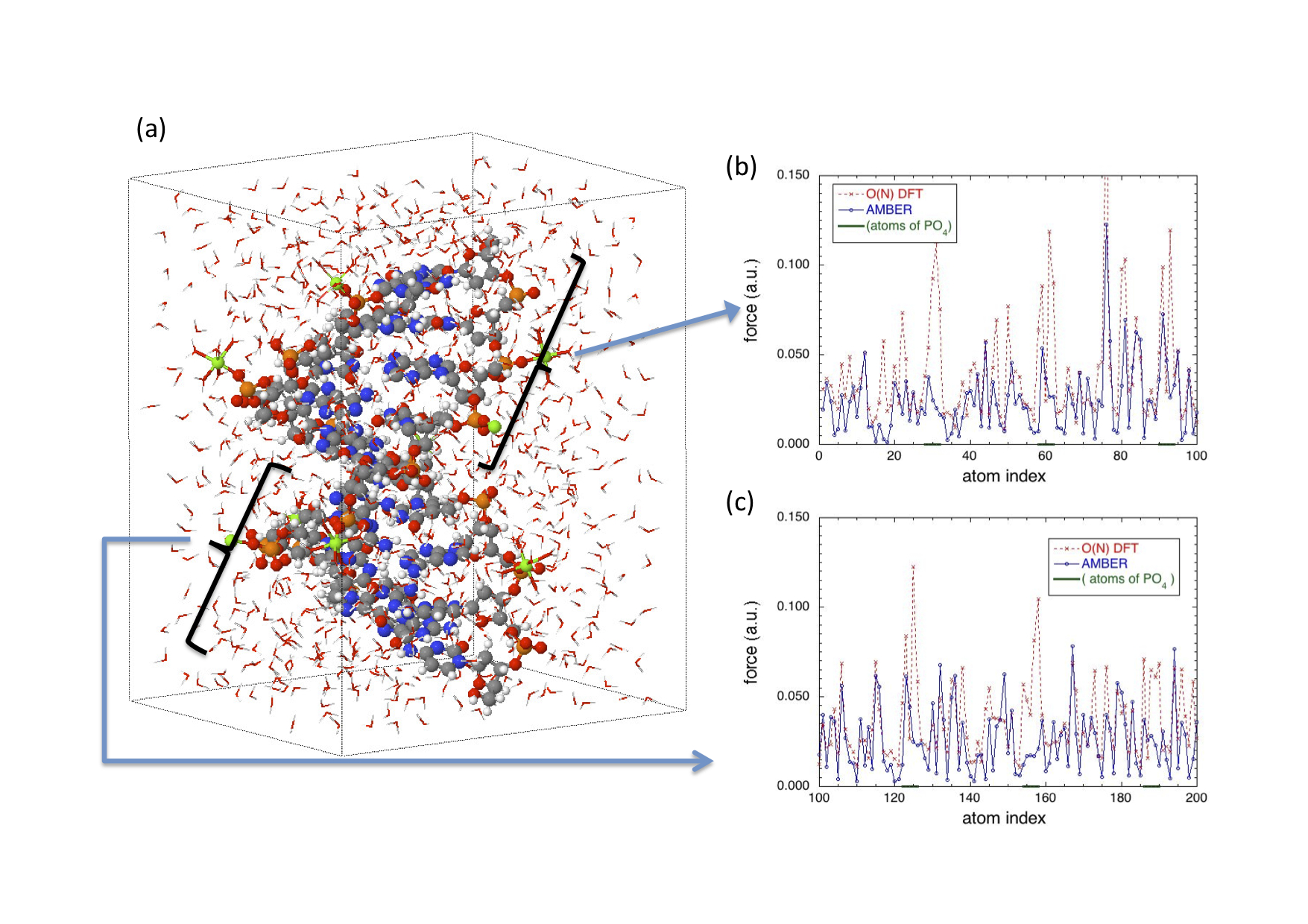}  
  \caption{(a) Snapshot of the structure of hydrated ten-mer of DNA, and the 
           calculated atomic forces on (b) 1st-100th and (c) 101st-200th atoms of DNA 
           by O(N) DFT and AMBER force field calculations. In (b) and (c), 
           green bars on the horizontal axis show the atoms of phosphoric acids.}
  \label{fig:DNA1}
\end{figure}

\begin{figure}[h]
  \centering
  \includegraphics[width=0.5\textwidth]{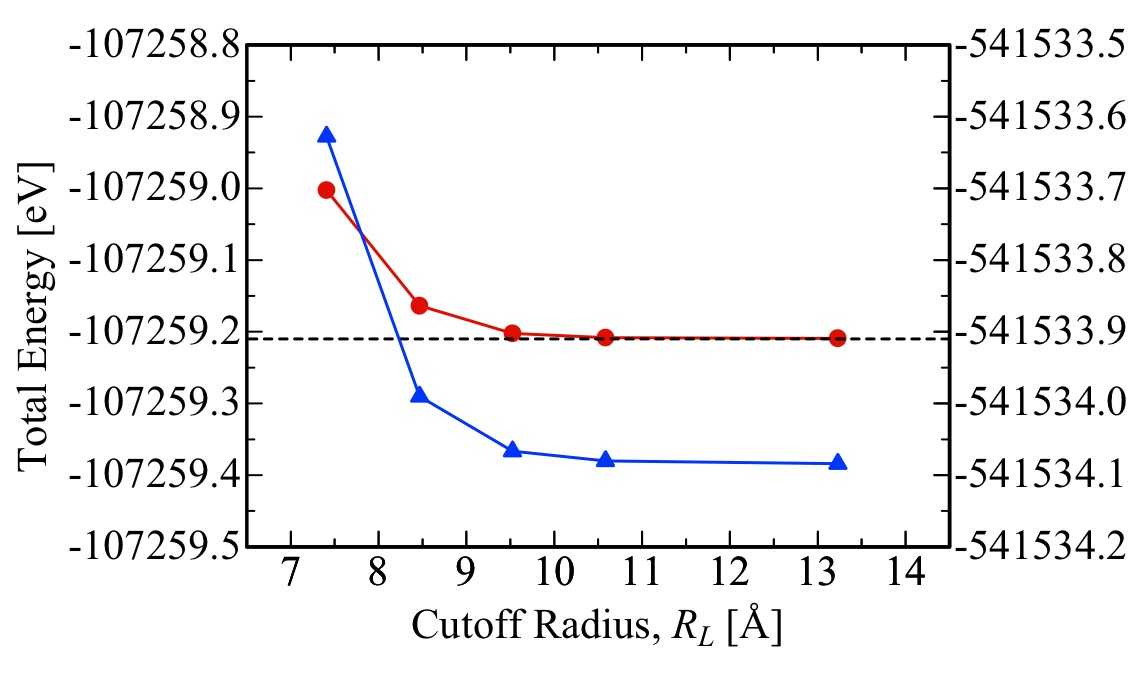}  
  \caption{Dependence of total energy on density matrix cutoff, $R_L$,
  for DNA without water molecules (red line and circles, left axis) and full
  system (blue line and triangles, right axis).  The dashed line shows
  the exact diagonalisation result for DNA without water.  From Ref.~\cite{Otsuka:2008ri}.}
  \label{fig:DNA2}
\end{figure}

Biomolecular systems usually have large HOMO-LUMO gaps and thus the
electronic structure is expected to be well localised. 
In this respect, the systems should be generally suitable for $\mathcal{O}(N)$ DFT studies.
This aspect was clearly demonstrated in a theoretical study on a test
DNA system using the \textsc{Conquest} and the DMM method\cite{Otsuka:2008ri}
The system investigated in their study is a B-DNA decamer (5'-d(CCATTAATGG)$_2$-3'), with 932 water molecules and 9 Mg$^{2+}$ counter ions. 
Its structure is shown in Fig.~\ref{fig:DNA1}(a), including 3,439 atoms in total. 
The atomic positions were prepared from a snapshot taken from an MD simulation with the AMBER classical force field. 
They showed that the DFT method can be applied to a system of this size, and the accuracy of the DMM method 
is reported to be surprisingly good. 
Figure~\ref{fig:DNA2} shows the dependence of total energy on the cutoff length $R_L$, calculated in a
non-self-consistent way with a minimal basis set of PAOs.
Here, both dry and hydrated DNA systems are calculated and the dry system is made by removing 
all water molecules from the system shown in Fig.~\ref{fig:DNA1}(a).
Since the dry system includes only 643 atoms (634 atoms for DNA and 9 Mg atoms), it was possible
to also employ an exact diagonalisation method to test the convergence
of the linear scaling method.  The results are shown in
Figure~\ref{fig:DNA2}.  
The energy obtained by diagonalisation is shown by a horizontal dotted line, and the total energy 
calculated with various cutoffs using the DMM method is plotted by a red line with circles. 
These results clearly illustrate that the total energy by the DMM method converges very rapidly. 
The error at $R_L$=8.47~\AA\ is already 0.046 eV (7.2 $\times 10^{-5}$ eV/atom) and, if we 
increase $R_L$ to 9.53~\AA, the error becomes only 0.0078 eV (1.2 $\times 10^{-5}$ eV/atom).
This rapidly convergent behaviour can also be observed in the calculation of a DNA system hydrated with many water 
molecules. The total energy of this system calculated as a function of $R_L$ is plotted by a blue line with 
triangles in Fig.~\ref{fig:DNA2}. Note that the energy scale for this system (right in Fig.~\ref{fig:DNA2}) is same as the one for the 
dry system (left in the figure), though shifted. The convergence of this system is also very rapid and 
the total energy at $R_L$=13.23~\AA\ can be considered as a well converged value. 
Then, the error at $R_L$=8.47~\AA\ is 0.094 eV, which corresponds to 2.7 $\times 10^{-5}$ eV/atom.
If we use $R_L$=9.53~\AA\ the error becomes 0.017 eV for 3439 atoms (4.9 $\times 10^{-6}$ eV/atom).
With such accuracy, it is possible to discuss the difference of the total energy induced by a local reaction 
in a system containing several thousand atoms. 
There are reports showing that similar accuracy can be obtained also in other linear-scaling methods\cite{Ozaki:2010bh}.
These results suggest that $\mathcal{O}(N)$ DFT methods will be able to provide quite accurate results in molecular biology.

The FMO method (described in Sec.~\ref{sec:divide-conquer}) is efficient and accurate for biomolecular systems,
and already has many published examples\cite{Fedorov:2007dq}.
One of the largest problems tackled with the approach was a calculation of active sites on influenza A viral 
haemagglutinin\cite{Sawada:2010zt}.  
The calculation used a polarisable continuum model and about 24,000 atoms of protein. 
They found that the binding of a class of chemicals known as sialosides is not regulated by allosteric effects 
(in other words the number of ligands bound does not affect binding affinity); this result can be used in design of drugs 
for coping with influenza pandemics. 
One of the advantages of the FMO method is that they can use DFT,
as well as Hartree-Fock and post-Hartree-Fock methods, like the
MP2 method. Another advantage is that the method can easily calculate the quantities by which we
can discuss whether the interactions between two fragments are attractive or repulsive. 
This is called pair interaction energy decomposition analysis, and can give useful information 
for the ligand-protein interactions.
There has been an attempt to apply the FMO method to silicon systems\cite{Ishikawa:2006pd}, but we note here that 
an investigation of the MTA fragment method for 2D $\pi$-conjugated systems\cite{Yeole:2010tw} found that large fragments were 
needed for accuracy in these types of system.

In the actual research on biological systems using DFT, it may be a serious problem that 
DFT functionals such as LDA or GGA cannot express van der Waals interactions correctly. 
However, there have been many attempts to solve this problem. 
A new class of exchange correlation functional which includes the vdW
interactions has been proposed by Dion et al., called vdW-DF method. 
There are already many examples using this method and some revised version (vdW-DF2) has been 
recently proposed\cite{Lee:2010zr}. 
There is also a more efficient method, called DFT+D method, which has been 
applied to many biological or organic systems\cite{Grimme:2006ly,Grimme:2010ul}.
This is a simple method where total energy and forces are calculated by adding 
empirically parametrized interatomic potentials to the total energy calculated by DFT.
In the early version of the method, the parameters used in the method only depended on the species of atoms and
did not vary in different environments. 
Recently the methods to improve the transferability, by changing the parameters 
using the charge density calculated by DFT method or from the analysis of the local
coordination numbers, have been proposed\cite{Tkatchenko:2009wd,Grimme:2010ul}.
The results obtained by the new methods reported so far are encouraging.
These methods, vdW-DF or DFT-D, can be easily applied to the $\mathcal{O}(N)$ methods and 
there are already some reports for their implementation to the $\mathcal{O}(N)$ codes\cite{Hill:2009fv,Roman-Perez:2009fk}.
The vdW interactions are usually very weak and seem to be more effective for larger systems, 
thus more important in $\mathcal{O}(N)$ DFT calculations. 

Another serious problem in the DFT study on biological systems is that the simulation time of
molecular dynamics is very short. 
One candidate to overcome this problem at present is to use the `force matching' method\cite{Ercolessi:1994dq},
which aims to refine the classical interatomic potentials using the DFT results.
To employ this method in the complex biological systems, it is necessary that we can calculate 
the total energy and atomic forces for very large biological systems using DFT.

As mentioned previously, DMM method can calculate the atomic forces easily and accurately.
For the system in Fig.~\ref{fig:DNA1}(a), the total energy and atomic forces are recently 
calculated with DZP basis sets and compared with those by the AMBER force 
field\cite{Otsuka.unpublished,Miyazaki:2009nx}.
The calculated forces acting on the 1st to the 200th atoms, from a part of DNA, 
are shown in Fig.~\ref{fig:DNA1}(b) and (c).
In the figure, the green bars on the horizontal axis show the indices of the atoms
forming the phosphoric acids.
The result shows that the atomic forces calculated by these two 
methods agree well for most of the atoms.
However, we can see that the agreement for the atoms in the phosphoric acids
is much worse, compared with the forces on atoms of DNA bases.
The deviation in the forces on the phosphoric acid part seem to depend
on the position of the Mg counter-ion close to this part. 
We can expect that such DFT results would be useful to revise the accuracy of classical force fields.
All of these results suggest that $\mathcal{O}(N)$ DFT (or other quantum mechanics) methods will be able to 
play a significant role soon in the study of complex biological systems.

\subsubsection{Order-N DFT study on nanoscale structures of Ge islands on Si(001)}
\label{sec:order-n-dft-Ge-Si}
\begin{figure}[h]
  \centering
  \includegraphics[width=0.6\textwidth]{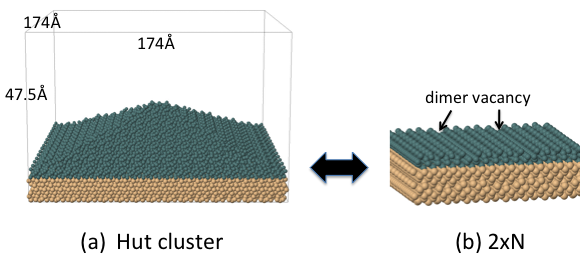}  
  \caption{(a) Structure of Ge hut clusters on Si(001). (b) Structure
    of the wetting layer of Ge on Si(001), showing the $2\times N$ reconstruction.}
  \label{fig:GeHut1}
\end{figure}

Another class of targets which can greatly benefit from $\mathcal{O}(N)$ DFT calculations are nano materials or nano science.
Here we would like to introduce the study on the energetics of a nano-structured system, 
Ge three dimensional islands grown on Si(001). 
The Ge/Si~(001) system has been extensively studied because it is a prototypical example of hetero-epitaxial 
Stranski-Krastanov growth. 
It is also technologically important because of the formation of organised quantum dots.
Many experimental studies have been reported so far, and they confirm the following results.
When Ge atoms are deposited on Si(001), growth initially occurs layer by layer, up to a critical thickness of 
about three monolayers (ML).
Strain due to the lattice mismatch is relieved by the formation of regularly spaced rows of dimer vacancies in this two-dimensional (2D) structure, resulting in the $2 \times N$ structure. 
Deposition of further Ge atoms leads to another strain-relief structure, 3D pyramid-like structure, called hut clusters\cite{Mo90},whose four facets are well established to be \{105\} surfaces.
The typical side length of the hut cluster is about 150 \AA.
Recently, all atom DFT calculations on the hut cluster including a substrate were performed using an $\mathcal{O}(N)$ technique 
to study the transition from the 2D to 3D structures\cite{Miyazaki:2007kr,Miyazaki:2008kx}. Here, we introduce this $\mathcal{O}(N)$ DFT study.

\begin{figure}[h]
  \centering
  \includegraphics[width=0.5\textwidth]{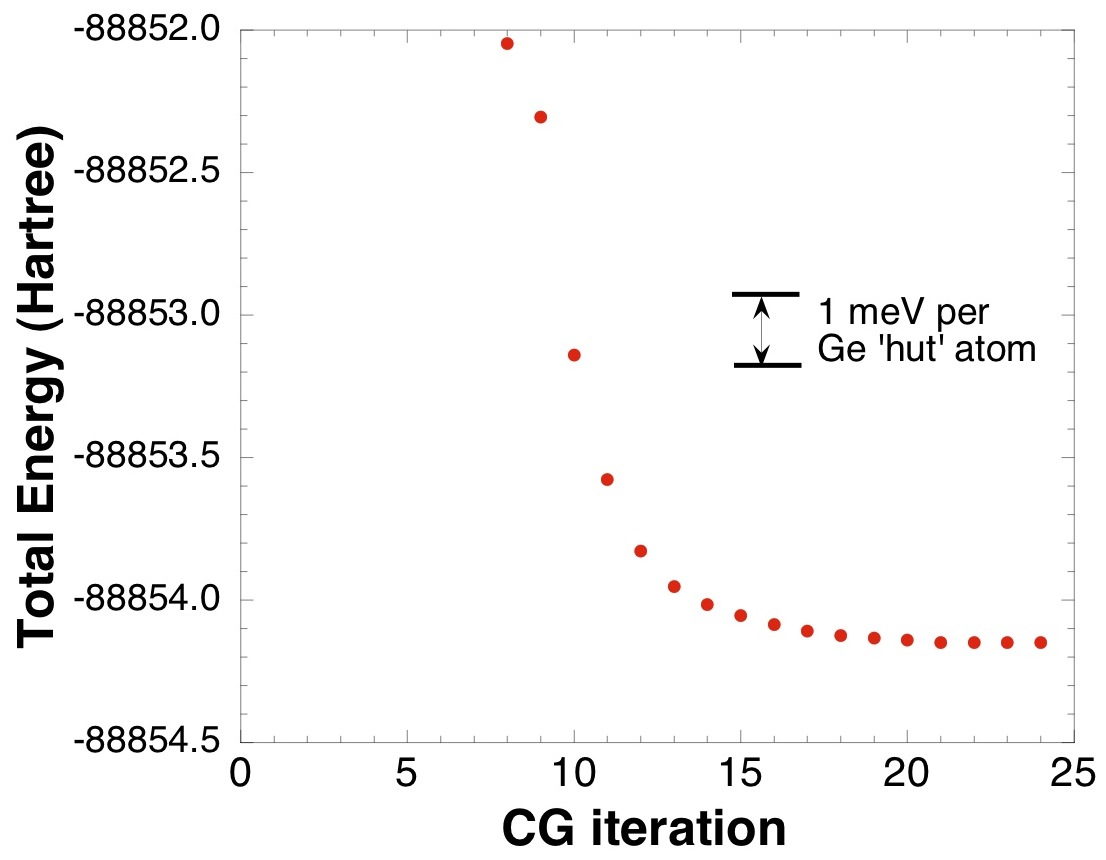}  
  \caption{Convergence of total energy during structural relaxation of 
           Ge hut cluster on Si(001), which contains over 20,000 atoms.
           From Ref.~\cite{Miyazaki:2008kx}}
  \label{fig:GeHut2}
\end{figure}

 The stability of the 3D structures grown on surfaces is usually governed by the competition between
 the release of the strain energy from the formation of a 3D structure
and the energy increase due to the larger surface area of the facets
on the surface.
So far, theoretical approach on the Stranski-Krastanov growth mode has usually used continuum elasticity theory 
for the first part, and employed DFT calculations for the latter. 
However, a unique situation exists in the Ge/Si(001) system.
The surface structure of the strained Ge~(105) was clarified by a combination of STM and DFT studies and
the DFT calculations show that the strained Ge~(105) surface is much more stable than the strained Ge(001) surface. 
Therefore, even though the surface area increases, the contribution from the surface energy is extremely small or 
may be lowered by the formation of facets.
If this is the case, we need other contributions to reproduce the 2D-3D transition. 
It is important to consider the energy contributions from the wetting layer, as well as the edges where the facets meet each other.
In addition, as the area of the facets of the experimentally observed Ge hut cluster is not large, the
evaluation of the surface energy using conventional DFT is also doubtful. 
For these reasons, the validity of previous theoretical approaches is uncertain, especially for small hut clusters. 
To overcome these problems, it is necessary to employ all-atom DFT simulations on this system, 
including the entire Ge hut cluster, wetting layer and Si substrate.  
Since the number of atoms exceeds a few thousands even for small hut clusters,
we need a linear-scaling technique to employ DFT calculations on such large systems and 
it has been recently shown that the calculations are possible with the \textsc{Conquest} code.

\begin{figure}[h]
  \centering
  \includegraphics[width=0.5\textwidth]{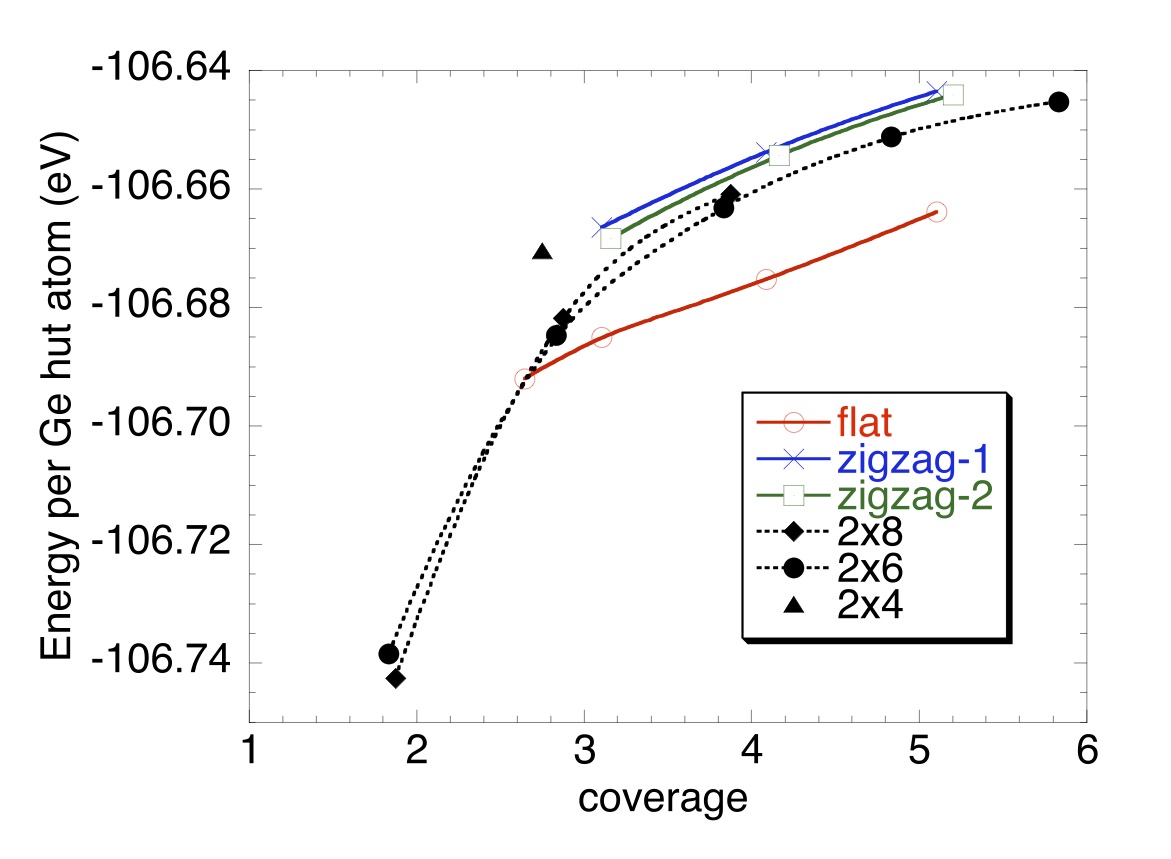}  
  \caption{Comparison of energetics for different 2D (dotted lines with filled symbols) and 
    3D structures (solid lines with open symbols or crosses) for Ge on Si(001). 
    From Ref.~\cite{Miyazaki:2008kx}}
  \label{fig:GeHut3}
\end{figure}

Before the work on full systems, the accuracy of the computational methods 
was thoroughly examined by the calculations on the strained Ge (105) surfaces\cite{Miyazaki:2007kr}.
It should be noted that even semi-empirical TB method, though it is based on the quantum mechanics, 
is found to be not accurate enough for the energetics of the surfaces in this system.
Using the results, $\mathcal{O}(N)$ DFT calculations on the 3D Ge hut clusters have been employed\cite{Miyazaki:2008kx}. 
At the non self-consistent level, structural optimisation on systems of different sizes
were employed using a standard CG method.
The largest system calculated in this work, shown in Fig.~\ref{fig:GeHut1} , contains $\sim$ 23,000 atoms.
As we can see in Fig.~\ref{fig:GeHut2}, the structure optimisation is robust and accurate enough even for 
such large systems. 
For the 3D hut clusters, three structural models having different facet and edge structures are investigated. 
Furthermore, the structure of 2D $2 \times N$ reconstructions ($N =$ 4, 6 and 8) and its total energy are 
calculated for comparison using the same calculation condition.
The energy of these structural models as a function of coverage is illustrated in Fig.~\ref{fig:GeHut3}.
It shows that the 2D structure is more stable for small coverage of Ge atoms, but the 3D hut structure
becomes stable when the coverage exceeds 2.7 ML. 
Interestingly, this coverage agrees with the experimental value showing the 2D-3D transition. 
This $\mathcal{O}(N)$ DFT study has succeeded to clarify the energetics of the 3D hut cluster systems, 
but the kinetic aspects are also important to simulate the actual growth.
In this respect, since the $\mathcal{O}(N)$ DFT study can treat the entire system, it is also possible
to work on the dynamical aspects by putting additional Ge atoms on the optimised structures. 
Such works had been unavailable so far and we expect many fruitful information would be obtained by 
$\mathcal{O}(N)$ DFT studies in the near future. 

\subsubsection{Other examples.}
\label{sec:other-examples}

Here, we survey other applications. Of course, it is impossible to
show all examples, but we try to show various areas of applications
using different $\mathcal{O}(N)$ methods.  

First of all, an excellent problem for linear scaling methods is a one dimensional system; there is also considerable interest in the transport properties of molecules\cite{Tao:2006uq,Joachim:2005kx}.  One approach to calculating transport for large systems\cite{Biel:2008ys} divides the system into layers with local coupling between them, and then uses normal DFT calculations to find the electronic structure of the layers.  The resulting method is linear scaling (and related to divide-and-conquer techniques, Sec.~\ref{sec:divide-conquer}).  This approach was applied to defects in carbon nanotubes (CNT), and the localisation lengths associated with them.  By examining systems with up to 25 defects they predict that Anderson localisation will be observed in CNT at room temperature.  The method allows systems hundreds of nanometers long to be simulated.

Divide-and-conquer-like algorithms have found a relatively wide application.  
The ONIOM method\cite{Humbel:1996km} was used, along with successive geometry updates on different fragments, to relax the 150,000 atom photosystem I trimer\cite{Canfield:2006rp}.  The resulting relaxed structure allowed quantitative identification of the location of key hydrogen atoms, as well as the insertion of missing atoms and correction of misaligned features from X-ray diffraction data.
A similar approach, with divide-and-conquer embedded within a multiscale modelling framework\cite{Vashishta:2006yj}, 
allows for massively parallel deployment. Tests have been carried out on\cite{Shimojo:2008gf}: alumina up to 11,796,480 atoms 
on 131,072 Blue Gene/L processors (though with a rather coarse grid of 0.4\,$a_0$); MD for 432 atoms of Rb,  
which yields good agreement to X-ray pair distribution data; MD on 512 atoms of graphene to look at vibrational spectra; 
and calculations on the electron affinity of a CdSe nanorod with 432 atoms.  
The same method has been used to simulate a thermite reaction with 1152 atoms (Al/Fe$_2$O$_3$), 
performing MD at 2000K for 5ps.  The key result was a metal-oxygen flip mechanism that enhances diffusivity\cite{Shimojo:2009qo}.

\begin{figure}[h]
  \centering
  \includegraphics[width=0.8\textwidth]{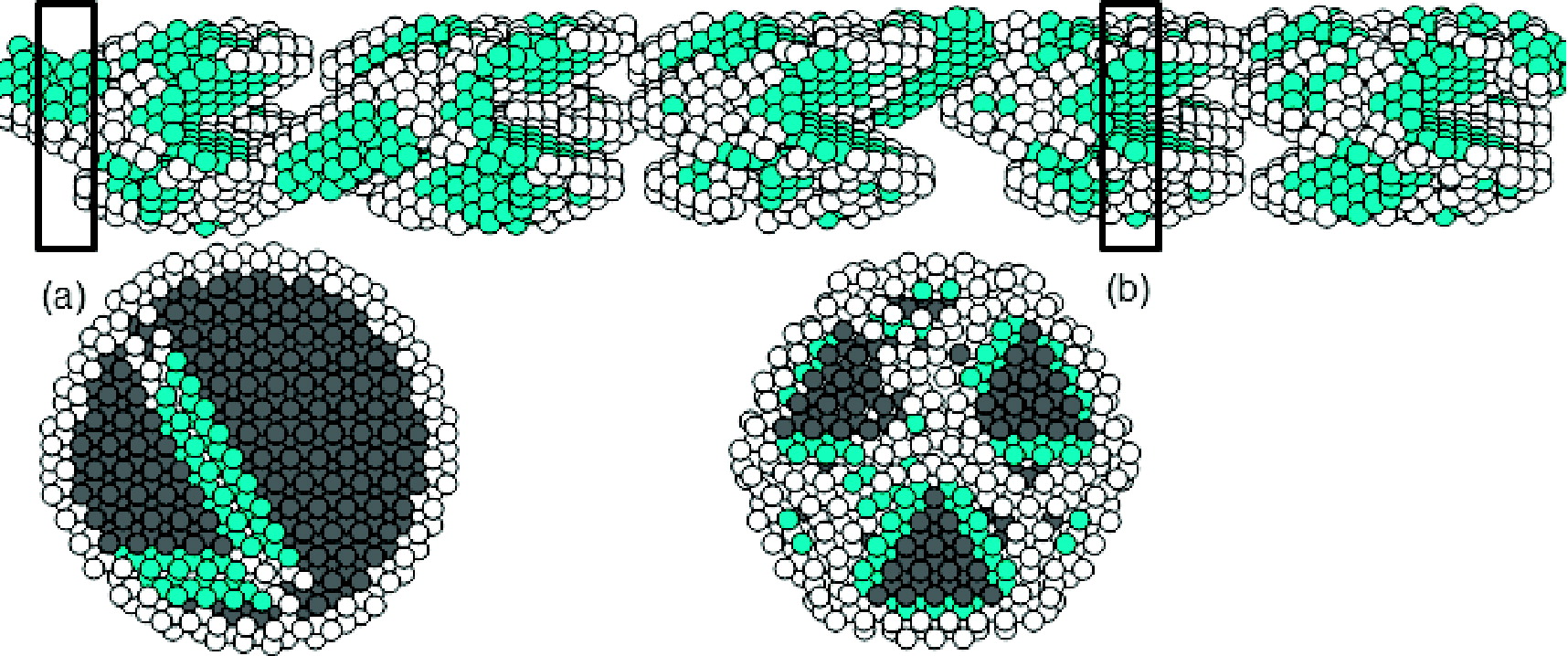}
  \caption{Slip planes of a 4 nm diameter Al nanowire formed upon loading
    one step past the elastic limit.  The top image shows the non-fcc
    interior atoms for the entire 20 nm long nanowire ([111] axis
    extending from left to right). The bottom two images show all the
    atoms in three-layer cross sections of the nanowire at (a) 4-10
    \AA\ and (b) 164-174\,\AA\ along the wire length. Light blue atoms have hcp
    structure; gray have fcc structure; and white have unknown
    structure. Reprinted with permission from L. Hung and
    E. A. Carter, J. Phys. Chem. C \textbf{115}, 6269 (2011)\cite{Hung:2011ly}. Copyright 2011 American Chemical Society.}
  \label{fig:AlNW}
\end{figure}

Orbital-free DFT (Sec.~\ref{sec:orbital-free-dft}) is generally used for metallic systems.  
The PROFESS code has been used extensively to model Al, such as the energetics and mobility of 
vacancies\cite{Ho:2007ul}, Al nanowires\cite{Ho:2007ul,Hung:2011ly}, and crack tips in Al\cite{Hung:2011pd}.
In the first example, vacancy formation and migration energies are calculated using cells 
up to 500 atoms for tri-vacancies. They find that while nearest-neighbour vacancy pairs are unstable, 
next-nearest-neighbour vacancy pairs are stable, and predict that vacancy clusters preferentially grow through
next-nearest-neighbour vacancies.
For the second example, Al nanowires (up to 16,770 atoms) with 1-8 nm diameter and up to 20nm long
are stretched to examine elastic and plastic behaviours. 
They find that the elastic deformation is qualitatively similar, but quantitatively different 
with respect to the diameter; thinner nanowires are more compressed relative to the bulk fcc structure.
On the other hand, clear size dependent behaviour is observed in the plastic region.
Partial slip as mechanism for plastic deformation is only seen for 4nm wires and above, while
amorphous deformation is seen in narrower wires.  These are
illustrated in Fig.~\ref{fig:AlNW}.
In the third example, they also calculate the system with the embedded atom model (EAM) method.
With the OFDFT they treated the system up to 7,800 atoms. 
They find qualitative differences of the OFDFT result to EAM for one orientation, in particular 
in how emission of twinning partial dislocations changes crack length; the difference is likely to be 
down to the surface energies being incorrect for EAM.  
There is also a quantitative difference in the onset of emission of
partial dislocations with load.

For the study of vacancies in aluminium, there is another OFDFT study which combined finite element modelling 
and coarse graining with OFDFT\cite{Radhakrishnan:2010yo}. 
In this method, far from the region of interest, they use the energy of distorted bulk Al based on the local environment 
to coarse-grain.  The authors find that 10$^3$--10$^4$ atoms are needed for convergence of monovacancy formation energy.  
Interestingly, they find a change of sign in interaction energy for the di-vacancies, when they increase the
size of the system from 32 atoms to larger.
However, this result does not seem to be consistent with the previous result by PROFESS.
In particular, the results for mono-vacancy formation energy, while giving almost the same value as 
the PROFESS result\cite{Ho:2007ul}, show different behaviour with cell size. 
There may be some effects from the coarse graining approximation,
however much of the difference comes from the boundary
conditions used, particularly for the electrostatics, and not from the
OFDFT component (which seems to be consistent between the two implementations)\cite{Hung:2011dp}.

The orbital minimisation method (Sec.~\ref{sec:direct-iter-appr}) has
been applied to molecular dynamics (MD) calculations of liquids.  It
has been used to test the effect of cell size on diffusivity of liquid
water\cite{Fernandez-Serra:2004jb}: the diffusivity of liquid water
calculated from DFT is too low  over long time scales.  Using MD
simulations with up to 128 water molecules, the study found no
appreciable size effects (between 32 to 128 molecules), nor any
effects due to parameters chosen (including the localisation radius
and basis chosen).  The final, detailed study used exact
diagonalisation with a DZP basis optimised for water, with basis set
superposition errors corrected.  

\begin{figure}[h]
  \centering
  \includegraphics[width=0.5\textwidth]{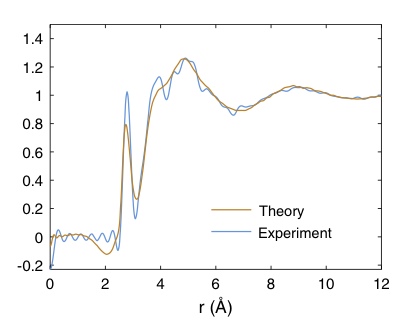}
  \caption{X-ray weighted radial distribution functions of liquid ethanol 
   calculated from ab initio O(N) molecular dynamics simulations (brown line) 
   and obtained from experiments(blue line). From Ref.~\cite{Tsuchida:2008ai}.}
  \label{fig:RDF_ethanol}
\end{figure}

The augmented OMM has been used to perform challenging and
scientifically relevant simulations with an $\mathcal{O}(N)$ code.  It
has been applied to molecular dynamics simulations of liquid
ethanol\cite{Tsuchida:2008ai}; this system requires large system sizes
(of the order of 10$^3$ atoms to ensure the structure of hydrogen
bonded chains is correct).  Using seven localisation regions for each
molecule (centred on the bond centres with one on the oxygen) with
radii of 12a$_0$ energy is conserved extremely well, and the 
computation cost is reduced by a factor of 4.6 compared with a conventional method.
The comparison
with experimental data is impressive: the radial distribution function
shown in Fig.~\ref{fig:RDF_ethanol} compares well; the self-diffusion coefficient is close
(8.2$\times$10$^{-6}$\,cm$^2$s$^{-1}$ for simulation compared to
1.1$\times$10$^{-5}$\,cm$^2$s$^{-1}$ from experiment); and the red
shift in O-H vibration mode due to hydrogen bonds in the liquid agrees
as well ($\sim$350cm$^{-1}$ in simulation compared to $320\pm
10$\,cm$^{-1}$ in experiment). 
The same method has been applied to calculations of Li$^+$ conductivity in
LiBH$_4$\cite{Ikeshoji:2011fk}.  Using 100 unit cells (each with two
LiBH$_4$ units, for a total of 1,200 atoms) and localisation regions
of 10~a$_0$ for Li and 14~a$_0$ for B,  the CPU time was reduced by a
factor of 2.5 with no reduction of accuracy.  The authors found that
the high ionic conductivity results from a metastable interstitial
site generated by a splitting of the units in the $c$ direction, which
is a new mechanism for ion conductivity.

The LNV method is implemented in both \textsc{Conquest} and
\textsc{onetep}; applications of \textsc{Conquest} have already been
described above in Sec.~\ref{sec:on-calc-biol} and
\ref{sec:order-n-dft-Ge-Si}.  A study of the formation energy of
vacancies in alumina with \textsc{onetep}\cite{Hine:2010fj}
($\alpha$-Al$_2$O$_3$) found, as would be expected, significant
simulation cell size dependence.  Using cells from 120 atoms up to
3240 atoms, a linear correlation between defect formation energy and
Madelung energy was found, and used to extrapolate an infinite cell
size result.  The simulation used a density matrix range of 24\,$a_0$
and radii of 8\,$a_0$ for the Wannier functions, though the
convergence of the energy with these values is not shown.  The same
approach has been used to calculate the electronic structure of
silicon nanorods with hydrogen passivation on the
surface\cite{Zonias:2010xh}.  The calculations used a kernel radius of
24\,$a_0$ and a Wannier function radius of 7\,$a_0$.  Nanorods
containing up to 1,648 atoms were modelled, though the geometry
optimisation was performed using an \emph{ab initio} tight binding
code.  The density of states and states around the band gap are found
using exact diagonalisation.

Krylov subspace methods (Sec.~\ref{sec:recursion}) have been applied
with tight binding Hamiltonians (rather than \emph{ab initio}) to study
cleavage in silicon. The cleavage of nano-crystalline
silicon\cite{Hoshi:2003oq} (for systems up to 100,000 atoms) showed
formation of (100) facets with a partial reconstruction,  and explored
the effects of size difference and competition between bulk and
surface energy.  Cleavage of bulk Si along (111)
plane\cite{Hoshi:2005nx} used 11,096 atoms (periodic in one direction
only). The formation of a (2$\times$1) reconstruction as well as steps
during cleavage was observed, with a similar interplay between
surface and bulk terms. 

As we have mentioned above, there are still not many scientific
applications using $\mathcal{O}(N)$ DFT methods.  However, as we can see from
the examples surveyed in this section, there are now real simulations
using $\mathcal{O}(N)$ DFT methods and the number of applications is growing
rapidly.  As the codes develop in robustness they will be more widely
used, and more experience will emerge in important areas such as
convergence and basis sets.  This will in turn encourage confidence in
results and more applications.  

\section{Conclusions}
\label{sec:conclusions}

In recent years, the trend in computing has been a dramatic increase
in the numbers of cores on a processor, and to massive numbers of
processors in high-performance computing centres; the recent emergence
of hundreds or thousands of processors on graphics processing units
has taken this trend even further.  As computational science is driven
by the hardware base available, we have seen a strong movement towards
real-space basis methods as a route to exploit hardware efficiently,
though the eigenvalue solvers have retained traditional scaling rather
than shifting towards linear scaling.

Since the first DFT linear scaling methods were proposed fifteen to
twenty years ago, it can legitimately be said that development in
$\mathcal{O}(N)$ methods has been slow.  However, in recent years,
real progress towards applications has been made (for instance, see
the proceedings of a CECAM workshop held in 2007\cite{Bowler:2008ly}).  The reasons
for this slow development are easily understood: linear scaling
introduces more parameters and sources of instability; standard
methods have developed rapidly in efficiency and robustness;
parallelising linear scaling methods is complex.

Nevertheless, we have reached a point where approximate linear scaling
methods (such as divide-and-conquer and orbital-free DFT, which are
hard to pursue to high accuracy) are producing real applications, as
discussed in Sec.~\ref{sec:applications}.  Methods which have the capacity for
exact behaviour are now at the point where they are more efficient
than conventional methods for systems over about a thousand atoms, and
are starting to demonstrate real applications with predictive capability.

Linear scaling codes can seem more complex to use than standard codes; at
the moment there is certainly less expertise in their use and
appropriate convergence criteria compared to standard methods.  One
of the key advantages of a plane-wave basis set is the simplicity of
convergence: the cutoff energy offers a single, simple variational
parameter; linear scaling methods which use a variational basis set
(such as blips or psincs) have a directly equivalent grid
spacing. Atomic-orbital or Gaussian based codes, whether conventional
or $\mathcal{O}(N)$ methods, by contrast, have basis sets which are
hard to converge in a variational manner.  Linear scaling methods do
have a cutoff both on the density matrix and
on the localised orbitals, though again cutoffs on local orbitals are
not unique to linear scaling codes.  We note that these codes do \emph{not}
in general include an integral in reciprocal space over the first
Brillouin zone, and therefore \emph{remove} the need to converge the
k-point mesh.  Linear scaling methods do introduce an extra
convergence criterion during minimisation (one for the solution of
the density matrix and another for the localised orbitals).  So we see
that, while apparently more complex, there is no reason why linear
scaling methods cannot become as widely used as conventional methods.
With more researchers working on practical calculations using these
methods, the field will rapidly improve, in the same way that the
field of \emph{ab initio} calculations changed in the years after the
Car-Parrinello method was first proposed.

Just as in conventional approaches, there is no consensus on
appropriate basis sets (in particular whether atomic-like or
variational is better).  However, this should not hold the field back:
a variational basis can be characterised in the same way as
plane-waves or real-space grids, and the same cutoffs can be used.
There is a widely-held assumption that double zeta (or double valence)
plus polarisation basis sets are required for atomic-like
representations, though without the consensus achieved in quantum
chemistry on different basis set qualities and likely errors.  This
should come with maturity of the field.  Even within variational basis
sets, however, it is not yet clear how many orbitals are required when
we impose localisation on the basis set functions: the smallest size,
the same as the number of bands, carries potential convergence
problems; a number equivalent to a minimal basis gives a good
compromise between computational effort and variational freedom; some
calculations have found that larger numbers (e.g. equivalent to the
valence orbitals plus polarisation orbitals) are needed for accuracy,
while other calculations achieve accuracy with the same number of
orbitals as valence orbitals.  As more calculations are performed, a
deeper understanding will emerge.

We now consider the challenges faced by the linear scaling community
and future routes for development.  Naturally these are personal
choices, but they certainly represent important problems in the area.
The first challenge is that of accuracy: how accurately can linear
scaling methods reproduce exact methods, and with what accuracy can
important quantities be calculated ? The question of accuracy (and
convergence) becomes more important when considering energy
differences in large systems, which is a natural area for applications
of linear scaling codes: tight convergence will be required, for
instance, when comparing different structures in biomolecules.  There
have been some investigations already in this area (showing energy
difference convergence for Ge nanostructures on
silicon\cite{Miyazaki:2008kx} and for solvated
DNA\cite{Otsuka:2008ri}, comparing absolute energy convergence between exact and linear
scaling methods for bulk silicon\cite{Skylaris:2007lp}, and comparing
performance for purification and DMM methods for a linear
alkane\cite{Rudberg:2011ul}, as well as work on error control within
linear scaling methods\cite{Rubensson:2005vs})
but more of these studies are needed.  It is becoming clear that good
accuracy can be achieved, though it naturally increases the
computational time required; this accuracy is important for the future
of these methods.

The second challenge is that of metallic systems: there is no clear
route to linear scaling solution for systems with low or zero gaps and
extended electronic structure.  While methods such as orbital-free DFT
offer some hope, and certainly allow large system sizes to be
addressed, they do not at present give sufficient accuracy for
quantitative calculations.  There are approaches with reduced scaling
over standard approaches, though they will slow down at some point.
It may be that these methods offer the best route forward either until
fully linear scaling methods are developed or until better
orbital-free functionals are found.

The third challenge, which is faced by codes in many other areas, is
to make efficient use of new computing architectures, particularly
given the shift towards petascale computing.  Real-space methods in
general are well placed to adapt to multi-core and GPU-based
computing, but the communications patterns developed for previous
generations of high-performance computing are not necessarily best
suited for novel architectures.  As linear scaling methods often use
specific approaches developed for the problem, it may be harder to
adapt than for other approaches using standard packages.  However,
this is an area where linear scaling codes can be extremely
successful, and the effort should be made.

The fourth challenge is to improve functionality while maintaining
linear scaling behaviour.  Recent years have seen DFT improved by
adding features such as exact exchange and dispersion forces (also known
as van der Waals interactions).  Methods have been proposed to
implement these with linear scaling, though as always adapting them to
the approach used (and parallelising while maintaining linear scaling
and efficiency) poses problems.  Time-dependent DFT is certainly
possible with linear scaling, as are certain parts of quantum Monte
Carlo and MP2 calculations, but approaches such as GW cannot be
adapted in their present form (though the influence and portability of
Wannier functions is making many interesting approaches viable).
Embedding of more accurate methods into DFT (or approximate DFT
methods) may become important, and linear scaling approaches are ideal
for this, starting from the locality of electronic structure.

The final, and in many ways most important, challenge is that of
applications to large systems, as already highlighted.  Long
timescales pose a challenge to all atomistic simulation methods, and larger
systems give longer length scales which typically are associated with
slower response times.  However, this is a generic problem.  Weak
forces such as hydrogen bonding and van der Waals interactions are
also important in large systems, particularly biological systems.
While there are semi-empirical and \emph{ab initio} methods for
calculating these forces, accuracy and testing is vitally important.
For the problem of preparing input for, and analysing output from,
calculations with millions of atoms, this community can learn valuable
lessons from the molecular dynamics community, and use existing tools
from that area.  Finally, as this is a new field, it will take time to
understand which physical properties can be calculated reliably and
efficiently.

This survey of recent developments in linear scaling approaches has
shown that we stand at a fascinating point. In 1999, in a previous
review of linear scaling methods, Goedecker wrote, ``Even with $\mathcal{O}(N)$
algorithms it will not be possible in the foreseeable future to treat
systems containing millions of atoms at a highly accurate
density-functional level using large basis sets, as would be necessary
for certain materials science applications''\cite{Goedecker:1999pv}.
However, we now have the first, true DFT calculations on millions of
atoms\cite{Bowler:2010uq}, and fully converged, highly accurate
results on systems of this size will be obtained in the very near
future.  With this capability, there is a rich variety of systems and
phenomena which can be tackled with accurate, linear scaling DFT
techniques.

\ack The authors gratefully acknowledge many years of support and
encouragement from Professor Mike Gillan, and the hard work and
contributions of the \textsc{Conquest} developers: Michiaki Arita,
Veronika Br\'azdov\'a, Ian Bush, Rathin Choudhury, Chris Goringe,
Eduardo Hern\'andez, Conn O'Rourke, Takao Otsuka, Alex Sena, Umberto
Terranova, Milica Todorovi\'c, Lianheng Tong, Antonio Torralba and
Lionel Truflandier.  They are also grateful for many helpful
suggestions and useful data from colleagues in the field, including
Emilio Artacho, Emily Carter, Vikram Gavini, Stefan Goedecker, Peter
Haynes, Nick Hine, Linda Hung, Sohrab Ismail-Beigi, Nicola Marzari,
Anders Niklasson, Mark Rayson, Emanuel Rubensson, Elias Rudberg,
Annabella Selloni, Chris-Kriton Skylaris, Jose Soler, David
Vanderbilt, Lin-Wang Wang and Weitao Yang.  This work was partly
supported by the Royal Society, a Grant-in-Aid for Scientific Research
from MEXT and JSPS, Japan

\section*{References}

\bibliographystyle{jpcm}
\bibliography{tag_O(N),TMadd}

\end{document}